\documentclass[journal]{IEEEtran}
  
\usepackage{epsfig}
\usepackage{graphicx}
\usepackage{amsmath}
\usepackage{amssymb} 
\usepackage{multirow}
\usepackage{array}

\usepackage{adjustbox}  
\usepackage{diagbox}
 
\usepackage[T1]{fontenc}    
\usepackage{url}            
\usepackage{booktabs}       
\usepackage{amsfonts}       
\usepackage{nicefrac}       
\usepackage{microtype}      
\usepackage{graphicx}
\usepackage{caption}
\usepackage{subcaption}
\usepackage[ruled]{algorithm2e}
\newcommand{\vecb}[1]{\mbox{\boldmath{$#1$}}}

\newcommand{\setn}[1]{\mbox{$\lbrace  #1 \rbrace$}}

\newtheorem{theorem}{Theorem} 
 
\newtheorem{lemma}{Lemma}

\usepackage{adjustbox}  
\usepackage{diagbox}
\usepackage{xcolor} 
\usepackage{xspace}
\DeclareMathOperator{\Tr}{Tr}
\newcommand*{\eg}{\textit{e}.\textit{g}.\@\xspace}
\newcommand*{\ie}{\textit{i}.\textit{e}.\@\xspace}
\newcommand{\etal}{\textit{et al}. }
\makeatletter
\newcommand*{\etc}{%
    \@ifnextchar{.}%
        {etc}%
        {etc.\@\xspace}%
}

\newcolumntype{K}[1]{>{\centering\arraybackslash}p{#1}}
\newcolumntype{L}[1]{>{\raggedright\arraybackslash}p{#1}}
\newcolumntype{R}[1]{>{\raggedleft\arraybackslash}p{#1}}
\usepackage{titletoc}
 
\newcommand*{\@rowstyle}{}

\newcommand*{\rowstyle}[1]{
  \gdef\@rowstyle{#1}%
  \@rowstyle\ignorespaces%
}
 
\usepackage{cite}
\usepackage{fancyhdr}
\usepackage{kantlipsum}

\fancypagestyle{plain}{
    \fancyhf{}
    \fancyhead[C]{Conference on \LaTeX}     
    \fancyfoot[L]{This is a notice}

}
\usepackage{eso-pic}

\ifCLASSINFOpdf
\else
\fi
\begin{document}
  
\title{An Adaptive Markov Random Field for Structured Compressive Sensing}

\author{Suwichaya Suwanwimolkul, Lei Zhang, Dong Gong,  Zhen Zhang, Chao Chen, Damith C. Ranasinghe,\\ Qinfeng Shi,

\thanks{Manuscript received June 6, 2018; revised September 30, 2018; accepted October 17, 2018. Date of publication XXX, 2018; date of current version XXX, 2018. This research is partially supported by grant funding from the Department of State Development under the Collaboration Pathways Program, Government of South Australia (No. CPP39) and ARC discovery grant (No. DP160100703). The research of  Asst. Prof. Chao Chen is partially supported by grants NSF IIS-1718802 and NSF CCF-1733866. Suwichaya Suwanwimolkul and Lei Zhang contributed equally to this work. {\emph{(Corresponding author: Suwichaya Suwanwimolkul.)}}}
	
\thanks{S. Suwanwimolkul, L. Zhang, D. Gong, Damith C. Ranasinghe, and Qinfeng Shi are with the School of Computer Science, The University of Adelaide, South Australia 5005, Australia (e-mail: suwichaya.suwanwimolkul@adelaide.edu.au; lei.zhang@adelaide.edu.au; edgong01@gmail.com; damith@cs.adelaide.edu.au; javen.shi@adelaide.edu.au).}
		
\thanks{Zhen Zhang is with the Department of Computer Science, The National University of Singapore, Singapore (e-mail: zhen@zzhang.org)}
		
\thanks{Chao Chen is with Department of Biomedical Informatics, Stony Brook University, Stony Brook, NY, USA (e-mail: chao.chen.cchen@gmail.com).} 

\thanks{This paper has a supplementary document available at http://ieeexplore.ieee.org., provided by the authors. Contact [suwichaya.suwanwimolkul@adelaide.edu.au] for further questions.} 
}

\markboth{SUBMISSION TO IEEE TRANSACTIONS ON IMAGE PROCESSING}%
{Suwanwimolkul \MakeLowercase{\textit{et al.}}: An Adaptive Markov Random Field for Structured Compressive Sensing}

\maketitle

\begin{abstract}
Exploiting intrinsic structures in sparse signals underpins the recent progress in compressive sensing (CS). The key for exploiting such structures is to achieve two 
desirable properties: generality (\ie, the ability to fit a wide range of signals with diverse structures) and adaptability (\ie, being adaptive to a specific signal). Most existing approaches, however, often only achieve one of these two properties. In this study, we propose a novel adaptive Markov random field sparsity prior for CS, which not only is able to capture a broad range of sparsity structures, but also can adapt to each sparse signal through refining the parameters of the sparsity prior with respect to the compressed measurements. To maximize the adaptability, we also propose a new sparse signal estimation where the sparse signals, support, noise and signal parameter estimation are unified into a variational optimization problem, which can be effectively solved with an alternative minimization scheme. Extensive experiments on three real-world datasets demonstrate the effectiveness of the proposed method in recovery accuracy, noise tolerance, and runtime.

\end{abstract}


\IEEEpeerreviewmaketitle 

\section{Introduction} 
\IEEEPARstart{C}{ompressed} sensing (CS) is to recover a $k$-sparse signal  \mbox{$\vecb{x} \in \mathbb{R}^N$} from  $M$ linear measurements $\vecb{y} = \vecb{Ax}$ \cite{Donoho2006}, where $\vecb{A} \in \mathbb{R}^{M \times N}$ represents a random transformation matrix ($M < N$). Recent CS algorithms focus primarily on reducing the number of measurements $M$. Standard state-of-the-art CS algorithms can recover a $k$-sparse signal $\vecb{x}$ from  $\mathcal{O}(k\log N/k)$ noisy measurements\cite{Donoho2006}. To further reduce the number of measurements required, many researches started to exploit the structure (\ie,  correlations) of the sparse coefficients.  \\  

In real applications, sparse signals often exhibit diverse structures, and the structures may vary within the same dataset. To efficiently exploit the intrinsic structures of sparse signals, a desirable sparsity model should possess two important properties, namely, {\bf generality} (\ie, the ability to fit a wide range of signals with diverse structures) and {\bf adaptability} (\ie, being adaptive to a specific sparse signal). Two dominant classes of sparsity models that have been studied~\cite{baraniuk2010low} include deterministic structured sparsity models~\cite{yuan2006model,Eldar2008,Eldar2009,Kowalaski2009,Zhao2009,jacob2009group,
Baraniuk1998, Carin2009Wavelet,duarte2008wavelet,la2006tree,AddedResubRe2,
huang2011learning,bach2012optimization, hegde2015nearly, zhou2016technical} 
 and probabilistic structured sparsity models~\cite{Cevher12009, Wolfe2004,Garrigues2008,Peleg2012,Dremeau2012,  yu2012bayesian, yu2015,Fang2015,  Fang2D2016,Wang2015ISAR,AdaptiveCluster1,AdaptiveCluster2}. A brief review is in Section~\ref{section:related}. 

Deterministic structured sparsity models such as group sparsity~\cite{yuan2006model,Eldar2008,Eldar2009,Kowalaski2009,Zhao2009,jacob2009group}, hierarchical sparsity~\cite{Baraniuk1998, Carin2009Wavelet,duarte2008wavelet,la2006tree,AddedResubRe2} and graph sparsity~\cite{huang2011learning,bach2012optimization, hegde2015nearly, zhou2016technical} models often assume prior knowledge on the geometrical structure of sparse signals, and restrict the feasible set to those signals that comply with the assumed geometrical model. Therefore, these methods tend to exclude all the signals that violate the assumed structure. To avoid exclusion of candidate sparse signals and achieve small sample complexity, Cevher \etal~\cite{Cevher12009}  proposed the concept of Probabilistic RIP, and used Markov random fields (MRFs) to model 
structure of sparse signals. This opens up a new line of works~\cite{Wolfe2004,Garrigues2008,Cevher12009,Peleg2012,Dremeau2012}, termed probabilistic structured sparsity models. With the high  generality and expressiveness of the MRF, these methods often achieve state-of-the-art performance. However, the MRF used in these approaches is trained with data; thus, it lacks the adaptability to adjust to new signal structures. To improve the adaptability, recent approaches~\cite{  yu2012bayesian, yu2015,Fang2015, Fang2D2016,Wang2015ISAR,AdaptiveCluster1,AdaptiveCluster2} exploit cluster sparsity (extension from group sparsity) where they assume that the non-zero signal coefficients are grouped as clusters. In particular, the works~\cite{Wang2015ISAR,AdaptiveCluster1,AdaptiveCluster2} use  MRFs, but these MRFs contain only pairwise potentials. Similar to the other cluster sparsity models, they assume limit signal structures which is fixed and cannot adapt for the new signal structures. \\

\begin{figure}[t]  
\centering{\includegraphics[ width =3.25in,trim=1.2cm  6cm  1.5cm 1.2cm,clip]{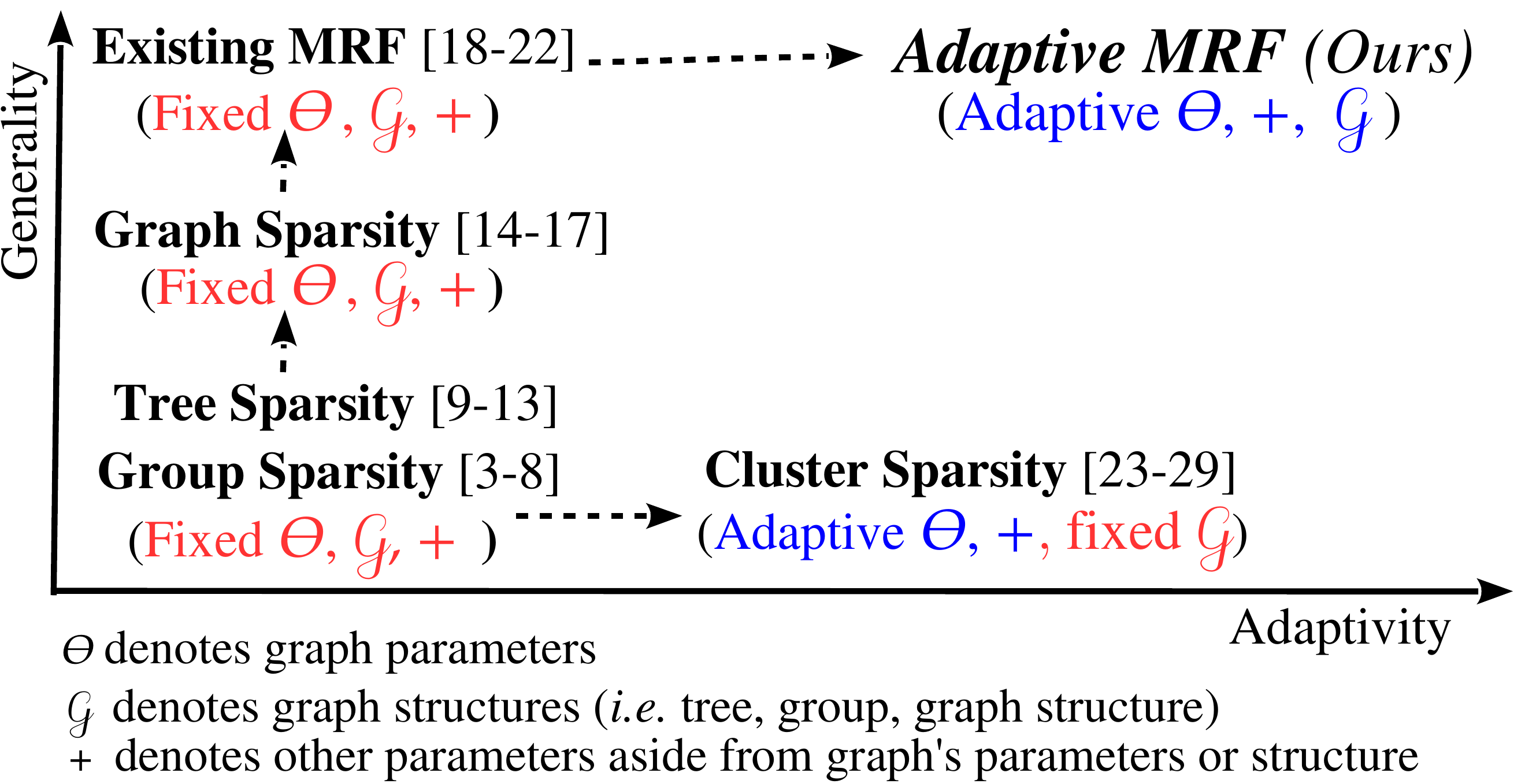} }  
\caption{Generality and adaptability in each line of work} \label{Fig_comparison}
\end{figure}

We present a novel graphical compressive sensing model that offers both {\bf generality} and {\bf adaptability}. We leverage the MRF~\cite{Wolfe2004,Garrigues2008,Cevher12009,Peleg2012,Dremeau2012} as the sparsity prior since it has been proven to be general and expressive enough to model various signal  structures. Unlike existing MRF-based methods~\cite{Wolfe2004,Garrigues2008,Cevher12009,Peleg2012,Dremeau2012}, our method uses a Bayesian principle to realize  \textit{adaptive MRF}  whose \textit{parameters} and  \textit{underlying graph} are updated according to measurements. Thus, the MRF parameters are adapted to represent the underlying structure of the sparse signals. Unlike the clustered  sparsity models~\cite{yu2012bayesian, yu2015,Fang2015, Fang2D2016,Wang2015ISAR,AdaptiveCluster1,AdaptiveCluster2}, the underlying  graph of our method can be updated for new signal structures.  Figure~\ref{Fig_comparison} summarizes  generality and adaptability of the proposed adaptive MRF in comparison with other approaches. Our adaptive MRF posses both the generality and adaptability which are inherited from the probabilistic MRF and the adaptive mechanism.

Figure~\ref{FigResult} demonstrates the improved performance by employing an adaptive MRF versus a fixed MRF as a prior in recovering a sample MNIST image. The unary and pairwise potentials of the adaptive MRF are refined to the structure of the digit number 2 over iterations. As the adaptive MRF being refined, the quality of reconstructed images improves. On the contrary, a fixed MRF that is obtained from  training process captures a universal pattern of the training images. The value of fixed MRF potentials appears in round shape. As the fixed MRF cannot adapt for the signal structure, the adaptive MRF achieves higher recovery quality both numerically and visually. \\

To exploit an MRF as a prior in signal recovery, most existing MRF  methods such as~\cite{Wolfe2004,Garrigues2008,Peleg2012} are based on the non-recursive two-step approach~\cite{Wolfe2004} that, first, estimates the support and, then, estimates the sparse signal~\cite{BayPersuit}. However, this can cause high computational time. Also, the error in the first step can propagate to the second step and can not be minimized later. Moreover, these methods employ homogeneous noise and signal parameters from training data, which may not well represent the actual noise and signal parameters. 

To address these problems, we propose to estimate sparse signal, support, and noise and sparse signal parameters jointly and iteratively, given an adapted MRF. However, by doing this, the whole signal estimation becomes a non-convex optimization problem over discrete and continuous variables--- support, sparse signals, noise and signal parameters (see Eq.~\eqref{sec3.1})---which is very difficult to solve in general. To tackle this non-convex problem, we propose to apply a latent Bayes model~\cite{Wipf2011, LeiZhang} to provide a new formulation (see Eq.\eqref{eq1.4}) where signal structure is considered. This brings in closed-form solutions for estimating sparse signal, noise and signal parameters. To  solve for the support efficiently, we propose to approximate non-linear, pairwise potentials derived from the new formulation Eq.\eqref{eq1.4} into linear, unary potentials which can be solved efficiently with any off-the-shelf graphical model inference tools.   \\

Therefore, we propose to leverage the adaptability of the MRF and develop a new sparse signal estimation to obtain the sparse signals with the adapted MRF.    We highlight our contributions as follows:

 \begin{figure}[t]  
\setlength{\tabcolsep}{2pt}  
 \begin{tabular}{K{6cm}K{2cm}}
\subcaptionbox{{\centering{Intermediate and final results}}. \label{FigGraph:adpt}} {
\hspace{1cm} 
\setlength{\tabcolsep}{2pt} 
\renewcommand{\arraystretch}{1}
\hspace{-2cm} \begin{tabular}{L{0.5in}cccc} 
&\multicolumn{4}{c}{\small{ {Adaptive MRF}}}\\
&\scriptsize{  1$^{st}$} & \scriptsize{ 2$^{nd}$} & \scriptsize{ 3$^{rd}$} & \scriptsize{ Final }    \\
\raisebox{+1.0\height}{\parbox{1.2cm}{\centering\scriptsize{MRF's Unary potential} } }  &  
{\includegraphics[width = 0.45 in]{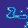}} &  
{\includegraphics[width = 0.45 in]{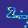}} &
{\includegraphics[width = 0.45 in]{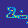}}  &
{\includegraphics[width = 0.45 in]{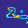}}  
 \\ 
\raisebox{+1\height}{\parbox{1.2cm}{\centering\scriptsize{MRF's Pairwise potential}}}     &
{\includegraphics[width = 0.45 in]{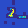}} &
{\includegraphics[width = 0.45 in]{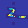}} &
{\includegraphics[width = 0.45 in]{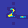}}&
{\includegraphics[width = 0.45 in]{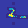}} \\ 
\raisebox{+1\height}{\parbox{1.2cm}{\centering\scriptsize{Recon. Images}}}     &
{\includegraphics[width = 0.45in]{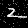}} &
{\includegraphics[width = 0.45 in]{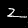}} &
{\includegraphics[width = 0.45 in]{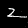}} &
{\includegraphics[width = 0.45 in]{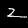}} \\ 
\raisebox{+1\height} {\parbox{1.2cm}{\centering\scriptsize{Error maps}}} &
{\includegraphics[width = 0.45 in]{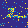}} &  
{\includegraphics[width = 0.45 in]{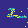}} &
{\includegraphics[width = 0.45 in]{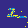}}  &
{\includegraphics[width = 0.45 in]{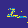}} 
 \\ 
\scriptsize{Accuracy} &\scriptsize{28.32~dB} &  \scriptsize{32.70~dB}  &  \scriptsize{32.85~dB} & \scriptsize{\textbf{32.95~dB}} \\
\raisebox{+1\height} {\scriptsize{(PSNR)}} & \multicolumn{4}{c}{\centering\scriptsize{(The higher PSNR the better)}}
\end{tabular}       }
&
 \subcaptionbox{{\centering {Final result}}. \label{FigGraph:fixed}} {
 \setlength{\tabcolsep}{2pt} 
\renewcommand{\arraystretch}{1}
\hspace{-0.8cm}
\begin{tabular}{K{2.5cm}} 
\small{{Fixed MRF}} \\
\\
 {\includegraphics[width = 0.45 in]
{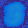}}   
 \\ 
{\includegraphics[width = 0.45 in]{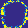}} 
\\
{\includegraphics[width = 0.45 in]{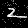}}
 \\  
 {\includegraphics[width = 0.45 in]
{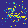}}   
 \\ 
\scriptsize{{30.15~dB}} \\
\\
\end{tabular}       }
 \end{tabular}   
\centering{ \caption{Comparison of the adaptive MRF vs. the fixed MRF on a sample of MNIST data. The top to the bottom rows: (i) the unary potential, (ii) the sum of pairwise potentials of adjacent pixels, (iii) reconstructed images, and  (iv) error maps. Our adaptive MRF is more tuned to digit number 2.  }  \label{FigResult} }   
\end{figure}

\begin{enumerate}

\item \textit{An Adaptive MRF} framework to adaptively estimate both the parameters and the underlying graph of the MRF to fit any signal structure. The improved performance due to adaptive MRF is demonstrated in Section~\ref{Effect_AdaptiveMRF}. \\

\item \textit{New sparse signal estimation algorithm} to jointly and iteratively estimate the support and the sparse signal, noise and signal parameters based on solving a new formulation Eq.\eqref{eq1.4}. Our algorithm offers improved accuracy and runtime over the existing methods~\cite{Peleg2012,Garrigues2008} (see Section~\ref{Adaptive_imp}).\\

\item \textit{Theoretical result} to demonstrate the essence of adaptive support prior in Section~\ref{Theoretical_Result}, \ie if the adapted support prior converges to the distribution of the test signal, then the feasible set is guaranteed to contain the test signal. Thus, the sample complexity of $\mathcal{O}(k)$ can be achieved.   \\

\item  The proposed method yields the state-of-the-art performance (see Section~\ref{Result_Comparison}) based on the evaluation with three benchmark datasets: i) MNIST, ii) CMU-IDB, and iii) CIFAR-10 in four different signal representations, \ie spatial, wavelet, discrete cosine transform (DCT), and principal component analysis (PCA) representations.   
  
\end{enumerate}

\section{Related work}~\label{section:related}
Here we review existing deterministic and probabilistic structured sparsity models that are most relevant to our work.  

\subsection{Deterministic structured sparsity} 
  
Deterministic structured sparsity models often assume the prior knowledge on the geometric structure of sparse signals~\cite{baraniuk2010low}. Three broad classes of models are developed as follows.  
 
\textbf{Group/Block sparsity models}~\cite{yuan2006model,Eldar2008,Eldar2009,Kowalaski2009,Zhao2009,jacob2009group} assume that signal coefficients in one group/block have to be either all zero or all non-zero. This property has been enforced by $l_1/l_2$ norms in early works and extended to overlapping group-sparsity. 
	
\textbf{Hierarchical sparsity models}~\cite{Baraniuk1998, Carin2009Wavelet,duarte2008wavelet,la2006tree,AddedResubRe2} represent signal coefficients as trees. For example, the wavelet transform of a piecewise smooth signal often exhibits the tree structure where a zero parent node implies zero offspring nodes~\cite{Baraniuk1998, Carin2009Wavelet,duarte2008wavelet,AddedResubRe2}. Another example is the $k$-sparse rooted sub-tree model~\cite{la2006tree}, where only non-zero element nodes form a sub-tree. 
	
\textbf{Graph sparsity models} ~\cite{huang2011learning,bach2012optimization, hegde2015nearly, zhou2016technical} organize signal coefficients in a general graph able to represent various types of sparsity patterns, including the above group and hierarchical sparsity models. Initially, graph sparsity models are employed as sparsity-induced regularization~\cite{huang2011learning,bach2012optimization}. Recently, a weighted graph sparsity model~\cite{hegde2015nearly, zhou2016technical} is employed to define the solution space of candidate supports.  \\

However, these deterministic structured sparsity models exclude all the signals that violate the assumed geometrical structure. Moreover, these models cannot adapt for new signal structures. In contrast, our adaptive MRF is able to fit a wide range of the sparse signal structures and can adapt for new signal structures.

\subsection{Probabilistic structured sparsity}
 The existing  work~\cite{Wolfe2004,Garrigues2008,Cevher12009, Peleg2012, Dremeau2012} employs Markov random fields (MRFs), a typical graphical model, able to represent various structures. The MRF is used to model the structure in signal support, and is employed as a prior in sparse signal recovery. Given the MRF, the support estimation requires an exhaustive search over all possible sparsity patterns, which is non-trivial. The existing work proposed different approaches to recover the sparse signals. Some of them \cite{Cevher12009,Dremeau2012} impose additional conditional independence assumptions to facilitate sparse recovery process. To estimate the support, the work~\cite{Cevher12009} exploits graph cut; meanwhile, the work \cite{Dremeau2012} adopts a mean-field approximation. To avoid imposing the assumptions, other approaches~\cite{Wolfe2004,Garrigues2008, Peleg2012} resort to non-recursive two-steps  approach, that is, the support is first estimated given the MRF, and then, the sparse signal is estimated based on the resulting support. Heuristic~\cite{Peleg2012} and stochastic Markov Chain Monte Carlo~\cite{Wolfe2004,Garrigues2008} approaches are proposed to estimate the support efficiently. All these methods learn the MRF from training, but they do not have the mechanism to adapt the MRF for different signal structures. On the contrary, the adaptive MRF is much more tuned to a specific signal structure.  

\subsection{Cluster Structured Sparsity}
Recent works~\cite{ yu2012bayesian, yu2015,Fang2015, Fang2D2016,Wang2015ISAR,AdaptiveCluster1,AdaptiveCluster2} improve the group sparsity model by allowing the sparsity model parameters to adaptively update according to measurements. These works  
often exploit two-state mixture models such as Gaussian-Gamma \cite{Fang2015, Fang2D2016} 
or beta-Bernoulli models~\cite{ yu2012bayesian, yu2015} which enable close-form updates. The works~\cite{Wang2015ISAR,AdaptiveCluster1,AdaptiveCluster2} exploits MRFs, but the MRFs contain only  the pairwise potentials. Variational Bayesian expectation maximization methods are employed to recover  sparse signal and estimate the model parameters. Nevertheless, these methods rely on a fixed neighborhood graph where each node is connected to all adjacent nodes in its neighborhood.  \\

In contrast to these methods, our adaptive MRF inference framework is based on the MRFs that can represent various types of sparse signals, and the underlying structure of the MRF can be automatically adjusted for new signal structures.

\section{Graphical compressive sensing}

In this study, we capture the structure of sparse signal $\mbox{\boldmath{$x$}}$ by modelling its support explicitly.
Let $\mbox{\boldmath{$s$}}  \in \lbrace -1,1 \rbrace^N$ indicate the support of $\mbox{\boldmath{$x$}}$  such that $s_i=1$ when $x_i\neq 0$ and $s_i=-1$ when $x_i=0$. Let $\mbox{\boldmath{$x$}} _{s} \in {\mathbb{R}^{k}}$ denote the non-zero coefficients of the $k$-sparse $\mbox{\boldmath{$x$}}$. Our goal is to estimate  $\vecb{s}$ and $\vecb{x}_s$ from the linear measurements $\vecb{y}$ corrupted by additive noise $\mbox{\boldmath{$n$}}$ as follows,
 \begin{equation}
\begin{aligned}
  {\vecb{y}}  =     \vecb{A}_s  \vecb{x}_s +  \vecb{n}.
\end{aligned}\label{sec2.1} 
\vspace{-0.1cm}
\end{equation}
Here $\mbox{\boldmath{$A$}}_s$ is the matrix with $k$ columns selected from the matrix $\vecb{A}$ according to $\vecb{s}$, and
 $\mbox{\boldmath{$n$}}$ is the Gaussian white noise, \ie, $\mbox{\boldmath{$n$}} \sim \mathcal{N}(0, \sigma_{n}\mbox{\boldmath{$I$}})$ where $\sigma_{n}$ is the noise variance and $\mbox{\boldmath{$I$}}$ denotes an identity matrix with a proper size. The corresponding likelihood over $\mbox{\boldmath{$y$}}$ can thus be formulated as
 \begin{equation} 
\begin{aligned}
& p(\mbox{\boldmath{$y$}}|\mbox{\boldmath{$x$}}_s,\mbox{\boldmath{$s$}}  ; \sigma_{n}) = \mathcal{N}(  \mbox{\boldmath{$A$}}_{s} \vecb{x}_s ,\sigma_{n} \vecb{I}).
\end{aligned}\label{sec2.2}
\end{equation}
Each observed measurement $y_i$ can be seen a noisy linear combination of non-zero sparse signal coefficients that are projected on measurement matrix atoms. The interdependencies among coefficients can be modelled through the prior defined on $\vecb{s}$. Specifically, we impose a graphical sparsity prior on $\mbox{\boldmath{$x$}}_s$ and $\mbox{\boldmath{$s$}}$  {(Section~\ref{Grap_sparsity})}. Subsequently, we show how to recovery sparse signal $\vecb{x}$ from the measurements $\vecb{y}$ by our new adaptive MRF inference framework  {(Section~\ref{Measurement_adaptive_inf})}.

\subsection{Graphical sparsity prior} \label{Grap_sparsity}
MRFs can represent a wide range of structures, \eg block and tree structures~\cite{wang2013markov}, to capture complex dependency between sparse signal coefficients by defining the probabilistic distribution over an undirected graph~\cite{Peleg2012,Ren2013,koller2009probabilistic,wainwright2008graphical}. Let $\mathcal{G} = (\mathcal{V},\mathcal{E})$ represents the undirected graph where $\mathcal{V}$ and $\mathcal{E}$ are the sets of the  nodes and the  undirected edges. $\vecb{\Theta}_{\mathcal{G}}  = \left\lbrace \mathcal{W}_{i}, \mathcal{W}_{i,j} \right \rbrace_{ i \in  \mathcal{V}, (i,j) \in \mathcal{E}}$  represents the set of potential  parameters associated with the probability distribution of the MRF.  We impose a graphical sparsity prior on $\mbox{\boldmath{$x$}}_s$ and $\mbox{\boldmath{$s$}}$ as follows. 

First, we define the prior of support $\mbox{\boldmath{$s$}}$ based on MRFs. \textit{Boltzmann machine (BM)} is commonly used as the probability distribution of the MRF to flexible represent signal structure. Each coefficient $ s_i$ of the support $\mbox{\boldmath{$s$}}$ is mapped to each node $ i \in \mathcal{V}$. Given the graph $\mathcal{G}$, the probability of the support $p( \mbox{\boldmath{$s$}}; \mbox{\boldmath{$\Theta$}}_{\mathcal{G}} )$ can be represented as follows, with a normalization constant $Z$,
\begin{equation}
\begin{aligned}
  \frac{1}{Z} \exp  \left(  \sum_{i\in \mathcal{V}} \mathcal{W}_{i}s_i + \sum_{(i,j)\in \mathcal{E}} \mathcal{W}_{i,j} s_i s_j  \right). 
\end{aligned}    \label{eq.2.6}
\end{equation}
where $\mathcal{W}_{i}$ defines bias (\eg, confidence) potential to each $s_i$; while $\mathcal{W}_{i,j}$ characterizes pairwise interaction between two variable nodes, \eg $ \xi^{(i,j)}$ weights dependency between $s_i,s_j$.   
  
In addition, we assume $\vecb{x}$ comes from a Gaussian distribution $\mathcal{N}(\vecb{0},\vecb{\Sigma}_x)$ where $\vecb{\Sigma}_x$ is a diagonal matrix. Given $\vecb{s}$, the probability of non-zero coefficients  is defined as 
 \begin{equation}
\begin{aligned}
p(\vecb{x}_s|\vecb{s})=\mathcal{N}(\vecb{0},\vecb{\Sigma}_{x,s}) 
\end{aligned}    \label{Ch3_Eq_non_zeroGauss}
\end{equation}
where $\vecb{\Sigma}_{x,s}$ is a diagonal matrix whose entries are chosen from $\vecb{\Sigma}_x$ according to $\vecb{s}$. This is to consider the underlying structure of the sparse signals. Then, $p(\vecb{x}_s|\vecb{s}) p( \mbox{\boldmath{$s$}}; \mbox{\boldmath{$\Theta$}}_{\mathcal{G}} )$ forms the graphical sparsity prior in this study. Therefore, $\vecb{\Sigma}_{x,s}$ is assumed to be a diagonal matrix to reduce computation in estimating the spares signal variance. Note that most of the previous works~\cite{Wolfe2004,Garrigues2008,Cevher12009, Peleg2012, Dremeau2012} also employ a similar assumption, but the sparse signal variance is obtained from training data.

\subsection{Adaptive Markov random fields inference} \label{Measurement_adaptive_inf}
Provided that these optimum parameters $\hat{\sigma}_n$, $\hat{\mbox{\boldmath{$\Sigma$}}}_{x,s}$,  $\hat{\mbox{\boldmath{$\Theta$}}}$, and $\hat{\mathcal{G}}$ are given beforehand, the latent $\mbox{\boldmath{$x$}}_s$ and $\mbox{\boldmath{$s$}}$ can be estimated by solving a maximum a posteriori (MAP) problem as
 \begin{equation} \label{eq.2.7}
\begin{aligned}
\max\limits_{\mbox{\boldmath{$x$}}_s, \vecb{s}} p(\mbox{\boldmath{$x$}}_s, \vecb{s}|\mbox{\boldmath{$y$}}) \propto p(\mbox{\boldmath{$y$}}|\mbox{\boldmath{$x$}}_s, \mbox{\boldmath{$s$}};\hat{\sigma}_n)p(\mbox{\boldmath{$x$}}_s| \vecb{s}; \hat{\mbox{\boldmath{$\Sigma$}}}_{x,s})p(\mbox{\boldmath{$s$}};\hat{\mbox{\boldmath{$\Theta$}}}_{\mathcal{G}}).
\end{aligned}
\end{equation}
These parameters are often unknown in real applications. However, some existing work obtains the MRF parameters and its underlying graph from training data \cite{Garrigues2008,Peleg2012,Dremeau2012}. Although these parameters can well represent the common characteristics among training data, they cannot adapt to represent the characteristics of the testing data. \\

To address this,  we propose the adaptive MRF inference framework where all the  parameters--- the MRF parameters and its underlying graph  ${\mbox{\boldmath{$\Theta$}}}_{\mathcal{G}}$ and ${\mathcal{G}}$, and noise and signal parameters ${\sigma}_n$ and ${\mbox{\boldmath{$\Sigma$}}}_{x,s}$--- are adaptively updated according to given measurements. Therefore, our objective is to estimate the these parameters---${\sigma}_n$, $ {\mbox{\boldmath{$\Sigma$}}}_{x,s}$, and $ {\mbox{\boldmath{$\Theta$}}}_{\mathcal{G}}$---by solving
 \begin{equation} \label{eq.2.8}
\begin{aligned}
&\max\limits_{ {\vecb{s}}, \sigma_n, \mbox{\boldmath{$\Sigma$}}_{x,s},   \mbox{\boldmath{$\Theta$}}_{\mathcal{G}} } p(\mbox{\boldmath{$y$}} |  {\vecb{s}}, \sigma_n, \mbox{\boldmath{$\Sigma$}}_{x,s}, \mbox{\boldmath{$\Theta$}}_{\mathcal{G}}) {\kern 2pt} \propto  \\
&\int
p(\mbox{\boldmath{$y$}}|\mbox{\boldmath{$x$}}_s,  \hat{\vecb{s}},\sigma_n)p(\mbox{\boldmath{$x$}}_s| \hat{\vecb{s}}, \mbox{\boldmath{$\Sigma$}}_{x,s})p( \hat{\vecb{s}}|{\mbox{\boldmath{$\Theta$}}}_{\hat{\mathcal{G}}})\mathrm{d} \mbox{\boldmath{$x$}}_s,
\end{aligned}
\end{equation} 
which intrinsically maximizes the likelihood of measurements over all model parameters as well as the support. Solving Eq.~\eqref{eq.2.8} directly is intractable. To circumvent this problem, we reduce Eq.~\eqref{eq.2.8} into two subproblems as follows.  \\
 
\begin{figure}[t] 
\vspace{-0.3cm}
\centering{\includegraphics[ width =2in,trim=1.00cm  11cm  1cm 1.cm,clip] {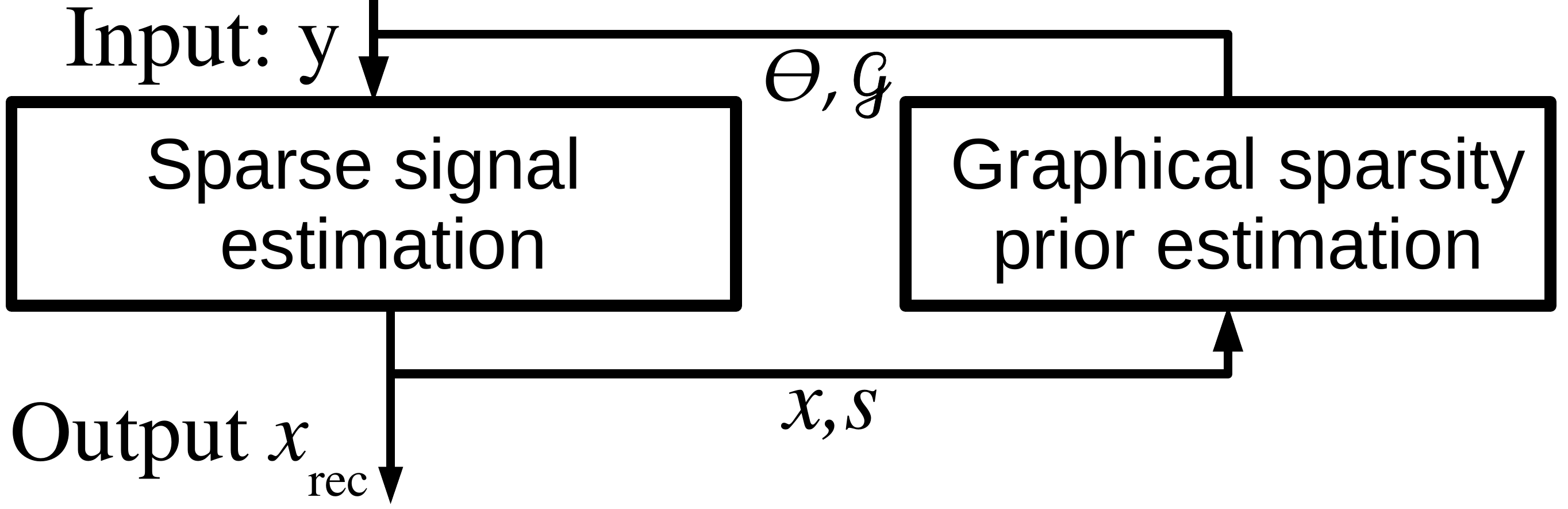} }  
 \caption{Visualization of our  Adaptive MRF inference framework } \label{Fig_model_visualization} 
\end{figure}

\subsubsection{Sparse signal estimation}  \label{SparseRecovery}
Given the graphical model parameters and its underlying graph, $\hat{\vecb{\Theta}}_{\hat{\mathcal{G}}}$ and  $\hat{\mathcal{G}}$, we first infer other parameters from the measurements by solving 

\begin{equation} 
\begin{aligned} \label{eq.2.9}
& \max\limits_{\mbox{\boldmath{$s$}}, \sigma_n, \mbox{\boldmath{$\Sigma$}}_{x,s}} p(\mbox{\boldmath{$y$}}|\vecb{s},\sigma_n, \mbox{\boldmath{$\Sigma$}}_{x,s}) \propto \\
& \int p(\mbox{\boldmath{$y$}}|\mbox{\boldmath{$x$}}_{s}, \mbox{\boldmath{$s$}}, \sigma_n)p(\mbox{\boldmath{$x$}}_{s}|\vecb{s},\mbox{\boldmath{$\Sigma$}}_{x,s})p(\mbox{\boldmath{$s$}}|\hat{\mbox{\boldmath{$\Theta$}}}_{\hat{\mathcal{G}}})\mathrm{d} \mbox{\boldmath{$x$}}_s.
\end{aligned}
\end{equation} 

The optimization problem in Eq.~\eqref{eq.2.9} can be equally reformulated as~\cite{Peleg2012,Tropp2007} :  
\begin{equation}
\begin{aligned} 
&\min\limits_{\vecb{s}, \sigma_n , \vecb{\Sigma}_{x,s}  }
-\log\int { p(\vecb{y}| \vecb{x}_s, \vecb{s},  {\sigma}_{n} ) p(\vecb{x}_s| \vecb{s} ,  \vecb{{\Sigma}}_{x,s} ) p(\vecb{s}; \hat{\vecb{\Theta}}_{\hat{\mathcal{G}}})} \mathrm{d} \mbox{\boldmath{$x$}}_s \equiv \\
& \frac{1}{2} \vecb{y}^T (\sigma_n + \vecb{A}_s\vecb{\Sigma}_{x,s}\vecb{A}^T_s )^{-1} \vecb{y}    +  \frac{1}{2} \log | \sigma_n \vecb{I} + \vecb{A}_s \vecb{\Sigma}_{x,s}\vecb{A}_s^T| \\
& - \log{p(\vecb{s}|\hat{\vecb{\Theta}}_{\hat{\mathcal{G}}})}. 
 \end{aligned}  \label{sec3.1}
 \end{equation} 

The existing work~\cite{Wolfe2004,Garrigues2008, Peleg2012} employed the two step-non-recursive approach~\cite{Wolfe2004,BayPersuit}: first, they attempt to solve Eq.~\eqref{sec3.1} for the support. Given the resulting support, they still have to estimate $\vecb{x}$ from Eq.~\eqref{eq.2.8}. However, this can cause error accumulation problem since the error in the first step cannot be minimized in the second step.  Moreover, the support estimation problem in Eq.~\eqref{sec3.1} is non-convex over discrete and continuous variables---support, sparse signals, noise and signal parameters---which is difficult to solve in general. Even fixing $\vecb{s}$, the remaining problem of Eq.~\eqref{sec3.1} is still non-convex, and there are no closed-form solutions for $\sigma_n$ and $\vecb{\Sigma}_{x,s}$. Therefore, these works~\cite{Wolfe2004,Garrigues2008, Peleg2012} resorts to employ the homogeneous noise and signal parameters from training data. \\ 
  
To tackle this non-convex problem, we propose to use a strict upper bound of Eq.~\eqref{sec3.1} based on a latent Bayes model~\cite{Wipf2011, LeiZhang}:

\begin{equation}
\begin{aligned}
& \vecb{y}^T (\sigma_n + \vecb{A}_s\vecb{\Sigma}_{x,s}\vecb{A}^T_s )^{-1} \vecb{y} \\ 
& =   \inf\limits_{\vecb{x}_x} 
 \frac{1}{  \sigma_n}      ( \vecb{y} -  \vecb{A}_s   \vecb{x}_{s} )^{T} (  \vecb{y} -  \vecb{A}_s   \vecb{x}_{s}  )  +   \vecb{x}_{s}^{T} \vecb{\Sigma}_{x,s}^{-1} \vecb{x}_{s}.    
\end{aligned}
\end{equation} 
With this bound, the cost function Eq.~\eqref{sec3.1} can be transformed into a new cost function as 

\begin{equation}
\begin{aligned}
L(\vecb{x}_s, \vecb{s},&{\sigma_n} ,\vecb{\Sigma}_{x,s} )  \\
= & \frac{1}{2 \sigma_n}      ( \vecb{y} -  \vecb{A}_s   \vecb{x}_{s} )^{T} (  \vecb{y} -  \vecb{A}_s   \vecb{x}_{s}  )   +  \frac{1}{2} \vecb{x}_{s}^{T} \vecb{\Sigma}_{x,s}^{-1} \vecb{x}_{s}    \\
 & + \frac{1}{2} \log{| {\sigma}_n \vecb{I} + \vecb{A} _s \vecb{\Sigma}_{x,s}  \vecb{A}_s^{T}| } - \log{p(\vecb{s};\hat{\vecb{\Theta}}_{\hat{\mathcal{G}}})}.
 \end{aligned}
 \label{eq1.4}
\end{equation}

It can be proved that the resulting $\vecb{s}$, $\sigma_n$, $\vecb{\Sigma}_{x,s}$ from  Eq.~\eqref{eq1.4} are equivalent to that from solving Eq.~\eqref{sec3.1}~\cite{Wipf2011,LeiZhang}. Additionally, given a fixed support $\vecb{s}$, the minimization of Eq.\eqref{eq1.4} becomes a convex optimization problem. This enables the closed-form solutions for  $\sigma_n$ and $\vecb{\Sigma}_{x,s}$. Moreover, $\vecb{x}$, $\vecb{s}$, $\sigma_n$, and $\vecb{\Sigma}_{x,s}$ are jointly estimated in a single framework. \\
 
Note that the previous latent Bayes model~\cite{Wipf2011,LeiZhang} does not consider the structured sparsity prior. Since the minimization of  Eq.~\eqref{eq1.4} involves several unknown variables, we apply an alternative minimization scheme to reduce this optimization problem into several subproblems.  With structured sparsity prior being considered, we derive several new formulations for the sub-optimization problems to efficiently estimate sparse signal, support, and signal covariance. 
The details on the optimization are provided in Section~\ref{SparseSigEstimation}.  \\
 
\subsubsection{Graphical sparsity prior estimation} \label{Graphlearning}
Given the estimated sparse signal, support, noise and signal variances, \ie $\hat{\mbox{\boldmath{$x$}}}_s$, $\hat{\mbox{\boldmath{$s$}}}$, $\hat{\sigma}_n$, and $\hat{\mbox{\boldmath{$\Sigma$}}}_{x,s}$,  we have
\begin{equation} \label{eq.2.10}
\begin{aligned}
p( \mbox{\boldmath{$y$}}| \mbox{\boldmath{$\Theta$}}_{\mathcal{G}} ) & \propto  \int p(\mbox{\boldmath{$y$}}| {\mbox{\boldmath{$x$}}}_s, {\sigma}_n)
p( {\mbox{\boldmath{$x$}}}_s|\vecb{s},{\mbox{\boldmath{$\Sigma$}}}_{x,s})
p_{\mathcal{G}}({\mbox{\boldmath{$s$}}}|\mbox{\boldmath{$\Theta$}}_{\mathcal{G}})\mathrm{d} \mbox{\boldmath{$x$}}_s \\
& =  p(\mbox{\boldmath{$y$}} |  \hat{\vecb{s}},  \hat{\sigma}_n,  \hat{\mbox{\boldmath{$\Sigma$}}}_{x,s})p(\hat{\mbox{\boldmath{$s$}}}|\mbox{\boldmath{$\Theta$}}_{\mathcal{G}}) \\
& \propto p(\hat{\mbox{\boldmath{$s$}}}|\mbox{\boldmath{$\Theta$}}_{\mathcal{G}}), 
\end{aligned}
\end{equation}

Let the underlying structure of the graphical model $\hat{\mathcal{G}}$ be given, the graphical model parameters $\mbox{\boldmath{$\Theta$}}_{\hat{\mathcal{G}}}$, therefore, can be inferred by solving the following maximum likelihood (ML) problem: 
\begin{equation} \label{eq.2.11}
\begin{aligned}
 \hat{\mbox{\boldmath{$\Theta$}}}_{\hat{\mathcal{G}}} = \max\limits_{\mbox{\boldmath{$\Theta$}}_{\hat{\mathcal{G}}}  } p(\hat{\mbox{\boldmath{$s$}}}|\mbox{\boldmath{$\Theta$}}_{\hat{\mathcal{G}}}) 
\end{aligned}
\end{equation}
which encourages $\mbox{\boldmath{$\Theta$}}_{\hat{\mathcal{G}}}$ (\ie graphical sparsity prior) to be adaptive to the distribution of the latent support signal. The  ML problem can be solved by many parameters learning approaches such as~\cite{Besag1974,HintonConstrativeDivergence}. The graph $\mathcal{G}$ can be estimated from structure learning approaches such as~\cite{koller2009probabilistic}. However, performing the structure learning in every iteration could result in extremely high computation. Thus, we use a graph update procedure, instead. The details on solving the ML problem and the graph update procedure are provided in Section~\ref{Ch2_GraphOptimize}.  Figure \ref{Fig_model_visualization} illustrates the proposed Adaptive MRF inference framework.  The estimation problems Eq.~\eqref{eq.2.9} in Section~\ref{SparseRecovery} and Eq.~\eqref{eq.2.10}  in Section~\ref{Graphlearning} are alternatively optimized until convergence, where we obtain the final result.

\subsection{Optimization} \label{Optimization}
Here, we will first focus solving the sparse signal estimation problem in Eq.~\eqref{eq.2.9} in Section~\ref{SparseSigEstimation}, given the MRF. Then, we will focus on the MRF parameter estimation based on the estimated sparse signal Eq.~\eqref{eq.2.11} in Section~\ref{Ch2_GraphOptimize}. \\
 
\subsubsection{Sparse signal estimation} \label{SparseSigEstimation}
Here, we mainly focus on optimizing Eq.~\eqref{eq1.4} to obtain all involved unknown variables, given the graphical sparsity prior: the parameters $\hat{\vecb{\Theta}}_{\hat{\mathcal{G}}}$ and the underlying graph $\hat{\mathcal{G}}$, as follows:  
\begin{equation}
\begin{aligned}
& \lbrace \hat{\vecb{x}}, \hat{\vecb{s}}, {\hat{\sigma}}_{n},  \hat{\vecb{\Sigma}}_{x,s} \rbrace
& = \min\limits_{  \vecb{x}_s, \vecb{s},   \sigma_n , \vecb{\Sigma}_{x,s}  }   L(\vecb{x},\vecb{s},{\sigma}_{n} ,\vecb{ \Sigma }_{x,s}).
 \end{aligned}
 \label{eq1.5}
\end{equation}
Since the optimization problem Eq.~\eqref{eq1.5} involves several unknown variables, we apply an alternative minimization scheme to reduce the problem Eq.~\eqref{eq1.5} into several subproblems, each of which only involves one variable and often can be solved directly. 
Here, we present the estimation of sparse signal, noise and signal parameters which gain the closed-form solutions, and provide the formulation to efficiently estimate for the support. 
 The detail derivation is provided in Appendix.  These subproblems are then optimized alternatively until convergence. \\

\paragraph{Optimization over $\vecb{s}$} \label{Opt_s}
 Given $\vecb{x}$, ${\sigma}_{n}$, and $\vecb{\Sigma}_{x,s}$, the subproblem over the support $\vecb{s}$ can be given as
\begin{equation}
\begin{aligned}
& \min\limits_{\vecb{s}}   \quad
 \frac{1}{2 \sigma_n}   \vecb{x}_{s} ^{T}   \vecb{A}_s   ^{T}  \vecb{A}_s  \vecb{x}_{s}
   - \frac{1} {\sigma_n}  \vecb{y}^{T}  \vecb{A}_s \vecb{x}_s
 + \frac{1}{2} \vecb{x}_{s}^{T}  \vecb{\Sigma}_{{x,s}}^{-1}   \vecb{x}_{s}   \\
& + \frac{1}{2} \log | \sigma_n \vecb{I}  +  \vecb{A}_s \vecb{\Sigma}_{{x,s}}  \vecb{A}_s^{T} |
-\log{p(\vecb{s};\hat{\vecb{\Theta}}_{\hat{{\mathcal{G}}}})}.
\end{aligned}
\label{eq1.7}
\end{equation}
The minimization problem in Eq.~\eqref{eq1.7} can be viewed as an MAP problem over an MRF. Solving Eq.~\eqref{eq1.7}  is  computationally extensive because the logarithmic and the pairwise terms require exhaustive search over all possible support patterns.  In particular, when the coefficients of the estimated sparse signals $\vecb{x}$ are all non-zero, the first term $ \vecb{x}_{s} ^{T}   \vecb{A}_s   ^{T}  \vecb{A}_s  \vecb{x}_{s}$  becomes a fully connected graph. 

To address these problems, we derive a new support estimation formulation  Eq.\eqref{eqAcc3} where the logarithmic and quadratic terms are approximated into a linear function (unary potential) with  respect to the support.  First, we  approximate the logarithmic term by using the upper bound of the determinant of a positive definite matrix, which is the determinant of the diagonal entries of $(\sigma_n \vecb{I}  +  \vecb{A}_s \vecb{\Sigma}_{{x,s}}  \vecb{A}_s^{T})$:      
\begin{equation}
\begin{aligned}
\log | \sigma_n \vecb{I}  +   \vecb{A}_s \vecb{\Sigma}_{{x,s}}  \vecb{A}_s^{T} |   
&  \leq  \sum_{i \in \mathcal{V} }  { \log[ \vecb{\Sigma}_x]_{i,i}} \\
& + \log [(\sigma_n  \vecb{\Sigma}_{{x}}^{-1}   +    \vecb{A}^{T} \vecb{A})]_{i,i},  
\end{aligned}
\label{sup_eq_sub2.1_determinant}
\end{equation}
where  $\mathcal{V}$ is the set of non-zero support coefficients. The notation $ [\vecb{M}]_{i,i} $ refers to the $i-$th diagonal entry of the matrix $\vecb{M}$. Then, we employ Hadarmard product to explicitly represent the support. Details are provided in Appendix~\ref{ApproxSupport}. The optimization problem in Eq.~\eqref{eq1.7} can be equivalently formulated  as 
\begin{equation}
\begin{aligned}
& \min\limits_{\vecb{v} \in \lbrace 0,1 \rbrace^N }  
\frac{1}{2 \sigma_n} \vecb{v}^T  ( \vecb{X}^{T} \vecb{A}^{T} \vecb{AX} +  \sigma_n \vecb{X}^{T} \vecb{\Sigma}_{x}^{-1} \vecb{X}   )   \vecb{v} \\
& +  ( - \frac{1}{\sigma_n} \vecb{y}^{T}   \vecb{A} \vecb{X} + \vecb{p}^{T}    + \vecb{q}^{T}  )  \vecb{v}  -\log{p(2\vecb{v}-1;\hat{\vecb{\Theta}}_{\hat{{\mathcal{G}}}})},
\end{aligned}
\label{eq1.7.1}
\end{equation}
where  $\vecb{v} =  \frac{1}{2}(\vecb{s} + 1)$ , $\vecb{p} = \frac{1}{2}\log(\textbf{diag}  \left\lbrace \vecb{\Sigma}_{x} \right\rbrace ) $;  $\vecb{q} = \frac{1}{2} \log \left( \textbf{diag} \left\lbrace  \sigma_n \vecb{\Sigma}^{-1}_{x} +  \vecb{Q} \right\rbrace\right)$;  $\vecb{Q}$ is a diagonal matrix whose diagonal entries are  the diagonal entries of  $\vecb{A}^{T} \vecb{A}$; and  $\vecb{X}$ is a diagonal matrix with diagonal coefficients from $\vecb{x}$. The cost function of~Eq.~\eqref{eq1.7.1} is the upper bound of Eq.~\eqref{eq1.7}.     
 
Then, we exploit the fact that the measurement matrix $\vecb{A}$ is nearly orthogonal~\cite{foucart2013mathematical} to approximate the quadratic function into a linear function:
\begin{equation}
\begin{aligned}
\vert\vert \vecb{A}_s^*\vecb{A}_s - \vecb{I}  \vert\vert_{2 \rightarrow 2} \leq \delta_s, 
\end{aligned}
\label{eqAcc1}
\end{equation}           
where $\vecb{I}$ is an identity matrix, $\vert\vert \cdot \vert\vert_{2 \rightarrow 2}$ is the operator norm, and $\delta_s$ is a small constant  corresponding restricted isometric constant.  The first term in Eq. \eqref{eq1.7.1} can be approximated as follows:   
\begin{equation}
\begin{aligned}
\vecb{v}^T (\vecb{X}^{T} \vecb{A}^{T} \vecb{AX} + & \sigma_n \vecb{X}^{T} \vecb{\Sigma}_{x}^{-1} \vecb{X})\vecb{v}   \approx \\ & \vecb{v}^T  \vecb{X}^{T} ( \vecb{I} +  \sigma_n   \vecb{\Sigma}_{x}^{-1}  ) \vecb{X} \vecb{v}. 
\end{aligned}
\label{eqAcc2}
\end{equation}

Thus, the signal support $\vecb{s}$ is estimated by solving the following optimization problem:  
\begin{equation}
\begin{aligned}
\min\limits_{\vecb{v} \in \lbrace 0,1 \rbrace^N }  
&(\frac{1}{2 \sigma_n}    \vecb{r}^T  
- \frac{1}{\sigma_n} \vecb{y}^{T}   \vecb{A} \vecb{X} + \vecb{p}^{T}    + \vecb{q}^{T}   )  \vecb{v} \\
&  -\log{p(2\vecb{v}-1;\hat{\vecb{\Theta}}_{\hat{\mathcal{G}}})},
\end{aligned}
\label{eqAcc3}
\end{equation}
where $\vecb{r}$ is a vector containing diagonal entry of the matrix $(\vecb{X}^{T} ( \vecb{I}  +  \sigma_n   \vecb{\Sigma}_{x}^{-1}  ) \vecb{X})$. As the pairwise terms in Eq.\eqref{eq1.7.1} reduces to a unary term, the Eq.~\eqref{eqAcc3} is much faster to evaluate. Thus, the MAP problem   Eq.~\eqref{eqAcc3} can be effectively solved by any off-the-shelf inference tools, \eg, dual  decomposition~\cite{DualDecom}, TWRS~\cite{Kolmogorov2006}, ADLP~\cite{Meshi2011}.  The computational complexity for solving  Eq.\eqref{eqAcc3} depends only on the underlying graph of the updated MRF defined by $\hat{\vecb{\Theta}}_{\hat{\mathcal{G}}}$ (see Section \ref{acceleration}). \\

\begin{algorithm}[t]
\SetKwInOut{Input}{Input}
\SetKwInOut{Output}{Output}
\Input{ A measurement signal $\vecb{y}$, $\vecb{A}$, and graphical model's parameters  $\vecb{\Theta}_{\mathcal{G}}$ and structure  $\mathcal{G}$} 
\textbf{Initialization}:  $\vecb{\Sigma}_{x} = \vecb{I}_{N \times N}$, ${\sigma}_{n} = 1 $,  $\vecb{x} = \vecb{0}$ , and $\vecb{s} = \vecb{1}$   \;
\While{a stopping criterion is not satisfied}{ 
 {1. Update the support $\vecb{s}$ by solving Eq.~\eqref{eqAcc3}  }\;
 {2. Update the covariance matrix $\vecb{\Sigma}_{x,s}$ as Eq.~\eqref{eq1.8.1.7} }\;   
 {3. Update the noise variance $\sigma_{n}$ as Eq.~\eqref{eq1.9.5} }\;
 {4. Update  the sparse signal $\vecb{x}_s$ as Eq.~\eqref{eq1.6.1} } \;}
\Output{$\vecb{x} $ whose non-zero coefficients are from $\vecb{x}_s$ } 
\caption{Sparse signal estimation.}
\label{Alg1}
\end{algorithm}

\paragraph{Optimization over $\vecb{\Sigma}_{x,s}$} 
We start from calculating $\vecb{\Sigma}_{x}$, then $\vecb{\Sigma}_{x,s}$ is chosen from the diagonal member of $\vecb{\Sigma}_{x}$ according to $\vecb{s}$. Let $\vecb{\nu} \in \mathbb{R}_{+}^{N}$ be a vector whose members are the diagonal entry of $\vecb{\Sigma}_{x}$. Given $\vecb{x}$,  $\vecb{s}$, and $\sigma_{n}$, we have the sub-problem over  $\vecb{\Sigma}_x$ as 
\begin{equation}
\begin{aligned}
\min\limits_{\vecb{\nu}}
\frac{1}{2}  \vecb{x}^{T}   \vecb{\Sigma}_{x}^{-1} \vecb{x}
+ \frac{1}{2} \log{|\sigma_n \vecb{I}  +  \vecb{AV}  \vecb{\Sigma}_{x}  \vecb{V}^{T} \vecb{A}^{T} |},
\end{aligned}
\label{eq1.8}
\end{equation}
where $\vecb{V}$ is a diagonal matrix with diagonal coefficients from $\vecb{v} = \frac{1}{2}(\vecb{s} + 1)$. The updated equation for the $i$-th entry of $\vecb{\nu}$ is 
\begin{equation}
\begin{aligned}
{\nu}_i^{new} =   x_i^2 + {\alpha}_i  .
\end{aligned}
\label{eq1.8.1.7}
\end{equation}
${\alpha}_i$ is the $i$-th entry of vector $ \vecb{\alpha} =  \textbf{diag}\lbrace (  \vecb{\Sigma'}_{x}^{-1} +    \frac{1}{\sigma_n} \vecb{V}^{T} \vecb{A}^{T}  \vecb{AV} )^{-1}  \rbrace$, and $\vecb{\Sigma'}_{x}$ is the resulted $\vecb{\Sigma}_{x}$ in previous iteration. 
Then, $ \vecb{\Sigma}_{x,s}$ is diagonal matrix where each diagonal coefficient ${\nu}_i^{new}$ are chosen according to $\vecb{s}$. \\

\paragraph{ Optimization over  $\sigma_n$}
Given $\vecb{x}_s$, $\vecb{s}$ and $\vecb{\Sigma}_{x,s}$, we have the sub-problem over $\sigma_n$ 
\begin{equation}
\begin{aligned}
\min\limits_{\sigma_n}  \quad
& \frac{1}{2 \sigma_n} (\vecb{y} -   \vecb{A}_s \vecb{x}_s   )^{T}  ( \mathbf{y} -     \vecb{A}_s \vecb{x}_s  ) \\ &  + \frac{1}{2} \log{|\sigma_n \vecb{I}  +   \vecb{A}_s \vecb{\Sigma}_{x,s} \vecb{A}_s^{T}| }.
\end{aligned}
\label{eq1.9}
\end{equation}
This problem gives rise to a closed-form solution for $\sigma_n$ as
\begin{equation}
\begin{aligned}
\sigma_n^{new} = \frac{1}{M} \sum_{i=1}^{M} \sqrt{\frac{  d_i^2}{ \eta_{i}}}.
\end{aligned}
\label{eq1.9.5}
\end{equation}
where $\eta_i$ is the $i$-th entry of vector $ \vecb{\eta} =  \textbf{diag} \lbrace ( \sigma_n \vecb{I}  + \vecb{A}_s   \vecb{\Sigma}_{x,s}   \vecb{A}_s^{T}   ) ^{-1}  \rbrace$, and $d_i$ is the $i$-th entry of $\vecb{d} = \vecb{y} - \vecb{A}_s\vecb{x}_s$.

\paragraph{Optimization over $\vecb{x}_s$}
Given $\vecb{s}$, $\sigma_n,$ and $\vecb{\Sigma}_{x,s}$, the subproblem for $\vecb{x}_s$ is 
\begin{equation}
\begin{aligned}
\min\limits_{\vecb{x}_s}
\frac{1}{\sigma_n}( \vecb{y} -  \vecb{A}_s \vecb{x}_{s})^{T} ( \vecb{y} -   \vecb{A}_s \vecb{x}_{s}  ) +  \vecb{x}_{s}^{T}   \mathbf{\Sigma}_{{x,s}}^{-1} \vecb{x}_{s},
\end{aligned} \label{eq1.6}
\end{equation}
which shows a closed-form updated equation as
\begin{equation}
\begin{aligned}
\vecb{x}_{s}^{new} =  ( \sigma_n  \mathbf{\Sigma}_{x,s}^{-1}  + \vecb{A}_s^{T} \vecb{A}_s)^{-1}{\vecb{A}_s}^{T}   \mathbf{y}.
\end{aligned}
\label{eq1.6.1}
\end{equation}

How to solve Eq.~\eqref{eq1.5} is summarized in Algorithm \ref{Alg1} where the sparse signal, support, and noise and signal parameters are jointly estimated in a unified framework.  In Algorithm~\ref{Alg1}, we solve the support estimation problem Eq.~\eqref{eq1.7.1}  in step 1 by performing graphical inference using the belief propagation implemented by  \cite{zhang2016pairwise}.

\begin{algorithm}[t]
\SetKwInOut{Input}{Input}
\SetKwInOut{Output}{Output}
 \Input{   Measurements $\vecb{y}$ and random matrix $\vecb{A}$} 
\textbf{Initialization }:    
Get $\vecb{x}$ from  Algorithm \ref{Alg1} where step 1  is removed and replaced  with a fixed $\vecb{s} = \vecb{1}$, $\vecb{b} = \vecb{0}$, $\vecb{W} = \vecb{0}$    \\
 \While{a stopping criterion is not satisfied}{ 
 {1. Obtain a binary vector $\vecb{b}$ from thresholding each of $\vecb{x}$, \ie, $b_i = 1$ if $\text{abs}(x_i)> \text{mean}( \text{abs}(\vecb{x}))$, and $b_i = -1$  otherwise   \;}
 {2. Calculate ${\mathcal{G}}$ from $\vecb{b}$ following Algorithm \ref{Ch3_Alg_graph} }  \;
 {3. Estimate $\vecb{\Theta}_{\mathcal{G}}$ from $\vecb{b}$ and ${{\mathcal{G}}}$ by solving Eq. \eqref{eq.2.11}   \;} 
 {4. Update $\vecb{x}$ by solving Eq.\eqref{eq1.5} with Algorithm \ref{Alg1}\; } 
}
\Output{Recovered $\vecb{x}_{rec}$.}    
\caption{Adaptive Markov random field inference for compressive sensing (Adaptive-MRF)}
\label{Alg2}
\end{algorithm}

\begin{algorithm}[t] 
\SetKwInOut{Input}{Input}
\SetKwInOut{Output}{Output}  
 \Input{Binary vector $\vecb{b}$}  
\textbf{Initialization }: $\mathcal{E}_i = \emptyset \quad \forall i = 1,...,N$, $\mathcal{E} =  \emptyset$, and the node set contains the node each of which corresponds to each coefficient in the binary vector  $\mathcal{V} = \setn{b_1, ..., b_N}$. \\
 \For{i = 1, ..., N} 
{ \For{each $ j \in \mathbb{N}_i$}{
      \If{$b_j = 1$ and the edge $(j,i) \not\in \mathcal{E}$ is not present}{$\mathcal{E}_i = \mathcal{E}_i \bigcup (i,j)$.}
     } 
	{$\mathcal{E} = \mathcal{E} \bigcup \mathcal{E}_i$.}  
  } 
 \Output{ $\mathcal{G} = (\mathcal{V},\mathcal{E})$.}   
 \caption{Update Graph $\mathcal{G}$}
 \label{Ch3_Alg_graph}
\end{algorithm}

Next, we turn to the MRF parameter estimation Eq.\eqref{eq.2.10} to update  the BM parameters and the underlying graph.\\

\subsubsection{MRF parameter estimation}
\label{Ch2_GraphOptimize}
Given the  resulting sparse signal $\vecb{x}$, we update the MRF graph and calculate the MRF parameter based on a binary vector  $\vecb{b}$ corresponding to the high-energy coefficients of $\vecb{x}$. Notice that the binary vector $\vecb{b}$ is not similar to the estimate support $\vecb{s}$, thus, avoiding overfitting to the previous MRF parameter estimation.\\

\begin{figure}[t]  
\centering{\includegraphics[ width =2in,trim=1.00cm  15cm  1cm 3.8cm,clip] {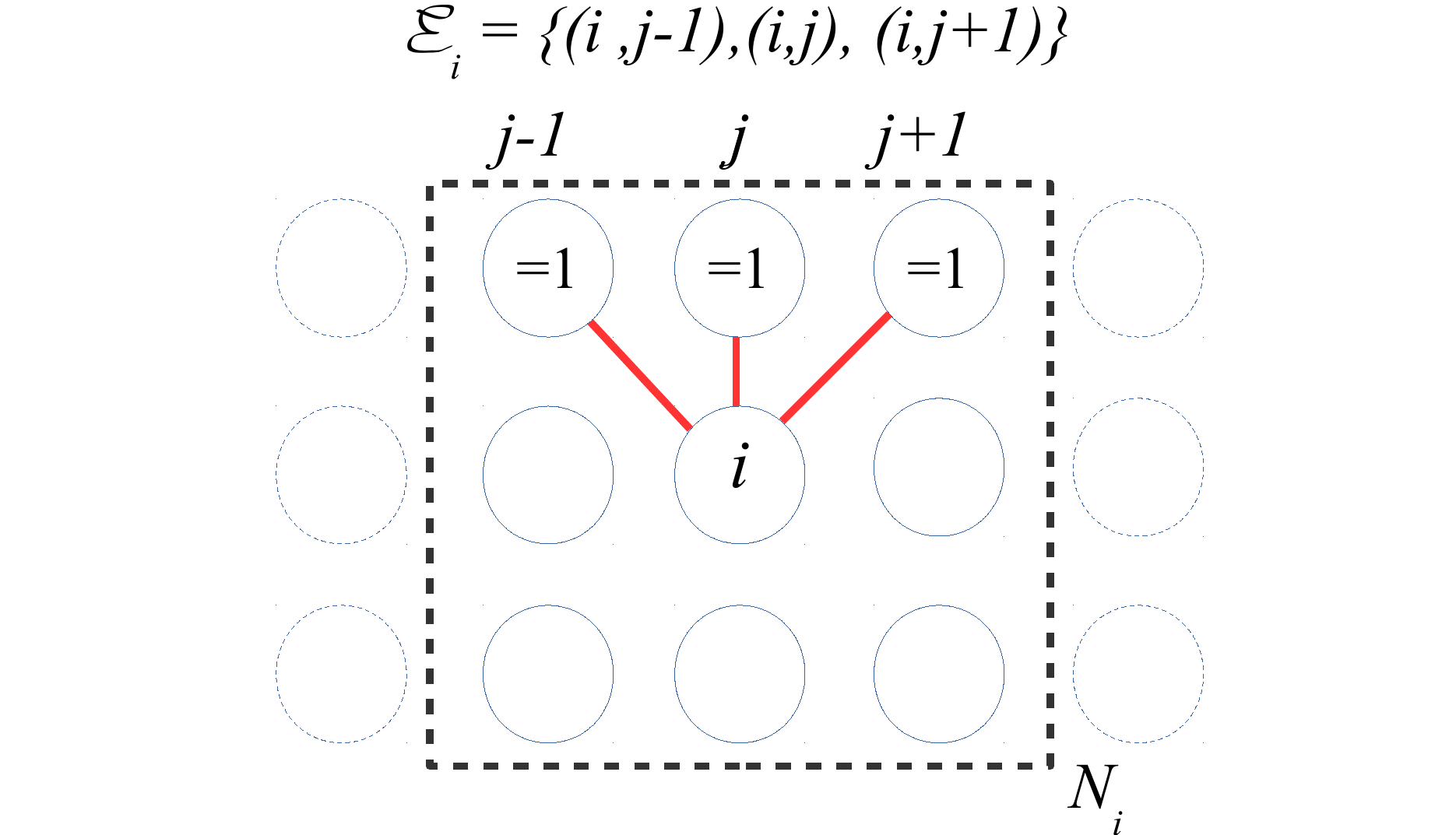} }  
\caption{Example of how edges in the local set $\mathcal{E}_i$ are updated. } \label{Ch3_NeighborGrah} 
\end{figure}

\paragraph{Graph update procedure $\hat{\mathcal{G}}$}
\label{Ch3_Graph_update}  
 In practice, we can simplify the graph estimation task, as suggested in~\cite{hegde2015nearly}, by forming a graph according to high energy coefficients in sparse signal which carry information about signal structure. Given the binary vector $\vecb{b}$, each of the binary coefficients is mapped to each node in a graph $\mathcal{G}$, and each edge is established from one node to adjacent nodes with value '1' within a predefined neighborhood $\mathbb{N}_i$. Figure~\ref{Ch3_NeighborGrah} illustrates how each edge in the graph is updated for capturing two-dimensional structure in an image.  Each pixel is mapped to a node in the graph. Let $b_{i}$  be the node of interest. As the adjacent nodes $b_{j-1}, b_{j}$, and $b_{j+1}$ in a radius of neighborhood $\mathbb{N}_i$ covering 8-neighborhood are equal to one, the edges $(i,j-1), (i,j),$ and  $(i,j+1)$ are included into the local edge set  $\mathcal{E}_i$ corresponding to the node $b_{i}$. Algorithm \ref{Ch3_Alg_graph} summarizes the graph update procedure.

As the edges of the graph are established between each node and its non-zero adjacent nodes, the updated (adaptive) graph can capture a finer detail than the fixed neighborhood graph  of the clustered structure sparsity models~\cite{ yu2012bayesian, yu2015,Fang2015, Fang2D2016,Wang2015ISAR,AdaptiveCluster1,AdaptiveCluster2}. In the fixed neighborhood graph, each node is connected to all of its adjacent nodes in its neighborhood. As our approach enables the flexible connection between nodes in $\mathbb{N}_i$, the updated graph can be more adapted toward the actual structure of signals. Additionally, the updated graph is more sparse than the fixed neighborhood graph; thus, it will also help the support estimation (Eq.~\ref{eqAcc3}) to be solved efficiently (see how this can improve the algorithm complexity in Section~\ref{acceleration}).

\paragraph{  BM parameter estimation $\mbox{\boldmath{$\Theta$}}_{\hat{\mathcal{G}}}$}
Given the binary vector $\vecb{b}$ and the graph $\mathcal{G}$, we solve the MAP problem Eq.~\eqref{eq.2.11} for the parameters $\mbox{\boldmath{$\Theta$}}_{\hat{\mathcal{G}}}$ using the pseudo-likelihood algorithm~\cite{Besag1974,parise2005learning} which requires low computational cost. The  pseudo-likelihood~\cite{Besag1974} is suitable for updated graph because it considers local dependency between $b_i$ and the adjacent neighbors in its  neighborhood. Thus, instead of considering the global likelihood over all nodes and edges in the graph, the pseudo-likelihood resorts to maximizes $  \prod_i p(b_i|\vecb{b}_{\mathcal{E}_i}, \vecb{\Theta}_{\mathcal{G}})$, where $\vecb{b}_{\mathcal{E}_i}$ are the adjacent neighbors connected to the node $b_i$ through edges defined by a local edge set $\mathcal{E}_i$.\\ 
 
The whole Adaptive MRF framework is summarized in Algorithm \ref{Alg2} where the sparse signal is updated in step 4, and the  MRF parameter estimation is performed in step 2 and 3. The sparse signal estimation is performed in Algorithm~\ref{Alg1}.  The alternative minimization scheme reduces the objective functions----the MRF parameter estimation  and sparse signal estimation---in each iteration. The objective functions can be proved to be bounded from below. Thus, the Adaptive MRF converges well as \cite{Wipf2011}. The empirical convergence of the proposed Adaptive MRF  is studied in Section~\ref{empirical_convergence}.

\section{Theoretical result} \label{Theoretical_Result} 
\subsection{Essence of adaptive support prior}
A main objective of this work is to adaptively estimate the MRF to capture the actual structure of sparse signals. In this section, we propose Theorem~\ref{Theory} that reveals the connection between adaptive support prior and probabilistic RIP (PRIP) condition~\cite{Cevher12009}. This is to motivate the essence of the adaptive signal prior in guaranteeing PRIP condition with the  lowest sample complexity of $\mathcal{O}(k)$. To show this, we start with reviewing the PRIP condition~\cite{Cevher12009}. Then, we present the Theorem~\ref{Theory}.

Let $\vecb{x}$ denote a  $k$-sparse, testing signal whose support $\hat{\vecb{s}} = \text{supp}(\vecb{x})$ is generated by a known probabilistic model $\mathcal{P}$. $\Omega_{k,\varepsilon}$ denotes the smallest set of candidate support captured by a learned model $\mathcal{M}$. 

\vspace{0.25cm}    
\begin{lemma}
\label{Ch2_lemma_PRIP}
\cite{Cevher12009}. PRIP condition: If the probability that the true support can be represented by a candidate support in $\Omega_{k,\varepsilon}$ is higher than $1 - \varepsilon$, \ie $ p(\hat{\vecb{s}} \in \Omega_{k,\varepsilon}) > 1 - \varepsilon$, a sub-Gaussian random matrix $\vecb{A} \in \mathcal{R}^{M \times N}$ satisfies the $(k, \varepsilon)$-PRIP with probability of at least $1-e^{-c_2 M}$ with $ M \geq c_1 (k + \log(|\Omega_{k,\varepsilon} |)) $, where $c_1, c_2 > 0$ depends only on the PRIP constant $\delta_k \in [0,1]$.  
\end{lemma}  

\vspace{0.25cm} 
Notice that the members of the set $\Omega_{k,\varepsilon}$ are chosen based on the support prior model $\mathcal{M}$, \eg a trained MRF. However, if $\mathcal{M}$ is learned based on the training data that cannot well represent the testing signals, then the necessary condition of the lemma can be violated. \\ 

To address this problem, we propose the concept of adaptive support prior to realize the smallest support set $\Omega_{k,\varepsilon}$ whose member can well represent the test signal. To do this, we study a sequence of random support vector ${S}_1, ..., {S}_n$ corresponding to the adapted support prior $\mathcal{M}_1, ..., \mathcal{M}_n$ such as Eq.~\eqref{eq.2.6}.

\vspace{0.25cm}
\begin{theorem} 
\label{Theory}
 For a fixed ground truth support $\hat{\vecb{s}} = \text{supp}(\vecb{x})$, if ${S}_1, ..., {S}_n$ converges to $\hat{\vecb{s}}$ in distribution, \ie $\lim_{n\rightarrow  \infty } \mathcal{M}_n(S_n) = \mathcal{P}(\hat{\vecb{s}})$, then we can show that $\lim_{n\rightarrow  \infty } p(\hat{\vecb{s}} \in \Omega^n_{k}) = 1$ where $\Omega^n_{k}$ is the ball containing the random variable support $S_n$ with center $c$ and radius $2\epsilon$.           
\end{theorem}
 \vspace{0.2cm}   
\textit{Proof.} Since $\hat{s}$ is a fixed ground truth support, then convergence in distribution implies convergence in probability. That is, if $\lim_{n\rightarrow  \infty } \mathcal{M}_n(S_n) = \mathcal{P}(\hat{\vecb{s}})$, then $\lim_{n\rightarrow  \infty } p( || S_n - \hat{\vecb{s}} || <  \epsilon) = 1$. Given that $\Omega^n_{k}$ is the ball containing the ensembles of the random variable support $S_n$  with center $c$ and radius $2\epsilon$, then $\hat{\vecb{s}} \in \Omega^n_{k}$ with probability one.   
  
Theorem~\ref{Theory} suggests that if the adapted support model  $\mathcal{M}_n(S_n)$ can represent the true probability $\mathcal{P}$,  then the set $\Omega^n_{k}$ always contains a candidate support that truly represents the testing signal support (\ie $\varepsilon = 0$). The smallest size of the set $\Omega^n_{k}$ is one. Therefore, the minimum measurements can achieve the theoretical sample complexity, \ie $ M \geq c_1 (k + \log(|\Omega_{k,0}|))$ where $|\Omega_{k,0}|$ is smallest (\eg $|\Omega^n_{k}|=1$). Thus, $M \approx \mathcal{O}(k)$.

\begin{figure*}[t]  
	\hspace{ -0.5cm}  
	\begin{tabular}{ccccc} 
		\setlength{\tabcolsep}{0.5pt}
		\renewcommand{\arraystretch}{1}
		{ \subcaptionbox{Ground truth face images \label{Fig: 2 Ground}}[2in]
		{\begin{tabular}{ccccc}
		{\includegraphics[width = 0.3 in]{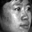}} 
		{\includegraphics[width = 0.3 in]{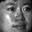}} 
		{\includegraphics[width = 0.3 in]{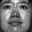}}
		{\includegraphics[width = 0.3 in]{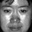}} 
		{\includegraphics[width = 0.3 in]{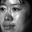}} \\ 
		{\includegraphics[width = 0.3 in]{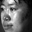}}    
		{\includegraphics[width = 0.3 in]{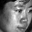}} 
		{\includegraphics[width = 0.3 in]{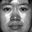}}  
		{\includegraphics[width = 0.3 in]{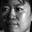}}    
		{\includegraphics[width = 0.3 in]{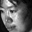}}  
		\end{tabular} }  }
		& \hspace{-1cm}  
		\subcaptionbox{Wavelet signal \label{Fig: 2 sparse_signal_wave} }[1.5in]{ \includegraphics[width = 0.7in ]{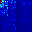}}  
		& \hspace{ -0.5cm}  
		\subcaptionbox{DCT signal \label{Fig: 2 sparse_signal_dct} }[1in]{   \hspace{-0.5cm}  \includegraphics[width = 1.1in,trim=0.5cm 8.2cm 0.8cm 8.cm,clip]{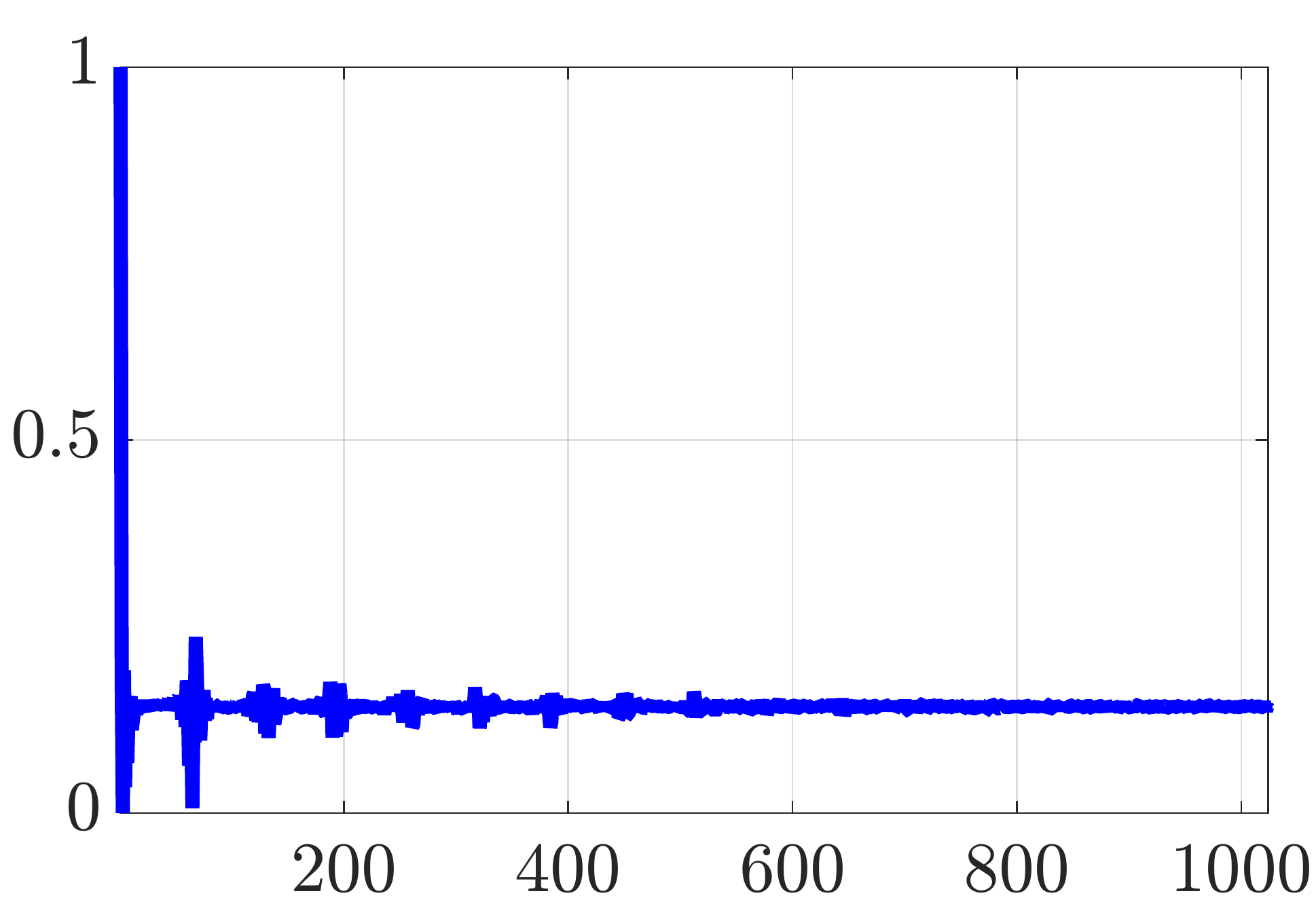}} 
		& \hspace{ 0cm}  
		\subcaptionbox{PCA signal \label{Fig: 2 sparse_signal_pca} }[1in]{  \hspace{-0.5cm}  \includegraphics[width = 1.1in,trim=0.5cm 7.2cm 0.8cm 7.5cm,clip]{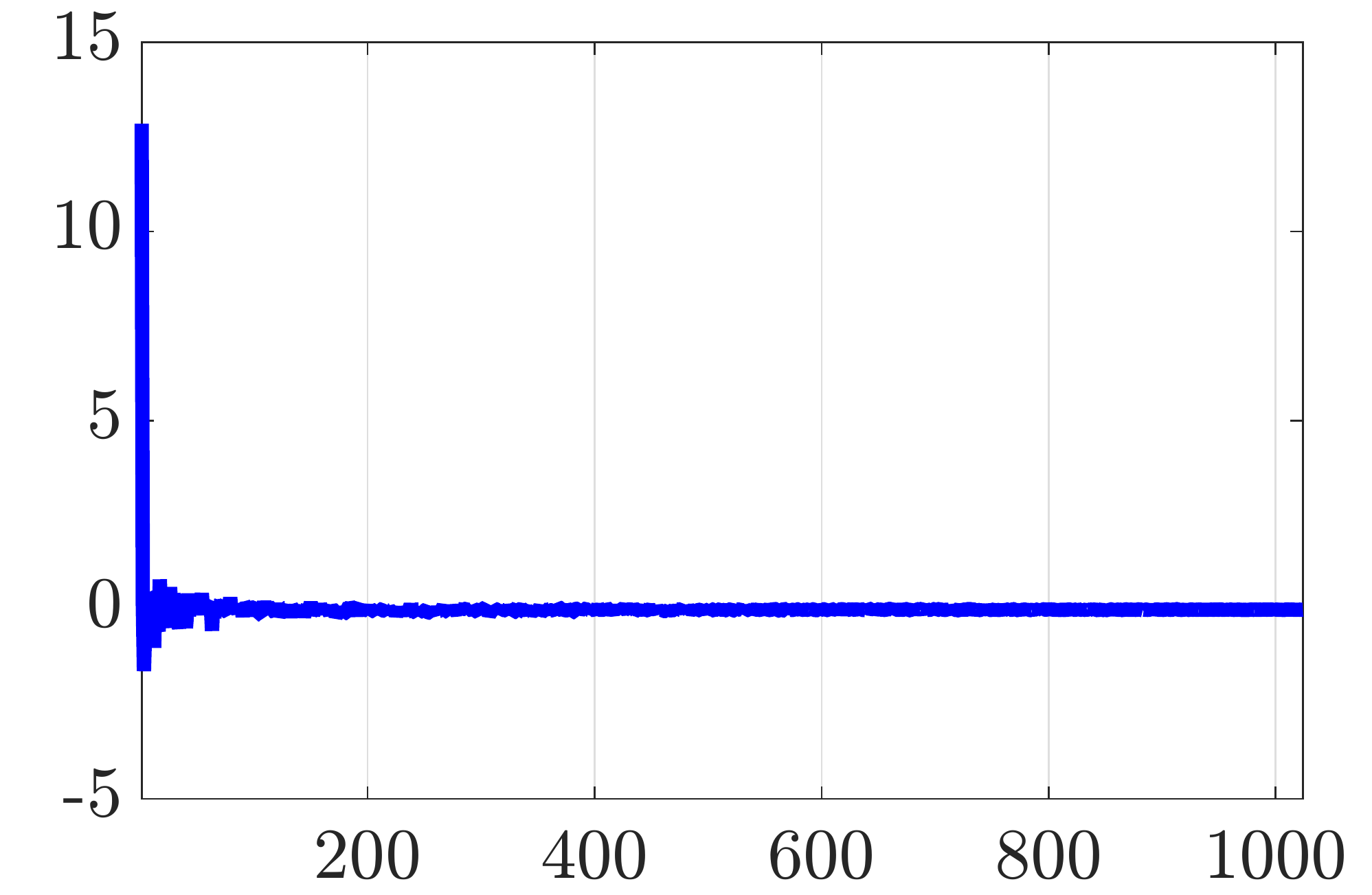}} 
		&  \subcaptionbox{Coef. decay \label{Fig:2 sparse_signal_decay} }[1in]{ \hspace{-0.5cm} \includegraphics[width = 1.1in,trim=0.5cm 8.2cm 0.8cm 8.cm,clip]{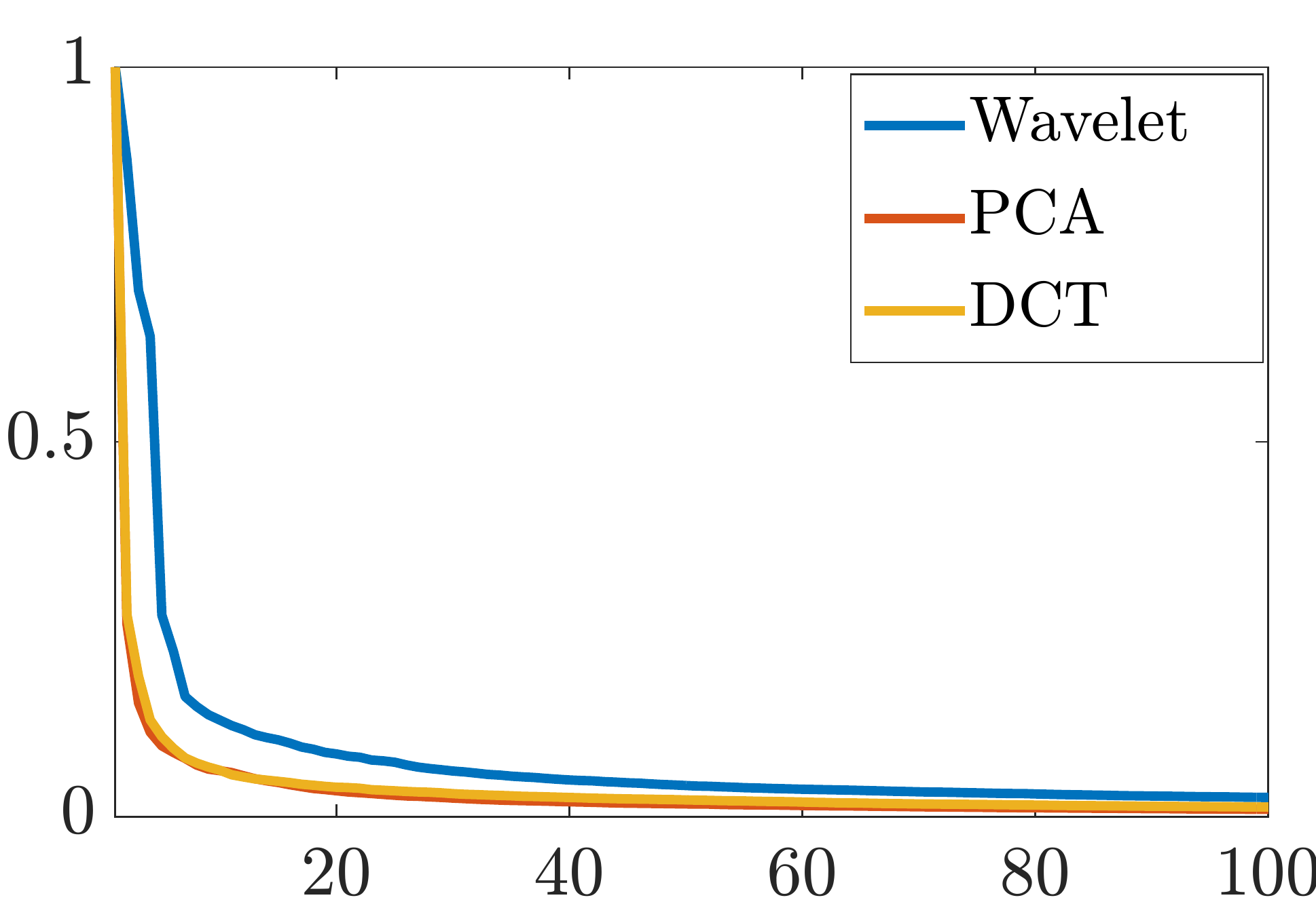}} 
	\end{tabular} 
\caption{CMU-IDB. (a) The ground truth face images. Examples of (b) wavelet signal, (c) DCT signal, (d) PCA signal, and (e) the decay of sparse signal coefficients in wavelet, DCT, and PCA domain.} \label{Fig:2} 
\end{figure*}

\begin{figure*} [t]
	\hspace{ -0.5cm}  
	\begin{tabular}{ccccc} 
		\setlength{\tabcolsep}{0.5pt}
		\renewcommand{\arraystretch}{1}
		{\subcaptionbox{Ground truth natural images \label{Fig: 3 Ground}}[2in]
		{\begin{tabular}{ccccc}
			{\includegraphics[width = 0.3 in, angle =-90]{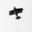}} 
			{\includegraphics[width = 0.3 in, angle =-90]{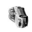}}
			{\includegraphics[width = 0.3 in, angle =-90]{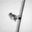}} 
			{\includegraphics[width = 0.3 in, angle =-90]{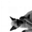}} 
			{\includegraphics[width = 0.3 in, angle =-90]{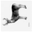}}     \\
			{\includegraphics[width = 0.3 in, angle =-90]{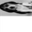}}  
			{\includegraphics[width = 0.3 in, angle =-90]{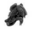}}  
			{\includegraphics[width = 0.3 in, angle =-90]{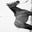}}    
			{\includegraphics[width = 0.3 in, angle =-90]{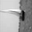}} 
			{\includegraphics[width = 0.3 in, angle =-90]{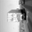}}
		\end{tabular} } } 
		&  \hspace{-1cm} \subcaptionbox{Wavelet signal \label{Fig: 3 sparse_signal_wave} }[1.5in]{ { \includegraphics[width = 0.7 in ]{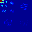}}}  
		& \hspace{ -0.5cm}   \subcaptionbox{DCT signal \label{Fig: 3 sparse_signal_dct} }[1in]{ \hspace{-0.5cm}  \includegraphics[width = 1.1in,trim=0.5cm 8.5cm 0.8cm 8.cm,clip]{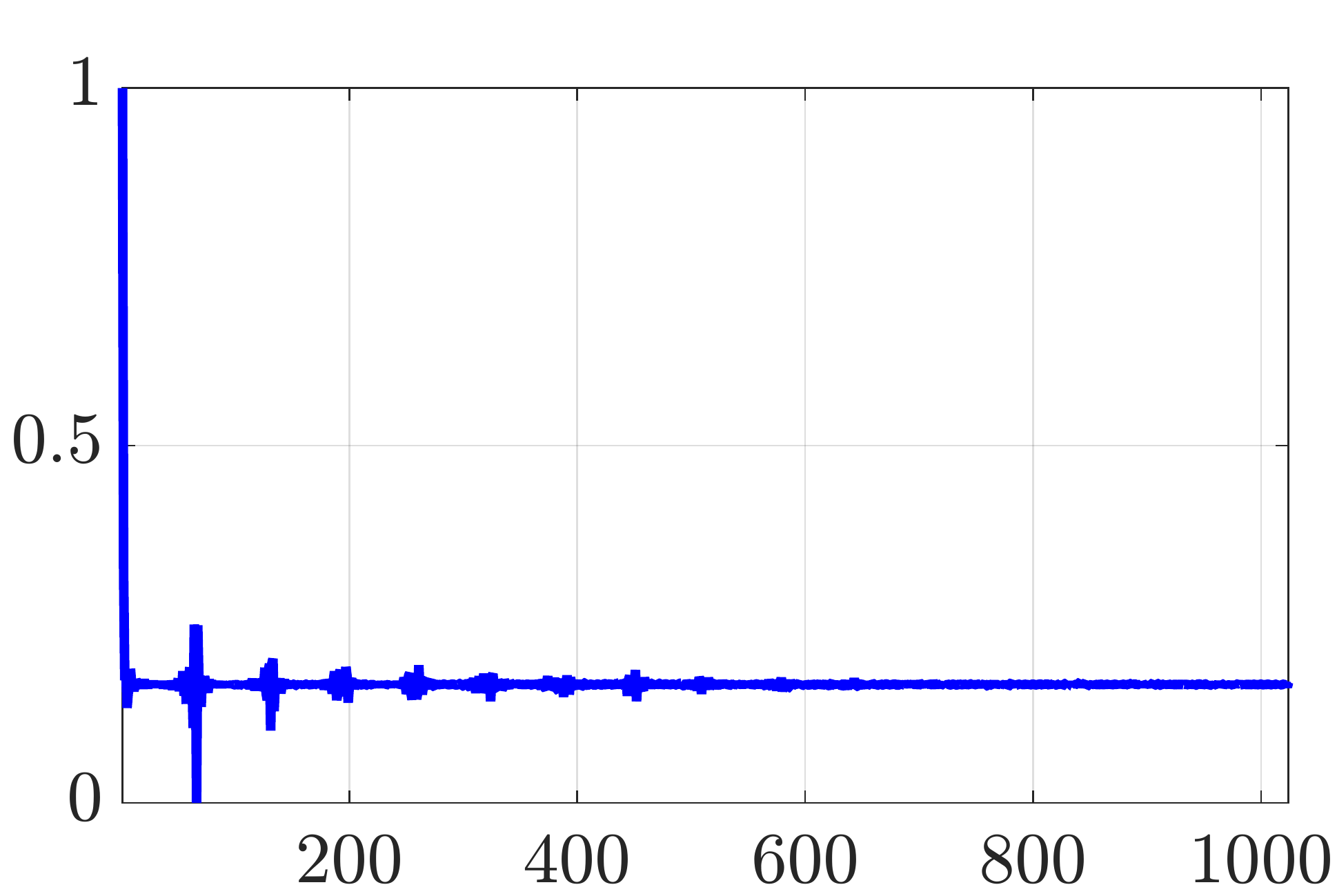}} 
		&   \subcaptionbox{PCA signal \label{Fig: 3 sparse_signal_pca} }[1in]{ \hspace{-0.5cm}  \includegraphics[width = 1.1in,trim=0.5cm 8.5cm 0.8cm 8.cm,clip]{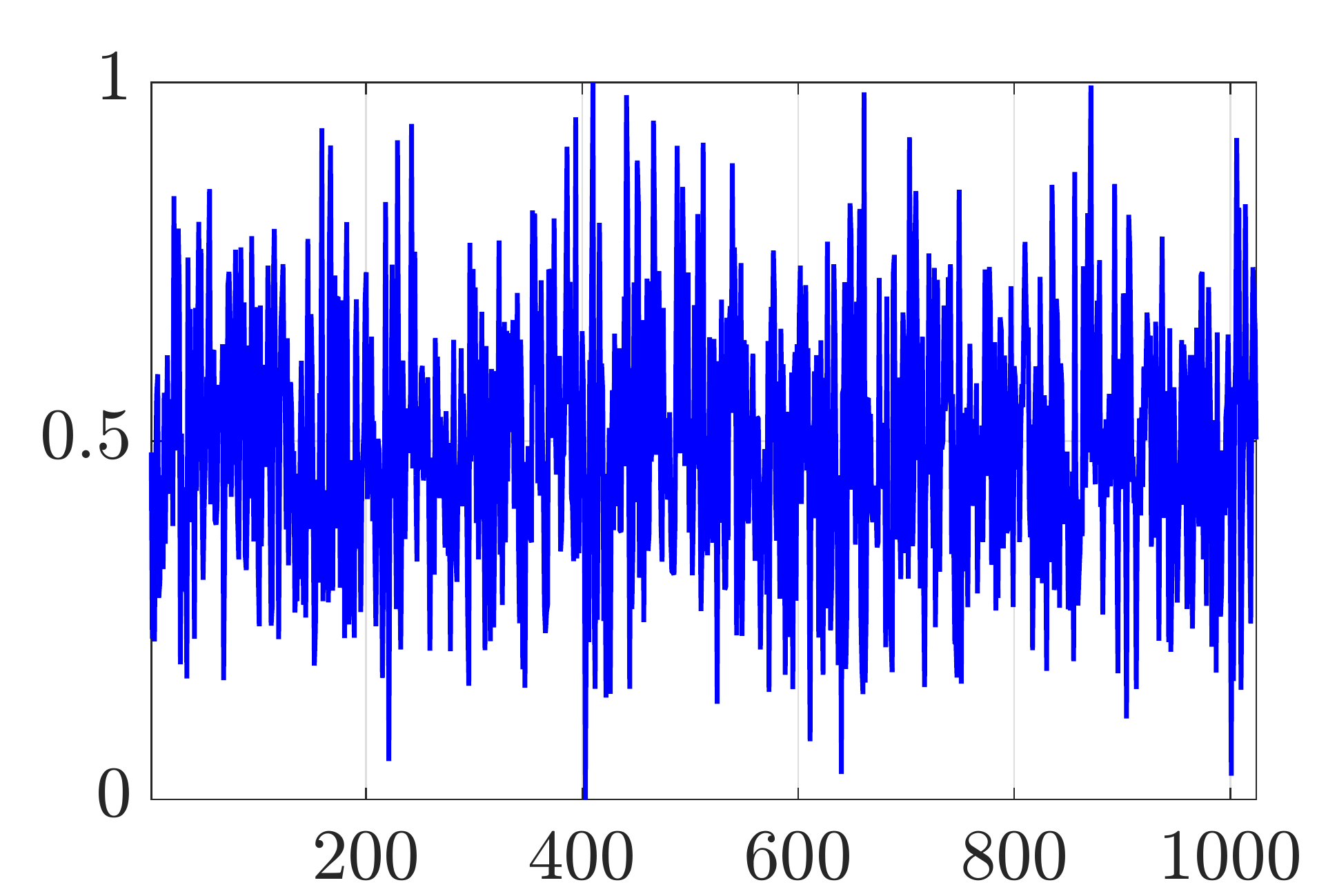}} 
		& \hspace{-0.2cm} \subcaptionbox{Coef. decay \label{Fig: 3 sparse_signal_decay} }[1in]{ \includegraphics[width = 1.1in,trim=0.5cm 8.2cm 0.8cm 8.cm,clip]{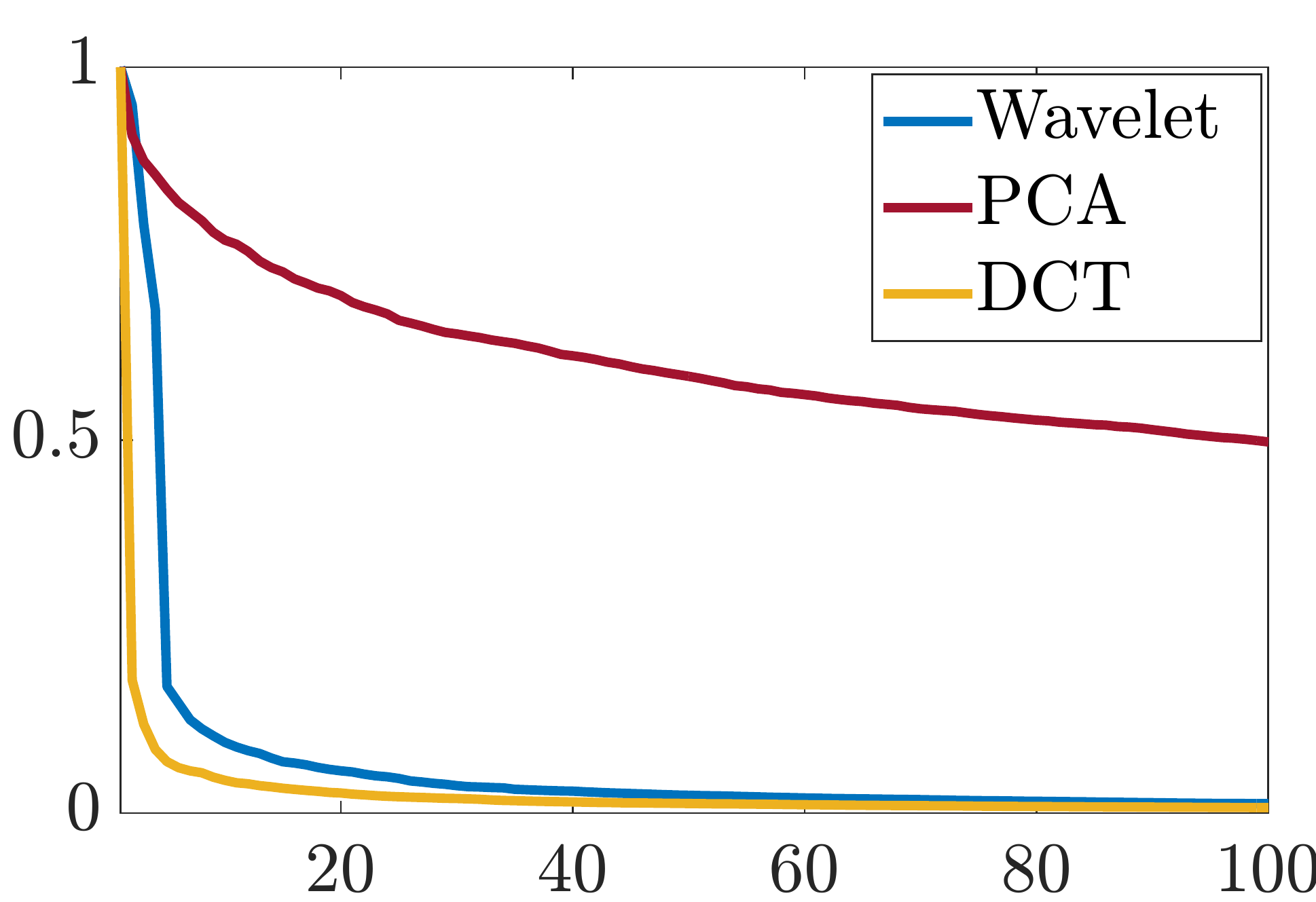}} 
	\end{tabular}
	\caption{CIFAR-10. (a) The ground truth natural images. Examples of (b) wavelet signal, (c) DCT signal, (d) PCA signal, and (e) the decay of sparse signal coefficients in wavelet, DCT, and PCA domain.} \label{Fig:3} 
\end{figure*}

\begin{figure}[t]  
	\begin{tabular}{cc}
	\setlength{\tabcolsep}{0.1pt} 
	\hspace{-0.8cm}\renewcommand{\arraystretch}{0.5}
	\begin{subfigure}{0.3\textwidth}
		{\hspace{0.5cm} \begin{tabular}{L{8cm}} 
				{\includegraphics[width = 0.3 in]{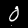}} 
				{\includegraphics[width = 0.3 in]{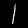}}  
				{\includegraphics[width = 0.3 in]{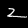}}   
				{\includegraphics[width = 0.3 in]{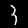}} 
				{\includegraphics[width = 0.3 in]{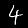}}  \\
				{\includegraphics[width = 0.3 in]{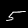}}   
				{\includegraphics[width = 0.3 in]{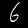}}  
				{\includegraphics[width = 0.3 in]{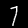}}  
				{\includegraphics[width = 0.3 in]{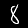}}   
				{\includegraphics[width = 0.3 in]{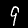}}  
			\end{tabular}} 
			\caption{Ground-truth digit images \label{Fig: 1 Ground}}
			\vspace{0.1cm}
		\end{subfigure}
		& \hspace{-0.8cm}\begin{subfigure}{1in}
			\raisebox{+0.8cm}{\subcaptionbox{ The image pixels decay   \label{Fig: 1Coeff}}[1.7in]{\includegraphics[width = 1.1in,trim=0.5cm 8.5cm 0.8cm 8.cm,clip]{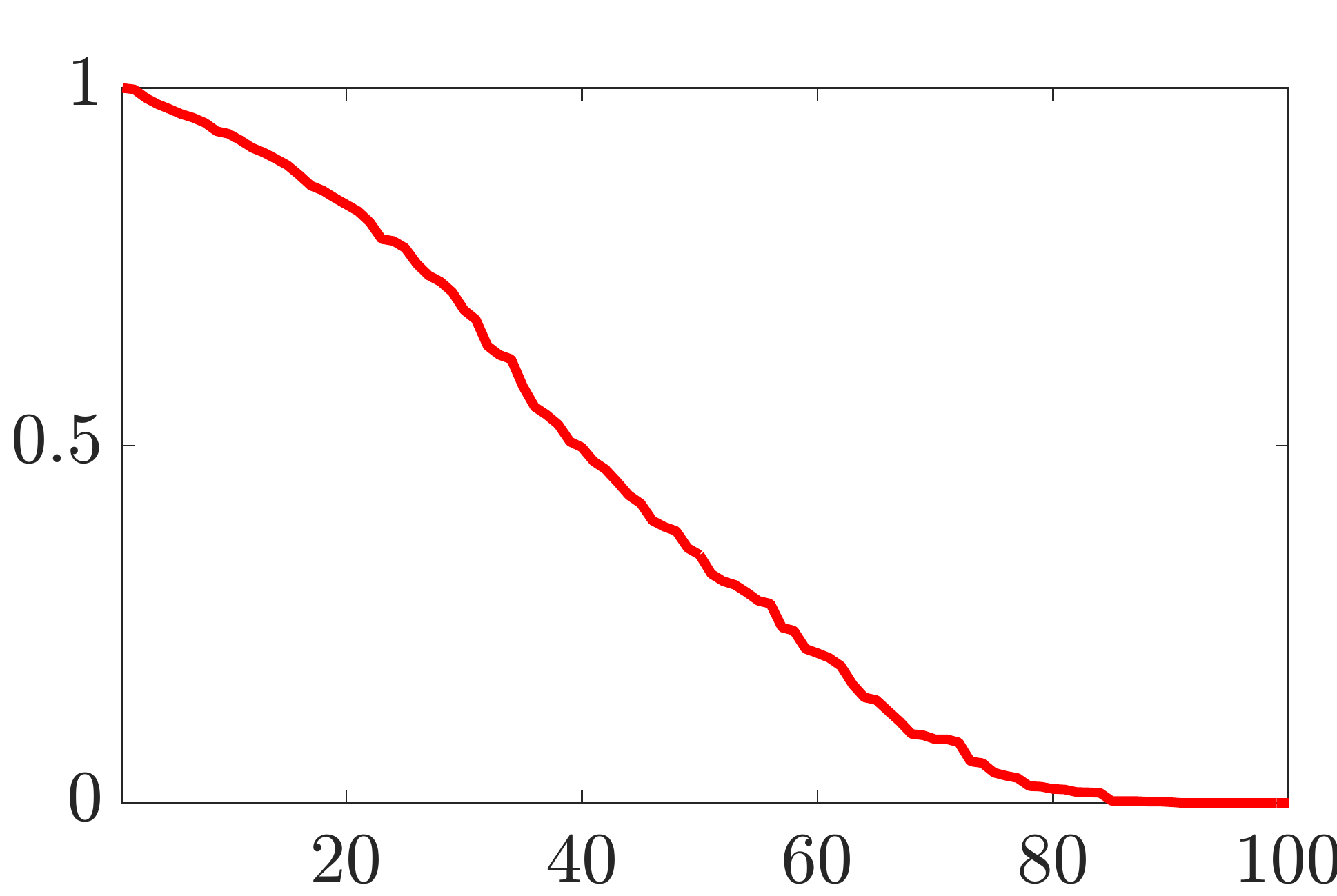}} }
		\end{subfigure}
	\end{tabular}
	\caption{MINST. (a) The ground truth handwritten digit images. (b)  The pixel coefficient's decay. } \label{Fig: 1}
\end{figure}     

\subsection{Algorithm complexity} \label{acceleration}
Dominant computation  is  
 the computation in the sparse signal estimation (Algorithm~\ref{Alg1}) involving    
  
\emph{1)} \textit{Matrix inversion} in  Eq.~\eqref{eq1.8.1.7}, in Eq.~\eqref{eq1.9.5}, and  in Eq.~\eqref{eq1.6.1} which   
costs $\mathcal{O}(N^3 + {M}^3 + \hat{k}^3)$. This cost can be  reduced  to $\mathcal{O}(2{M}^3)$ by employing matrix inversion properties. Let us first consider the matrix inversion in  Eq.~\eqref{eq1.8.1.7}. We can use the following matrix inversion property to reduce the cost of $\mathcal{O}(N^3)$ to $\mathcal{O}(M^3)$:  
\begin{equation}
\begin{aligned}
( & \mathbf{\Sigma'}_{x}^{-1}  + \frac{1}{\sigma_n }\vecb{V}^{T} \vecb{A}^{T} \vecb{A}\vecb{V})^{-1}  =   \mathbf{\Sigma'}_{x} - \\ &  \mathbf{\Sigma'}_{x}  \vecb{V}^{T}  \vecb{A}^{T} ( \sigma_n \vecb{I}  + \vecb{A}  \vecb{V}    \vecb{\Sigma'}_{x}  \vecb{V}^T \vecb{A}^{T}   ) ^{-1}  \vecb{A}  \vecb{V} \mathbf{\Sigma'}_{x}, 
\end{aligned} \label{Ch3_Eq_Complexity}
\end{equation}
where the matrix inversion requires $\mathcal{O}(M^3)$ and $M \ll N$. A similar technique used in Eq.~\eqref{eq1.6.1} can be used to transform the term $ (\sigma_n  \mathbf{\Sigma}_{x,s}^{-1}  + \vecb{A}_s^{T} \vecb{A}_s)^{-1}$ in Eq.~\eqref{eq1.6.1} such that it shares the similar term $ ( \sigma_n \vecb{I}  + \vecb{A}_s   \vecb{\Sigma}_{x,s}   \vecb{A}_s^{T}   ) ^{-1} $ to the matrix inversion  in Eq.~\eqref{eq1.9.5}; thus, the complexity of matrix inversion of these two equation is reduced from  $\mathcal{O}({M}^3 + \hat{k}^3)$ to $\mathcal{O}({M}^3)$. Consequently, the total complexity due to matrix inversion is $\mathcal{O}(2{M}^3)$.

\emph{2)} \textit{Matrix production} from the support  estimation Eq.~\eqref{eqAcc3}, the signal coefficient variance estimation  Eq.~\eqref{eq1.8.1.7}, in noise  estimation Eq.~\eqref{eq1.9.5}; and in estimating sparse coefficient Eq.~\eqref{eq1.6.1}.  The costs of the \textit{matrix production} can be  reduced by computing $\vecb{A}^{T} \vecb{A}$ off-line. Meanwhile, the value of $\vecb{A}^{T} \vecb{y}$ is needed to be computed only once and can be reused. After using the matrix property Eq.~\eqref{Ch3_Eq_Complexity}, the total complexity for matrix production is approximately $\mathcal{O}( 2 MN^2 + 4 M^2 N + 5N^2 + MN)$.

\emph{3)}  \textit{Support estimation} Eq.~\eqref{eqAcc3} which is solved by the belief propagation algorithm~\cite{zhang2016pairwise} that costs $\mathcal{O}(t_{max} |\mathcal{E}|)$ per iteration. $|\mathcal{E}|$ and $t_{max}$ denote the number of edges in $\mathcal{G}$ and the number iteration required by ~\cite{zhang2016pairwise}. Because $\mathcal{G}$ is constructed according to the graph update Algorithm~\ref{Ch3_Alg_graph} where the edge connects each signal coefficient to its non-zero neighbors in  $\mathbb{N}_i$ only, the complexity corresponding to the support estimation is low. That is, as $ |\mathcal{E}| = N|\mathbb{N}|$  where $\mathbb{N}$ is the largest set among $~\setn{\mathbb{N}_i}$, the complexity is linear with the number of nodes, \ie $\mathcal{O}(t_{max} N|\mathbb{N}|)$. 

Therefore, the total computational  complexity of Algorithm \ref{Alg1} is reduced to  $\mathcal{O}( 2{M}^3 + 2 MN^2 + 4 M^2 N + 5N^2 + MN + t_{max} N|\mathbb{N}|)$ in each iteration. Notice that matrix inversion and production takes up most of the computational cost; meanwhile, the cost of computing the support estimation is very low.   
The computation of Algorithm \ref{Alg1} is included in step 4 of Algorithm \ref{Alg2} (our Adaptive-MRF). The runtime performance of our Adaptive-MRF is provided in Figure~\ref{Run_Time_1} and Table~\ref{Tabel_long} (see Section~\ref{Comp_cost_exp}). Our method requires moderate runtime among other methods.

\section{Experimental Results and Analysis} 

We study the performance of the proposed Two-steps-Adaptive MRF through three different experiments: (i) To study the effectiveness of the adaptive mechanism, we study the performance of the adaptive MRF in comparison with a fixed MRF in Section \ref{Effect_AdaptiveMRF}; (ii) We study the performance of our proposed sparse signal estimation that solves Eq.~\eqref{eq1.4} in comparison with the existing MRF-based methods \cite{Peleg2012,Garrigues2008} that solve Eq.~\eqref{sec3.1} in adaptive MRF framework  in Section \ref{Adaptive_imp}; and (iii) Finally, we compare the performance of the proposed Adaptive MRF with state-of-the-art competitors in compressibility, noise tolerance, and runtime in Section \ref{Result_Comparison}. \\

We test the performance on three datasets--- MNIST~\cite{MNIST_dataset}, CMU-IDB~\cite{CMU_IDB_dataset}, and  CIFAR-10~\cite{CIFAR10_dataset}, detailed in Section \ref{Dataset}. The experiment setting and comparison methods are given in Section \ref{Exp_setting} and  \ref{Comparison}. More results on the algorithm convergence, datasets, and visual results are provided in \textit{supplementary materials}.

\subsection{Dataset} \label{Dataset}
We evaluate the performance on three datasets---\textit{ i) MNIST  images}~\cite{MNIST_dataset} contain few long-continued lines and are strictly sparse. \textit{ii) CMU-IDB face images}~\cite{CMU_IDB_dataset} contains facial features have more dense and diverse information than MNIST's. \textit{iii) CIFAR-10 natural images}~\cite{CIFAR10_dataset} are   more diverse and less synthesized than the previous two datasets. The selected images from these datasets are shown in Figure~\ref{Fig: 1 Ground}, \ref{Fig: 2 Ground}, \ref{Fig: 3 Ground}.  The MNIST images are strictly sparse, as shown in the pixel decay Figure~\ref{Fig: 1Coeff}. The compression process can be applied onto the signals directly. Meanwhile, the sparse representation of CMU-IDB  and  CIFAR-10 images can be obtained by employing i) wavelet transform,  ii) discrete cosine transform (DCT), and iii) principal component analysis (PCA). Examples of the sparse representation of CMU-IDB and CIFAR-10 images in wavelet, DCT, and PCA domains are in Figure~\ref{Fig: 2 sparse_signal_wave}, \ref{Fig: 2 sparse_signal_dct}, \ref{Fig: 2 sparse_signal_pca} and Figure~\ref{Fig: 3 sparse_signal_wave}, \ref{Fig: 3 sparse_signal_dct}, \ref{Fig: 3 sparse_signal_pca}. Most of these signal representations are compressible or sparse, except the PCA representation of CIFAR-10 images that is very dense which violates the sparsity assumption of compressive sensing.  

Therefore, we  mainly focus experimental results and analysis on (i) MNIST images,  (ii) the PCA representation of CMU-IDB images, and (iii) the wavelet representation of CIFAR-10 images.  MNIST images and wavelet representation demonstrates the signal structure modeling in 2D. PCA and DCT representations demonstrate the structure modeling in 1D. 

The experimental results on the recovery of DCT and wavelet representation of CMU-IDB images and DCT representation of CIFAR-10 images are provided in \textit{Section I-B in the supplementary material}.  

\subsection{Experimental settings} \label{Exp_setting}
 
 In the compression, the sparse signal $\mbox{\boldmath{$x$}}$ is sampled by a random Bernoulli matrix $\mbox{\boldmath{$A$}}$ to generate the linear measurements $\mbox{\boldmath{$y$}}$. The recovery performance is tested across different  sampling rates ($M/N$), \ie, 0.2, 0.25, 0.3, 0.35, and 0.4.  To simulate the noise corruption on measurements, $4$ different levels of Gaussian white noise are added into $\mbox{\boldmath{$y$}}$, which results in the signal to noise ratio (SNR) to be 5, 10, 20, and 30~dB\footnote{The noise level (in SNR) from 5~dB to 30~dB is the highest to the lowest noise corruption}.

\textbf{Algorithm Setting:} The proposed  Adaptive-MRF  (Algorithm \ref{Alg2}: the main algorithm) will stop when the minimum update difference of $\vecb{x}$ from step 4 is less than $10^{-3}$, or when the iteration reaches to five. In step 2, the graph update is performed where  $\mathbb{N}_i$ covers 8 neighboring nodes for capturing 2D structure in MNIST and wavelets. Meanwhile, $\mathbb{N}_i$ is set to cover two-adjacent nodes for capturing 1D structure in PCA and DCT signals. In step 3, the MRF parameters are  estimated using the package  \cite{parise2005learning} where the maximum iteration is set to 20.  In step 4, the sparse signal estimation is performed by Algorithm \ref{Alg1} that terminates  when the minimum update difference of $\vecb{x}$, \ie $\frac{|| \mathbf{x}^{prev} - \mathbf{x}^{new}||_2}{||\mathbf{x}^{prev}||_2},$ is less than $10^{-3}$, or when the iteration reaches 200.

\textbf{Evaluation criterion:} We demonstrate the proposed Adaptive-MRF performance on recovery accuracy, noise tolerance, and runtime performance.  The recovery accuracy is evaluated by peak signal to noise ratio (PSNR). The runtime performance is studied across different sampling rates ($M/N$).

\subsection{Comparison methods} \label{Comparison}

Our method is compared with $9$ state-of-the-art competitors: 
\begin{itemize}
\item  \textbf{Existing MRF-based methods}:  MAP-OMP\footnote{The graphical model, noise and signal variance parameters provided to MAP-OMP and Gibbs is from training data.}~\cite{Peleg2012}, Gibbs\footnotemark[2]
~\cite{Garrigues2008} ---whose support estimation are based on solving the optimization problem Eq.\eqref{sec3.1} with heuristic and stochastic approaches --- and LAMP \cite{Cevher12009} that imposes additional conditional independence assumptions;  
\item \textbf{Clustering structured sparsity-based methods}: MBCS-LBP\footnote{For both MBCS-LBP and Pairwise MRF, we use 8-neighbor system to capture local cluster structure}~\cite{Wang2015ISAR} and Pairwise MRF\footnotemark[3]~\cite{yu2015}; 
\item\textbf{Graph sparsity-based methods}: GCoSamp~\cite{hegde2015nearly} and StructOMP~\cite{huang2011learning} 
\item \textbf{Sparsity-based methods}: RLPHCS~\cite{LeiZhang} and   OMP~\cite{Tropp2007}.
\item We use \textit{the oracle estimator} suggested in \cite{Peleg2012} that uses \textbf{the  ground truth support}  to estimate the signal (via Eq.~\eqref{eq1.6.1}) with homogeneous noise and signal parameters from training. Note that all other methods do not have the access to the ground truth support. The oracle estimator has this unfair advantage.
\end{itemize}

All of the comparison methods, except Pairwise MRF~\cite{Wang2015ISAR}, are implemented by the code of the authors with tuned parameters to the best performance. For Pairwise MRF, we implemented the code ourselves. Here, 8-neighboring system is used as the local cluster structure, and we set the Pairwise MRF algorithm to terminate when the minimum update difference is less than $10^{-3}$, or when iteration reaches 200.

\subsection{Effectiveness of Adaptive-MRF}  
\label{Effect_AdaptiveMRF}
\begin{figure}[t]    
 	\begin{tabular}{ccc}
 		\multicolumn{3}{l}{\hspace{-1cm} {\includegraphics[width =3in,trim=0cm 15cm 0cm 13.cm,clip]{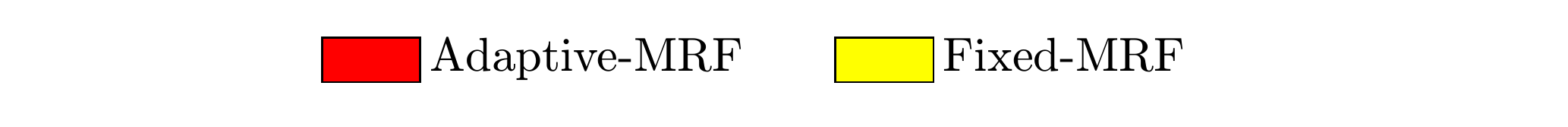}} \vspace{-0.05cm}}\\
 		\hspace{-0.75cm}    {\includegraphics[width =1.5in,trim=1.5cm  5cm  4cm 13.8cm,clip]{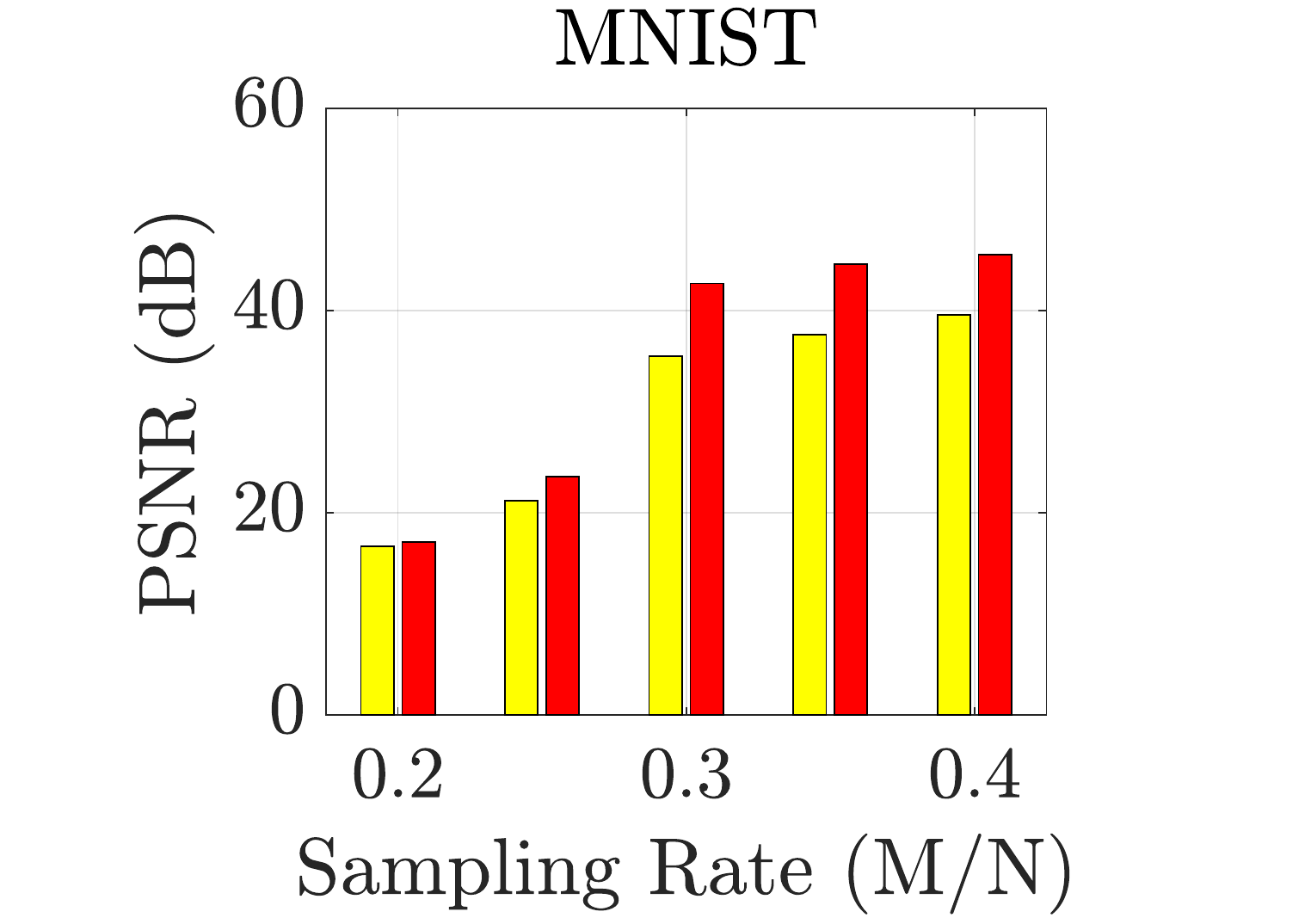}}  
 		&
 		\hspace{-1.4cm}   
 		{\includegraphics[width =1.5in,trim=1.5cm  5cm  4cm 13.8cm,clip]{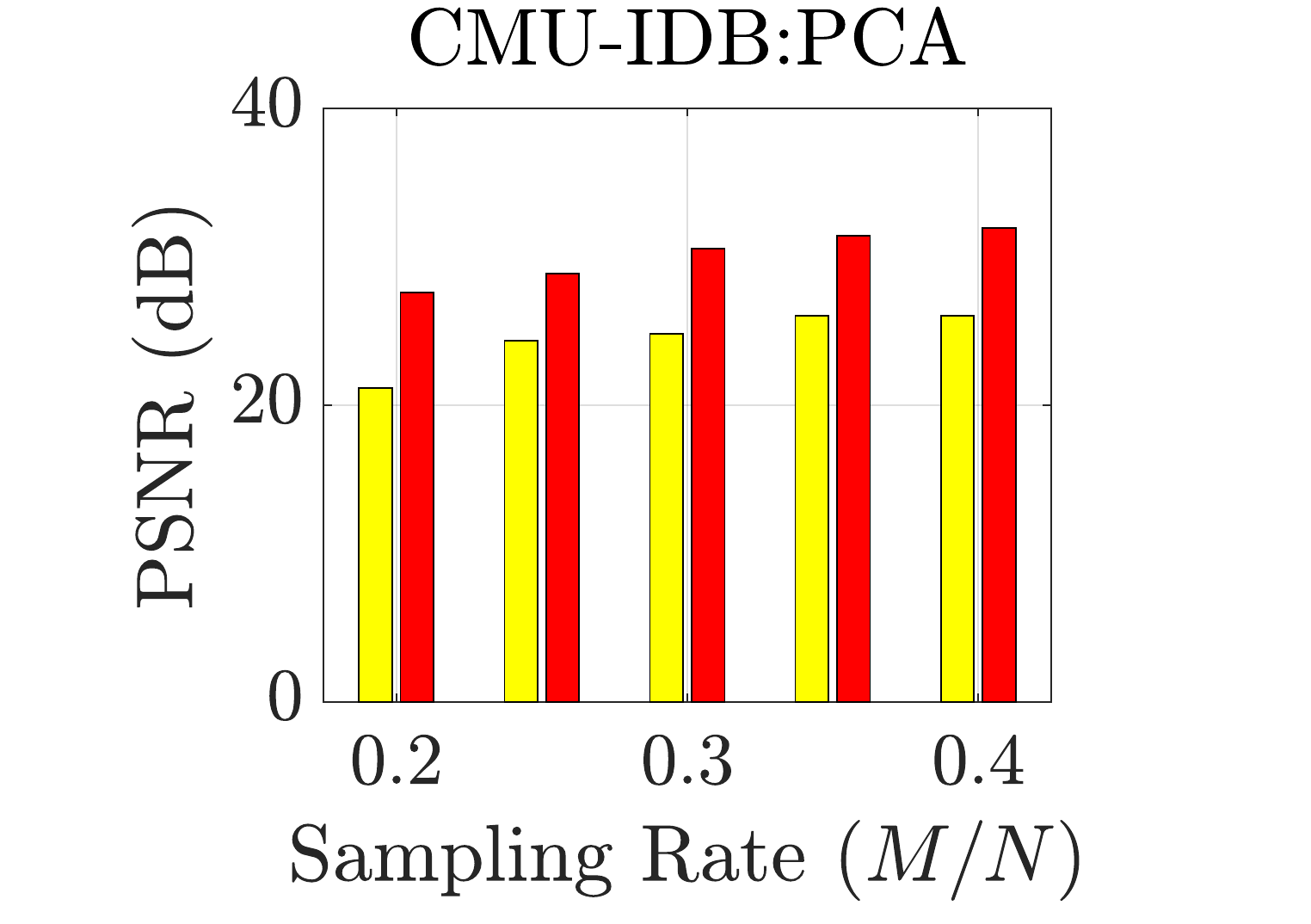}}     
 		&
 		\hspace{-1.4cm}   
 		{\includegraphics[width =1.5in,trim=1.5cm  5cm  4cm 13.8cm,clip]{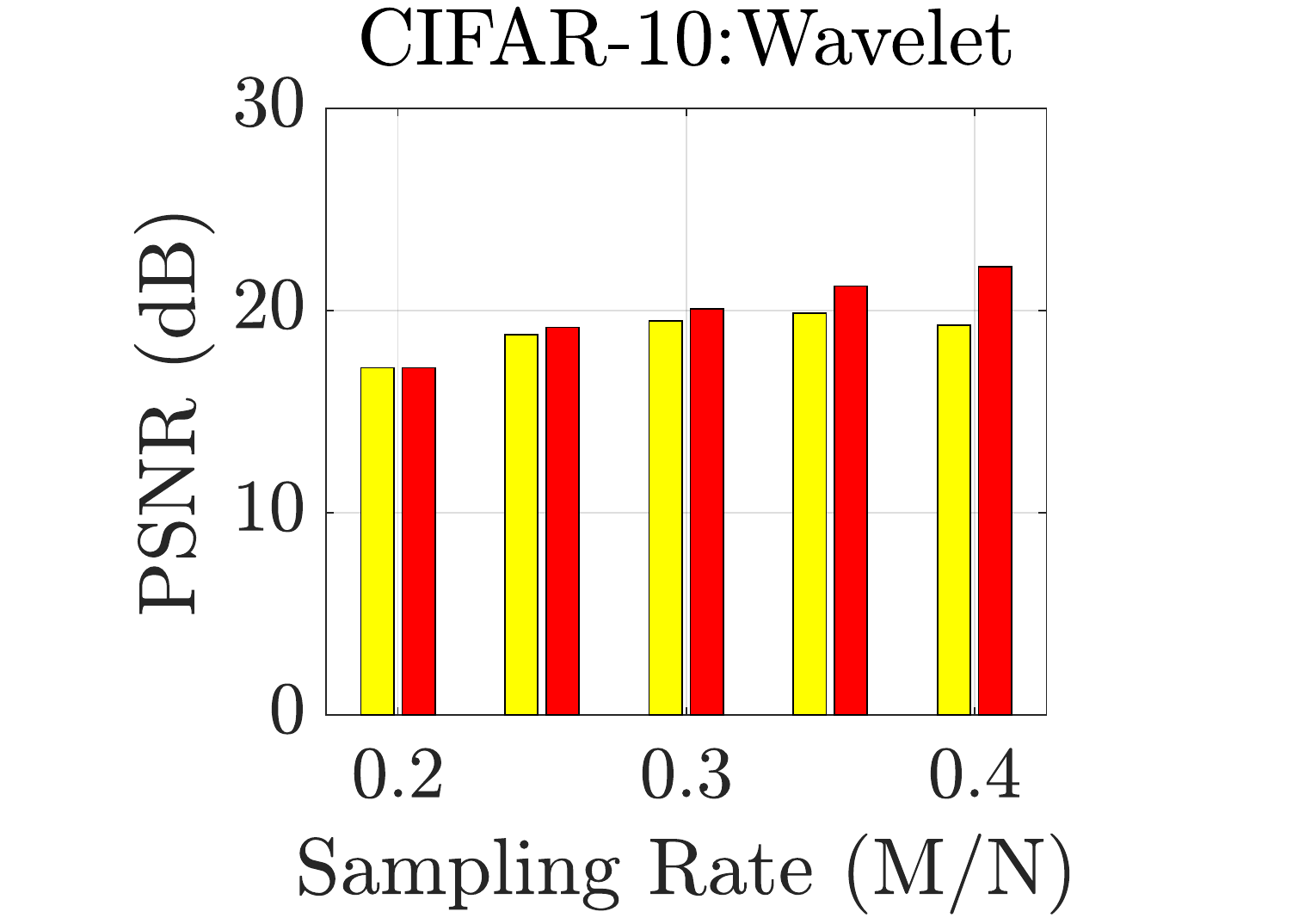}}
 	\end{tabular}
 	\caption{ Adaptive-MRF vs. Fixed-MRF on MNIST dataset,   CMU-IDB dataset, and CIFAR-10 dataset under noise level (SNR) of 30 dB. } \label{Adaptive_Fix_MRF}
 \end{figure}

To reflect the improved performance due to adaptive MRF inference framework, we compare the performance of the proposed Adaptive-MRF against the performance when an MRF is fixed. To employ the fixed MRF, we employ the sparse signal estimation (Algorithm \ref{Alg1}) where the MRF is obtained from training, which is fixed through out the sparse signal estimation. Thus, we denote it  as \textit{Fixed-MRF}. Figure \ref{Adaptive_Fix_MRF}  shows the bar graph of the average PSNR value across different sampling rates on the three datasets---MNIST images, PCA representation of CMU-IDB images, and wavelet representations of CIFAR-10 images--- at noise level (SNR) of 30 dB. It is clear that Adaptive-MRF outperforms Fixed-MRF in all cases. In particular, when sampling rate $(M/N)$ is higher than 0.2, Adaptive-MRF outperforms Fixed-MRF by at least 2 dB on MNIST images, 3 dB on CMU-IDB images, and 0.5 dB on CIFAR-10 images. 
 
\begin{figure}[t]  
	\begin{tabular}{cc}
		\multicolumn{2}{l}{  {\includegraphics[width = 3in,trim=4cm 13.7cm 3.5cm 13cm,clip]{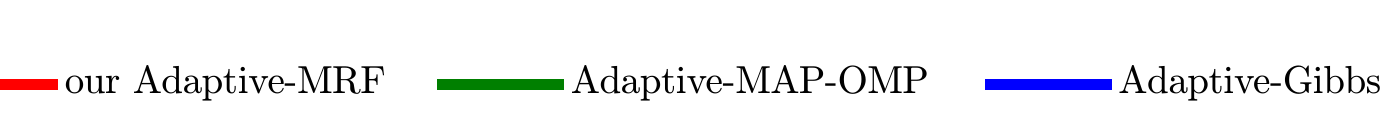}}  } \\
		\hspace{-0.5cm} \subcaptionbox{Accuracy\label{Adaptive_3alg:PSNR} }  {\includegraphics[width =2in,trim=2.5cm  9.5cm  2.5cm 9.cm,clip]{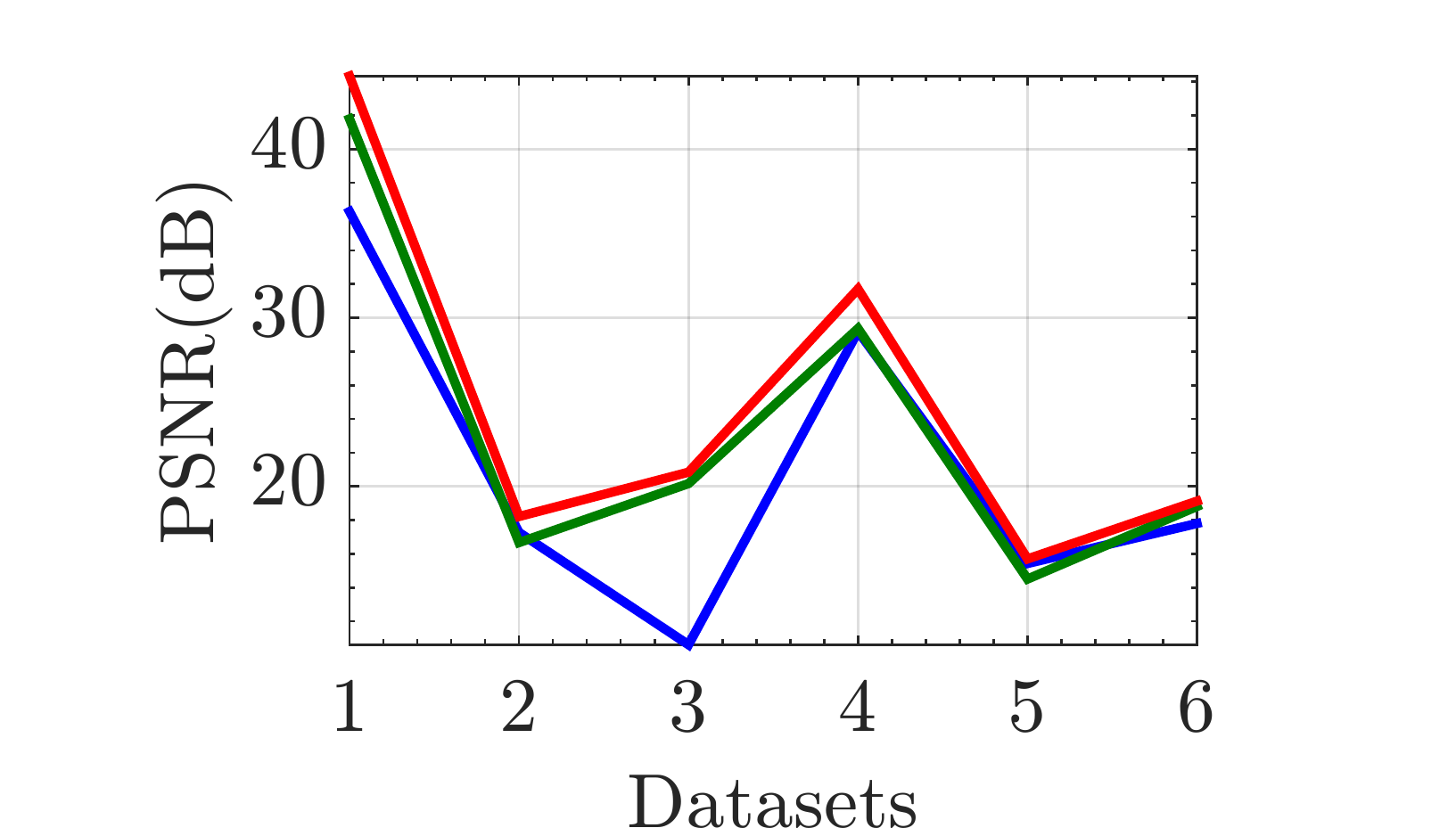}}    
		& \hspace{-1.5cm}  \subcaptionbox{Runtime\label{Adaptive_3alg:Time} }
		{\includegraphics[width =2in,trim=2.5cm  9.5cm  2.5cm 9.cm,clip]{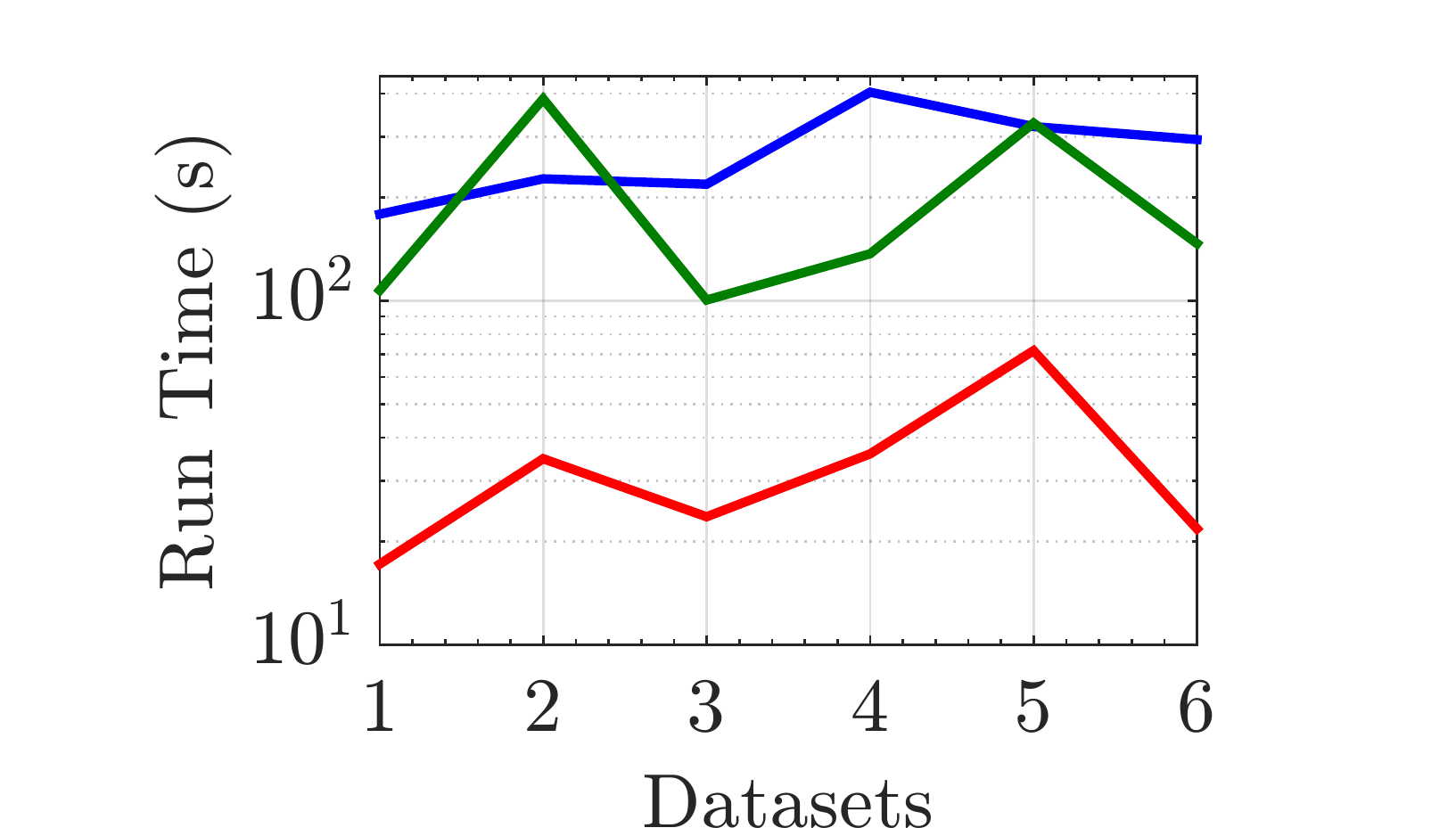}} 
	\end{tabular}  
	\caption{  Solving Eq.~\eqref{eq1.4} (our Adaptive-MRF) vs solving Eq.~\eqref{sec3.1} (Adaptive-Gibbs and Adaptive-MAP-OMP) in (a) recovery accuracy and (b) total runtime on 6 image sets: (1) MNIST, (2)(3)(4) CMU-IDB in wavelet, DCT, and PCA domains, and (5)(6) CIFAR-10 in wavelet and DCT domains. The sampling rate is 0.3 and noise level is 30 dB. } \label{Adaptive_3alg}  
\end{figure}    
\subsection{Effectiveness of the proposed sparse signal estimation.} \label{Adaptive_imp}  
In this section, we demonstrate the effectiveness of our sparse signal estimation (Algorithm~\ref{Alg1}) that aims to optimize the new formulation Eq.~\eqref{eq1.4} which allows the sparse signal, support, noise and signal parameters to be estimated jointly and recursively. Here, we compare our sparse signal estimation against Gibbs~\cite{Garrigues2008} and MAP-OMP~\cite{Peleg2012} that attempt to solve Eq.~\eqref{sec3.1} with non-recursive two-step approach that, first, estimates the support and, then, estimates the sparse signal. Gibbs and MAP-OMP also employ homogeneous noise and signal parameters from training data. 
All the algorithms are tested in the same adaptive MRF framework setting. Thus, we compare our Adaptive-MRF against Adaptive-Gibbs (Gibbs + the adaptive MRF framework) and Adaptive-MAP-OMP (MAP-OMP + the adaptive MRF framework).  The adaptive MRF framework performs at the main-loop. The sparse signal estimation performs at the inner-loop. The main loop is set to terminate when the number of iteration reaches to 3.  All the algorithms that are employed to perform sparse signal estimations and terminate when the iteration reaches 1000, or when minimum update differences between two consecutive estimate $\vecb{x}$ is less than $10^{-5}$.  

Figure~\ref{Adaptive_3alg} illustrates the recovery performance across six datasets (no. 1-6): no. (1) denotes MNIST handwritten images; no. (2)(3)(4) denote sparse representation  of CMU-IDB face images in wavelet, DCT, and PCA domain; and no.  (5)(6) denote sparse representation  of CIFAR-10 natural images in wavelet and DCT domains. The performance is tested at the sampling rate and noise level (SNR) of  0.3 and 30 dB, respectively. It is clear that  Adaptive-MRF requires the least runtime and provides the highest accuracy in all cases.

\begin{figure}[t]   
	\begin{tabular}{cc}
		\multicolumn{2}{l}{  {\includegraphics[width =3 in,trim=4cm 13.7cm  3.5cm 13cm,clip]{Figure/App_Result/EffectivenessofAdaptiveMRF/label.pdf}} \vspace{-0.15cm} } \\
		\hspace{-0.5cm} \subcaptionbox{Convergence of accuracy \label{Adaptive_Conv:PSNR} }  {\includegraphics[width =1.9in,trim=1cm  8.5cm  1cm 8cm,clip]{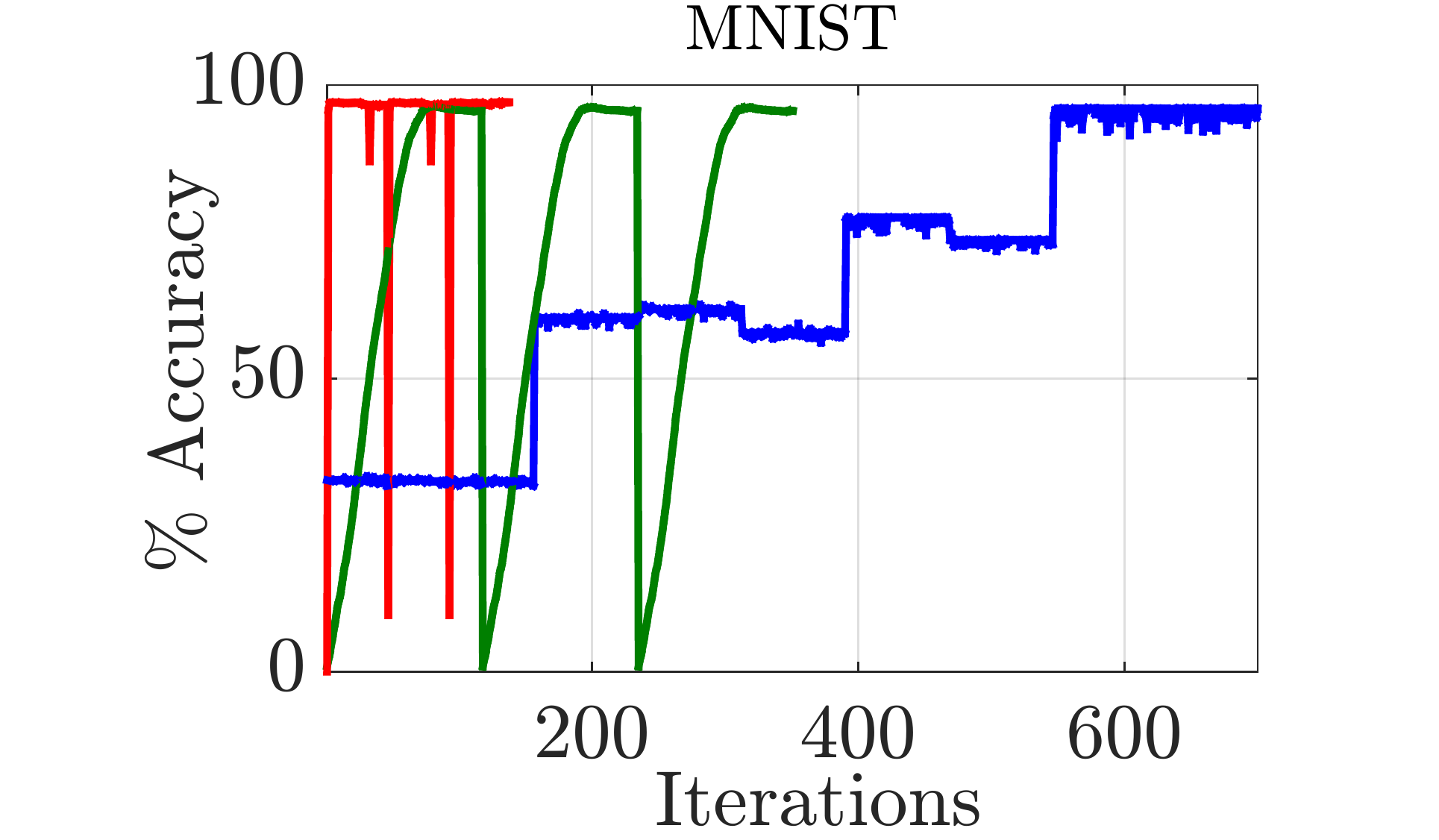}} 
		& \hspace{-1.0cm}  \subcaptionbox{Runtime per iteration\label{Adaptive_Conv:Time} }
		{\includegraphics[width =1.9in,trim=1cm  8.5cm  2cm 8cm,clip]{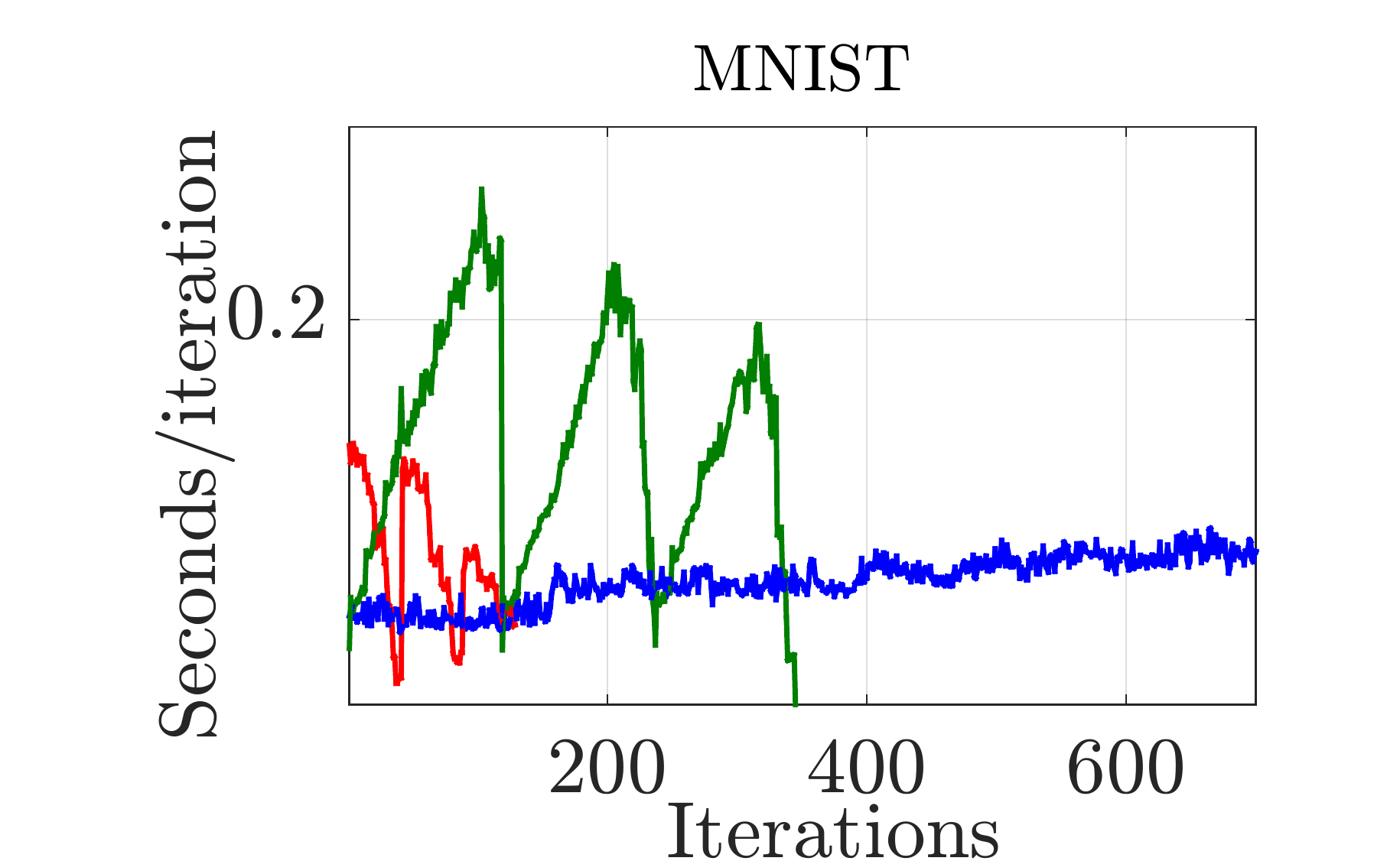}}   
	\end{tabular}
	\caption{  Solving Eq.~\eqref{eq1.4} (our Adaptive-MRF) vs solving Eq.~\eqref{sec3.1}  (Adaptive-Gibbs and Adaptive-MAP-OMP) in terms of (a) convergence and (b)  runtime per iteration on MNIST dataset.  The sampling rate  and noise level are 0.3 and 30 dB.   } \label{Adaptive_Conv} 
	
\end{figure}

Then, we further examine the convergence of average accuracy and runtime per iteration of our Adaptive-MRF against Adaptive-Gibbs and Adaptive-MAP-OMP. Figure \ref{Adaptive_Conv:PSNR} and  \ref{Adaptive_Conv:Time} further examine the convergence of average accuracy and runtime per iteration of our Adaptive-MRF  on MNIST  images. Here, \textit{iterations} on the horizontal axis denotes the total iterations of the sparse signal estimation that is performed at the inner-loop. It is clear that Adaptive-MRF takes much less iterations to converge. Note that there are three ripples on the runtime and accuracy curves of both Adaptive-MRF and Adaptive-MAP-OMP according to the setting where the main-loop performs three times. The ripples of Adaptive-Gibbs cannot be seen because it converges much slower (at around 2000 iterations). The ending of each ripple does not appear as a sharp vertical drop because they are resulted from averaging over 10 images. Adaptive-MRF converges the fastest with the highest accuracy because our sparse signal estimation jointly and recursively estimates sparse signal and support. More results on CMU-IBD and CIFAR-10 datasets are provided in Section I-C of the supplementary material which is consistent with the result in the main paper.

\begin{figure*}[t]   
	\begin{minipage}{1\textwidth}
		\hspace{2cm} \includegraphics[width = 5.7in,trim=4.5cm 12cm 4.0cm 5.5cm,clip]{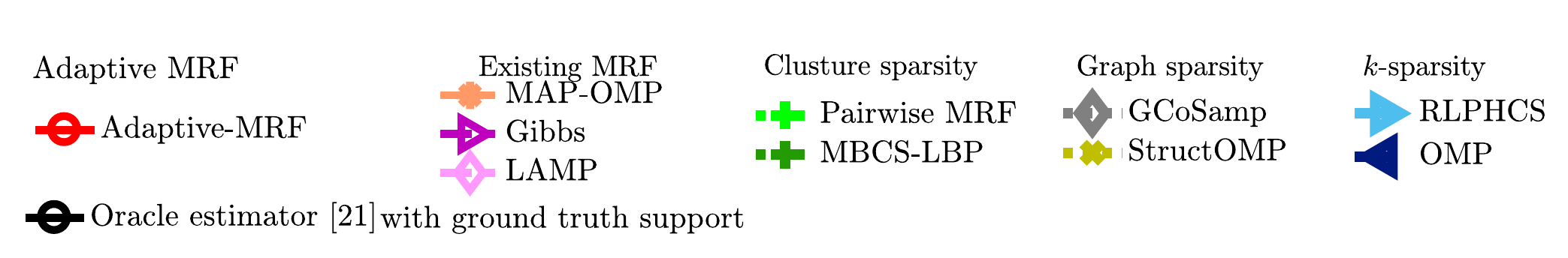}
	\end{minipage}  
\end{figure*}

\subsection{Performance evaluation} \label{Result_Comparison}
In this section, we compare the performance with several state-of-the-art methods over three  datasets: \textit{(1)} ten MINST images; \textit{(2)} ten CMU-IDB face images in PCA domain; and \textit{(3)} ten  CIFAR-10 natural images in wavelet domain.  The compression is performed on these sparse representations. Then, all methods are conducted to recover each image from a few linear measurements.  \\
  
\subsubsection{Compressibility}  
\begin{figure}[t] 
	\vspace{-0.5cm}  
	\begin{tabular}{ccc}   
		\hspace{-0.5cm}   {\includegraphics[width =1.55in,trim=3cm 8.5cm 4.5cm 10.5cm,clip]{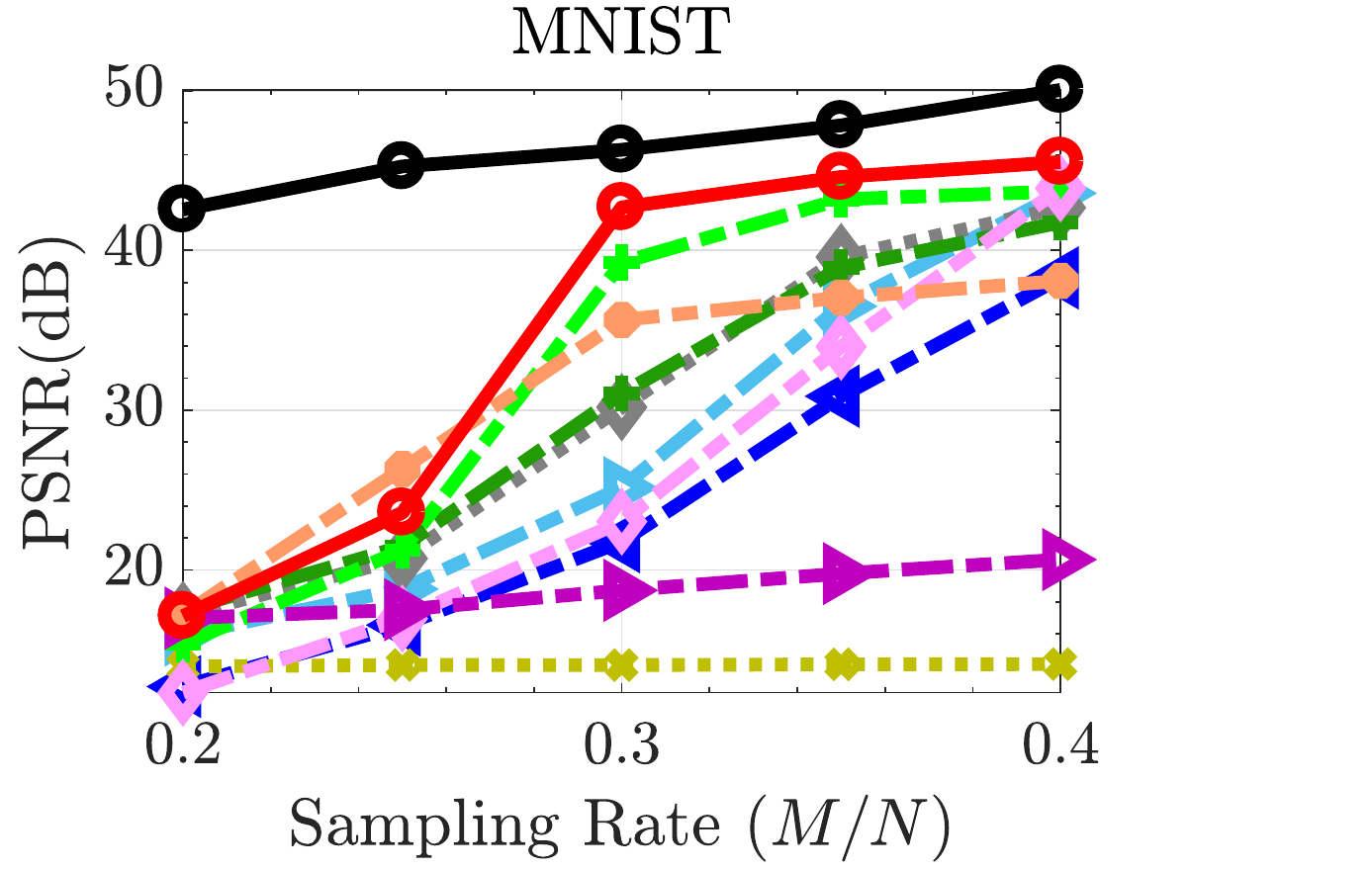}} 
		&
		\hspace{-1.1cm}  
		{\includegraphics[width =1.38in,trim=4.5cm 8.05cm 4cm 12.5cm,clip]{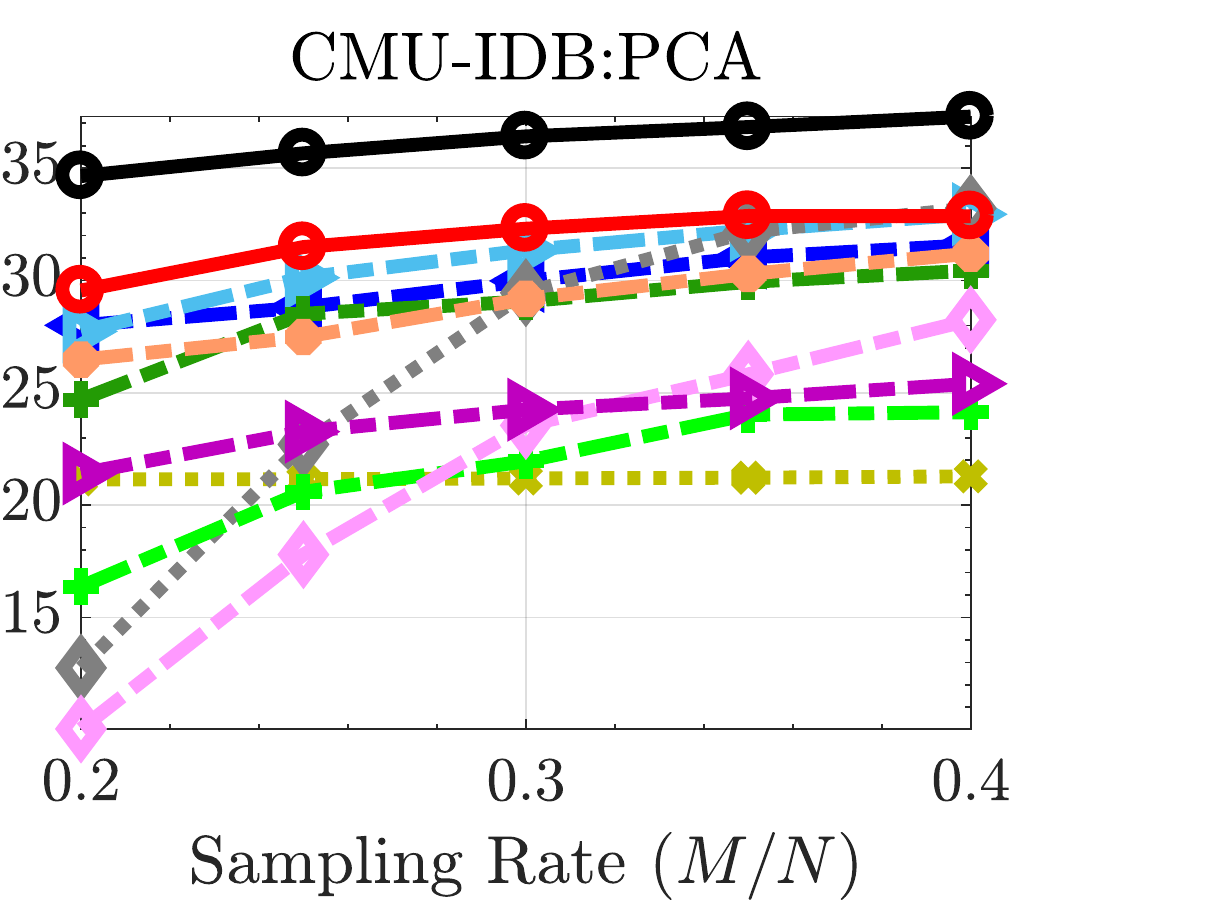}}     
		&
		\hspace{-1.1cm}  
		{\includegraphics[width =1.42in,trim=4.5cm 8cm 4cm 10.5cm,clip]{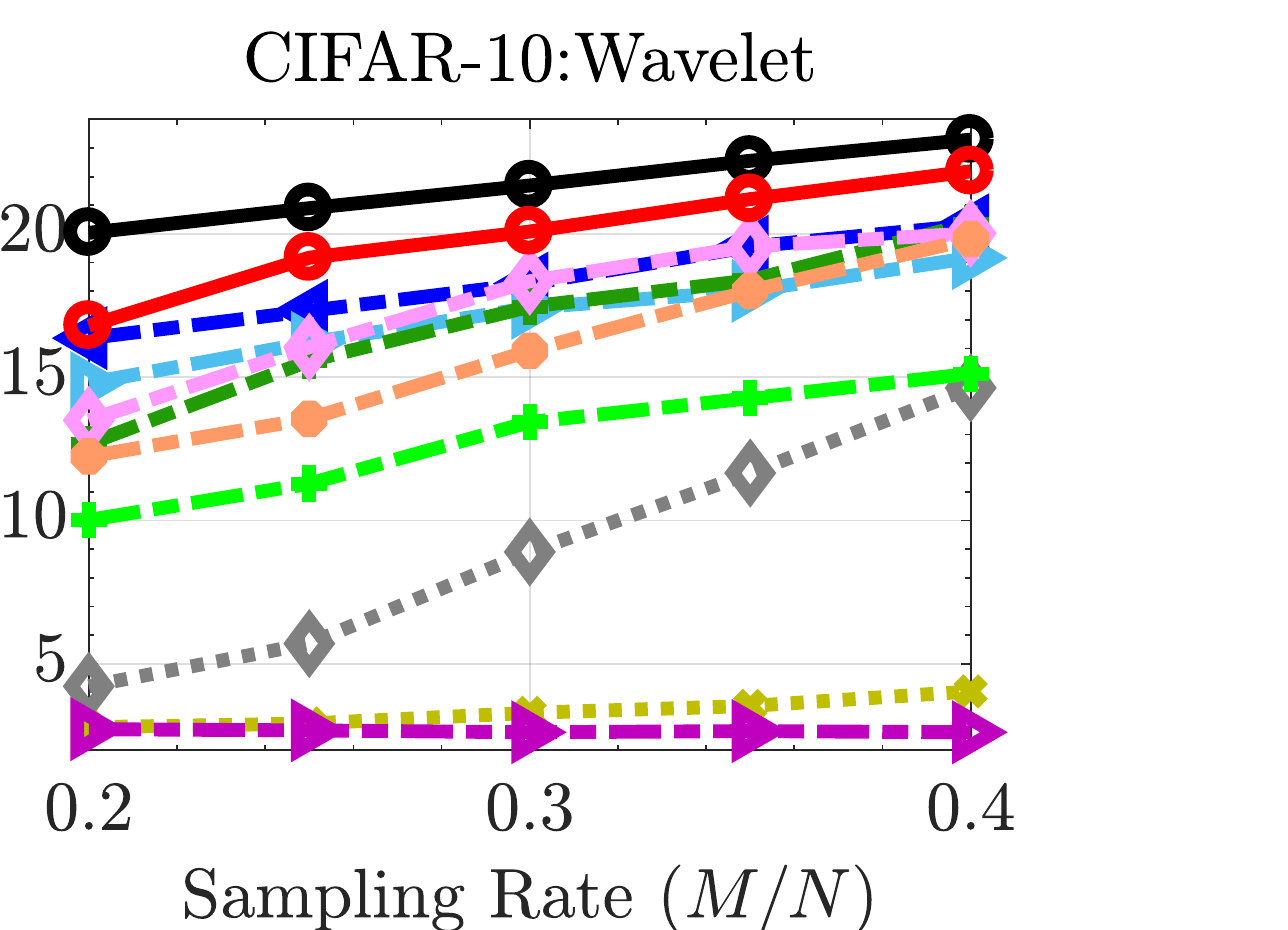}} 
	\end{tabular}
	\caption{Compressibility. The PSNR curves across different sampling rates on three datasets at noise level (SNR) of 30~dB. } 
	\label{Compressibility}
	\vspace{-0.2cm}
\end{figure}

To demonstrate the compressibility performance of the proposed Adaptive MRF, we evaluate  the recovery performance across different sampling rates ($M/N$). Figure \ref{Compressibility}  shows the average PNSR curves across different sampling rates on the three datasets. The noise level (SNR) is $30$~dB.  When the sampling rate is higher than  0.25, it exceeds other competitors by at least 1 dB on MNIST dataset and 1 dB on CIFAR-10 dataset. Meanwhile, on CMU-IDB, it offers the similar performance to other methods such as RLPHCS and GCoSamp that achieve the highest performance, but when the sampling rate is lower than 0.3 (less measurements), the proposed method outperforms the others by at least 0.25 dB.    
 
\begin{figure*}  [t]  
\setlength{\tabcolsep}{1pt} 
\renewcommand{\arraystretch}{0.5}
\begin{tabular}{cccccccccccc}
 \scriptsize{OMP} &  \scriptsize{RLPHCS} &  \scriptsize{StructOMP} &  \scriptsize{GCoSamp}   &
  \scriptsize{MBCS-LBP} &    \scriptsize{Pairwise MRF} &   \scriptsize{LAMP}   &  \scriptsize{Gibbs} &  \scriptsize{MAP-OMP} & 
  \scriptsize{Fixed-MRF} & \scriptsize{Adaptive-MRF} & \scriptsize{Ground Truth} \\
 &  &      & & &  & & &  & \scriptsize{(Our method)} & \scriptsize{(Our method) } &  \\ 
{\includegraphics[width =  0.5 in]{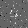}} & 
{\includegraphics[width =  0.5 in]{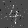}}  &
{\includegraphics[width =  0.5 in]{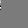}}  &
{\includegraphics[width =  0.5 in]{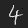}} 
 &
{\includegraphics[width =  0.5 in]{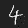}}  &
{\includegraphics[width =  0.5 in]
{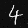}}  &
{\includegraphics[width =  0.5 in]{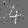}}  & 
{\includegraphics[width =  0.5 in]{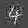}}  &
{\includegraphics[width =  0.5 in]{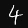}} &   
{\includegraphics[width = 0.5 in]{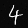}}  &
{\includegraphics[width =0.5 in]{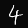}}  &
 {\includegraphics[width =  0.5 in]{Figure/App_Result/MNIST/Original/ori_label5}}
 \\ 
\scriptsize{12.99~dB} & \scriptsize{16.14~dB} &  \scriptsize{13.01~dB} & \scriptsize{32.93~dB}  &  \scriptsize{27.62~dB} &  \scriptsize{31.51~dB} &
\scriptsize{15.62~dB} &  \scriptsize{17.85~dB} & \scriptsize{42.42~dB} &  \scriptsize{41.88~dB} &  \scriptsize{\textbf{44.42~dB}} & 
\\
\end{tabular}
\centering
\caption{Visual results of some selected MNIST digit images (at $M/N$ = $0.3$, SNR = $30$~dB).} \label{Fig: 4}
\end{figure*}

\begin{figure*}  [t] 
\setlength{\tabcolsep}{1pt} 
\renewcommand{\arraystretch}{0.5}
\begin{tabular}{cccccccccccc}
 \scriptsize{OMP} &  \scriptsize{RLPHCS} &  \scriptsize{StructOMP} &  \scriptsize{GCoSamp}   &
  \scriptsize{MBCS-LBP} &    \scriptsize{Pairwise MRF} &   \scriptsize{LAMP}   &  \scriptsize{Gibbs} &  \scriptsize{MAP-OMP} & 
  \scriptsize{Fixed-MRF} & \scriptsize{Adaptive-MRF} & \scriptsize{Ground Truth} \\
 &  &    &   &  & & &  &  &  \scriptsize{(Our method)} & \scriptsize{(Our method) } &  \\ 
 {\includegraphics[width =  0.5 in]{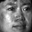}} &
{\includegraphics[width =  0.5 in]{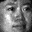}}  & 
{\includegraphics[width =  0.5 in]{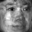}}  & 
{\includegraphics[width =  0.5 in]{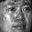}}  &
{\includegraphics[width =  0.5 in]{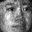}}  &
{\includegraphics[width =  0.5 in]{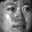}}  &
{\includegraphics[width =  0.5 in]{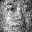}}  &
{\includegraphics[width =  0.5 in]{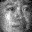}}  &
{\includegraphics[width =  0.5 in]{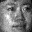}} &  
{\includegraphics[width = 0.5 in]{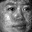}}  &
{\includegraphics[width = 0.5 in]{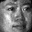}}  &
 {\includegraphics[width =  0.5 in]{Figure/App_Result/Face/Original/image2_Ori_c_th}}
 \\
    \footnotesize{32.32~dB} &  \footnotesize{32.86~dB} &  \footnotesize{26.68~dB}   & 
    \footnotesize{31.39~dB} &  \footnotesize{{28.25~dB}} &  \footnotesize{26.47~dB}   &  
    \footnotesize{22.81~dB} &    \footnotesize{27.46~dB} &   \footnotesize{{26.40~dB}} &    
    \footnotesize{31.42~dB} &   \footnotesize{{29.65~dB}} & \footnotesize{\textbf{33.50~dB}}  
\end{tabular}
\centering
\caption{Visual results of a selected CMU-IDB face images from PCA signal reconstruction (at $M/N$ = $0.3$, SNR = $30$ dB).} \label{Fig: 5_3}
\end{figure*}

\begin{figure*}  [t] 
\setlength{\tabcolsep}{1pt} 
\renewcommand{\arraystretch}{0.5}
\begin{tabular}{cccccccccccc}
 \scriptsize{OMP} &  \scriptsize{RLPHCS} &  \scriptsize{StructOMP} &  \scriptsize{GCoSamp}   &
  \scriptsize{MBCS-LBP} &    \scriptsize{Pairwise MRF} &   \scriptsize{LAMP}  &  \scriptsize{Gibbs} &  \scriptsize{MAP-OMP} & 
  \scriptsize{Fixed-MRF} & \scriptsize{Adaptive-MRF} & \scriptsize{Ground Truth} \\
 &  &      & & &  & & &  & \scriptsize{(Our method)} & \scriptsize{(Our method) } &  \\ 
  {\includegraphics[width =  0.5 in,angle =-90]{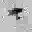}} &
 {\includegraphics[width =  0.5 in,angle =-90]{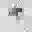}}  & 
 {\includegraphics[width =  0.5 in,angle =-90]{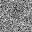}}  & 
 {\includegraphics[width =  0.5 in,angle =-90]{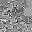}}  &
 {\includegraphics[width =  0.5 in,angle =-90]{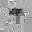}}  &
 {\includegraphics[width =  0.5 in,angle =-90] {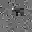}}  &
 {\includegraphics[width =  0.5 in,angle =-90] {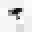}}  &
 {\includegraphics[width =  0.5 in,angle =-90] {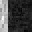}}  &
 {\includegraphics[width =  0.5 in,angle =-90]{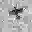}} &  
 {\includegraphics[width = 0.5 in,angle =-90]{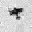}}  &
 {\includegraphics[width = 0.5 in,angle =-90]{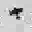}}  &
  {\includegraphics[width =  0.5 in,angle =-90]{Figure/App_Result/Tiny/Original/Im1}}
\\ 
\scriptsize{20.50~dB} & \scriptsize{18.36~dB} & \scriptsize{2.46~dB} & \scriptsize{4.31~dB} &  \scriptsize{16.55~dB} & \scriptsize{12.48~dB} & 
\scriptsize{20.26~dB} &    \scriptsize{1.17~dB}
 & \scriptsize{16.31~dB}  & \scriptsize{23.09~dB} & \scriptsize{\textbf{23.15~dB}}&  
\end{tabular}
\centering
\caption{Visual results of a selected CIFAR-10 images from wavelet signal reconstruction (at $M/N$ = $0.3$, SNR = $30$ dB).} \label{Fig: 6_3}
\vspace*{-0.5cm}
\end{figure*}

Next, we provide the visual results on selected images from MNIST, CMU-IDB, and CIFAR-10 in Figure~\ref{Fig: 4}, Figure~\ref{Fig: 5_3}, and Figure~\ref{Fig: 6_3}, respectively. Adaptive-MRF gives rise to the best results which contain more details  and  less noise than its competitors. The full visual results are provided in Section II in the supplementary material document. With the adaptive MRF, Adaptive-MRF offers the best performance in most cases. \\

\begin{figure}[t]  
	\begin{tabular}{ccc}    
		\hspace{-0.5cm}   {\includegraphics[width =1.55in,trim=3.5cm 8.0cm 4cm 10.5cm,clip]{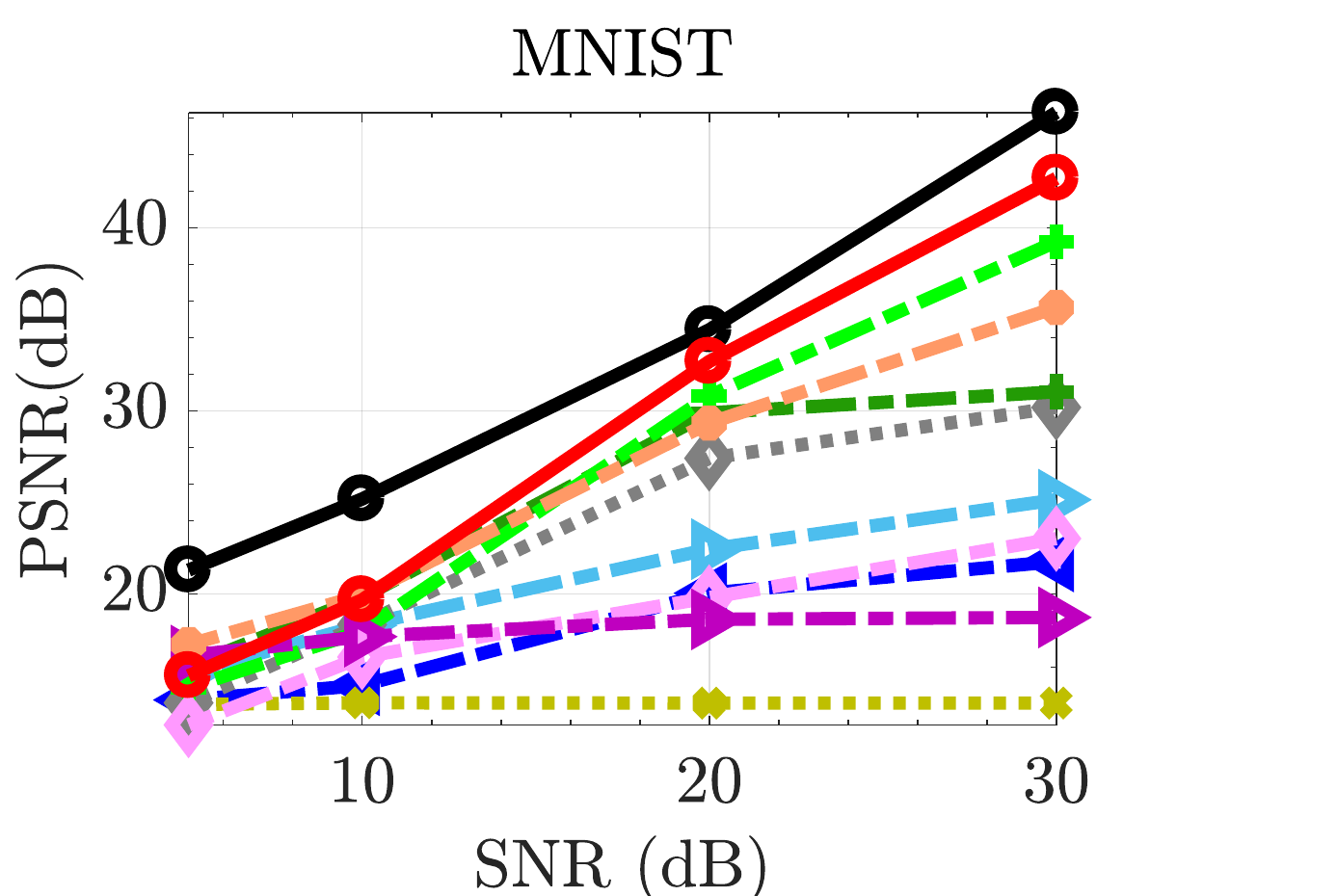}}   
		&
		\hspace{-1.1cm}   
		{\includegraphics[width =1.45in,trim=4.5cm 8.2cm 4cm 12.5cm,clip]{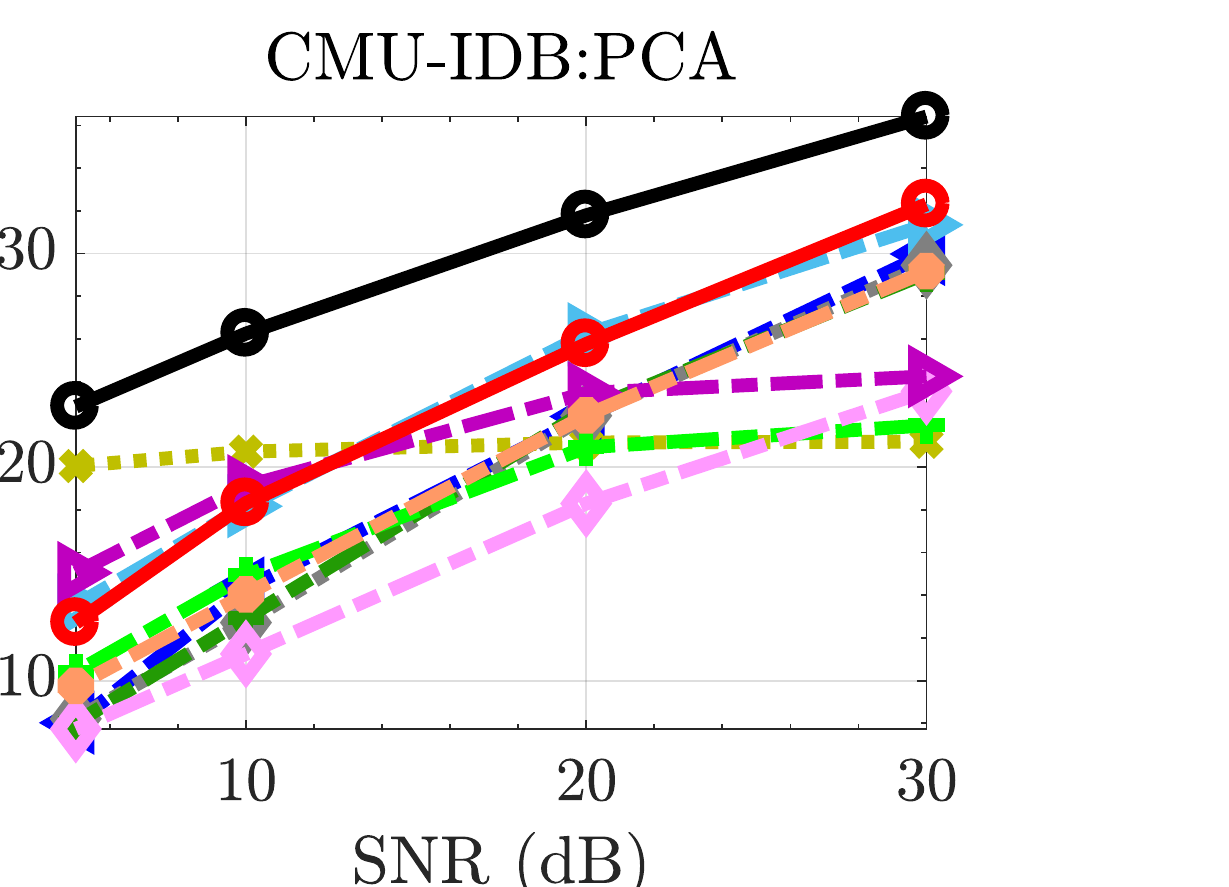}}     
		&
		\hspace{-1.3cm}  
		{\includegraphics[width =1.45in,trim=4.5cm 8.0cm 4cm 10.5cm,clip]{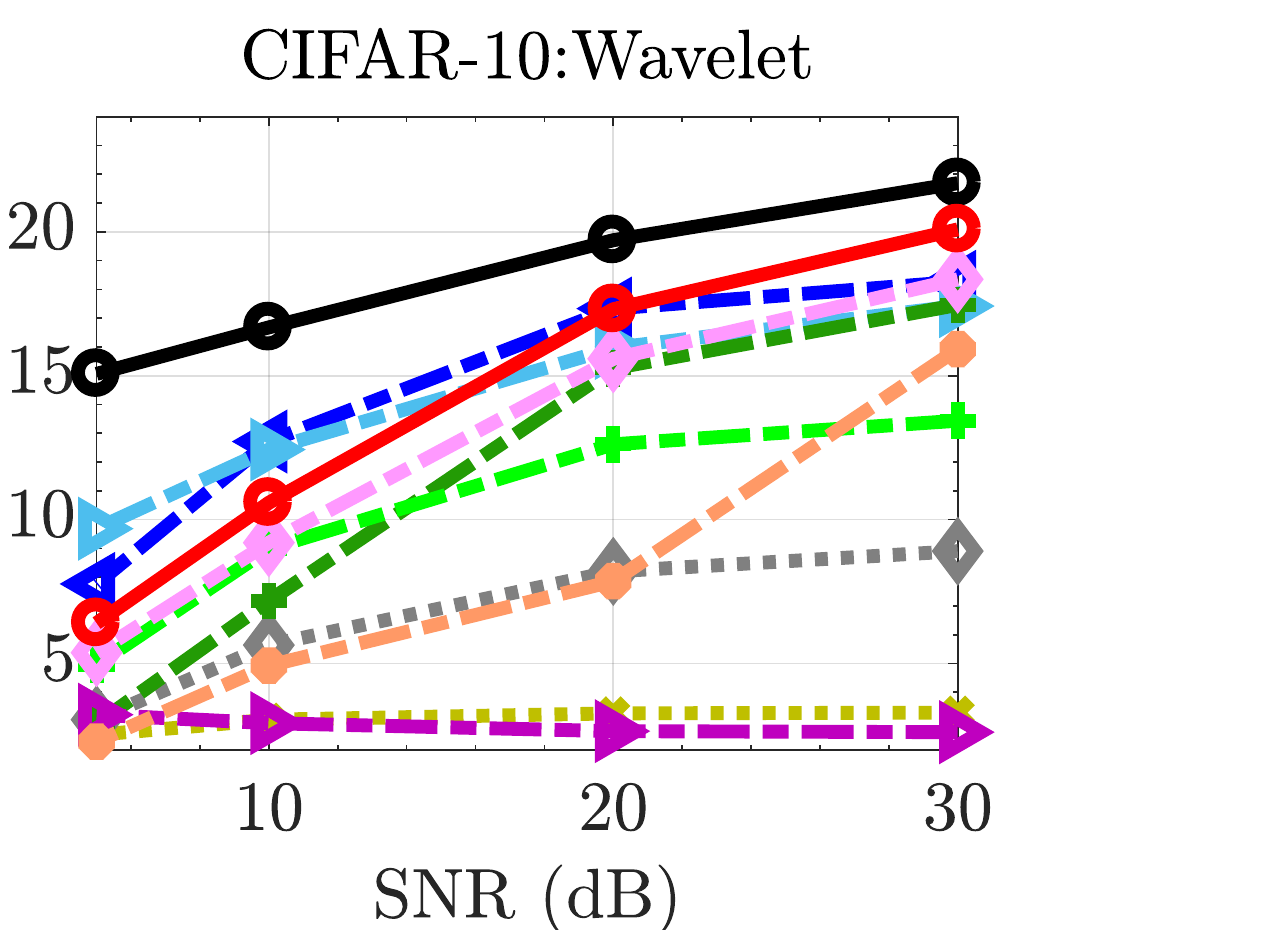}} 
	\end{tabular}
	\caption{Noise Tolerance. The PSNR curves across different noise levels (SNR) on three datasets at sampling rate of 0.3. } \label{Noise_Tolerance}
\end{figure} 
  
\subsubsection{Noise Tolerance}   
To demonstrate the noise tolerance performance, we test the performance of Adaptive-MRF across different noise levels (in SNR). Figure \ref{Noise_Tolerance}  provides the average PNSR curves across different noise levels (SNR) on the three datasets. The sampling rate is set to 0.3. Adaptive-MRF outperforms other competitors by at least 2 dB when noise level is higher than 5 dB on MNIST images. It also achieves the best performance in most cases on CMU-IDB face images and CIFAR-10 natural images. However, because the accuracy of an updated MRF is  based on the previously estimated sparse signal, Adaptive-MRF performance can be interfered by a certain noise level. For example, when noise level  is lower than $15$ dB, Adaptive-MRF is beaten by RLPHCS and OMP that do not exploit any signal structure in recovering CIFAR-10 natural images. 
This could be because the CMU-IDB face images and CIFAR-10 natural images contain more information and less structured than MNIST images. Thus,  the estimation for both images are more challenging, which results in less accurate estimation of MRF parameters. Nevertheless, when noise level becomes higher ($> 15$~dB), Adaptive-MRF outperforms the other methods.  \\

 \subsubsection{Computational Cost}  \label{Comp_cost_exp}
\begin{figure}[t] 
	\vspace{-0.7cm}
	\begin{tabular}{ccc}    
		\hspace{-0.5cm}   {\includegraphics[width =1.45in,trim=4.3cm 8.15cm 4.0cm 12.0cm,clip]{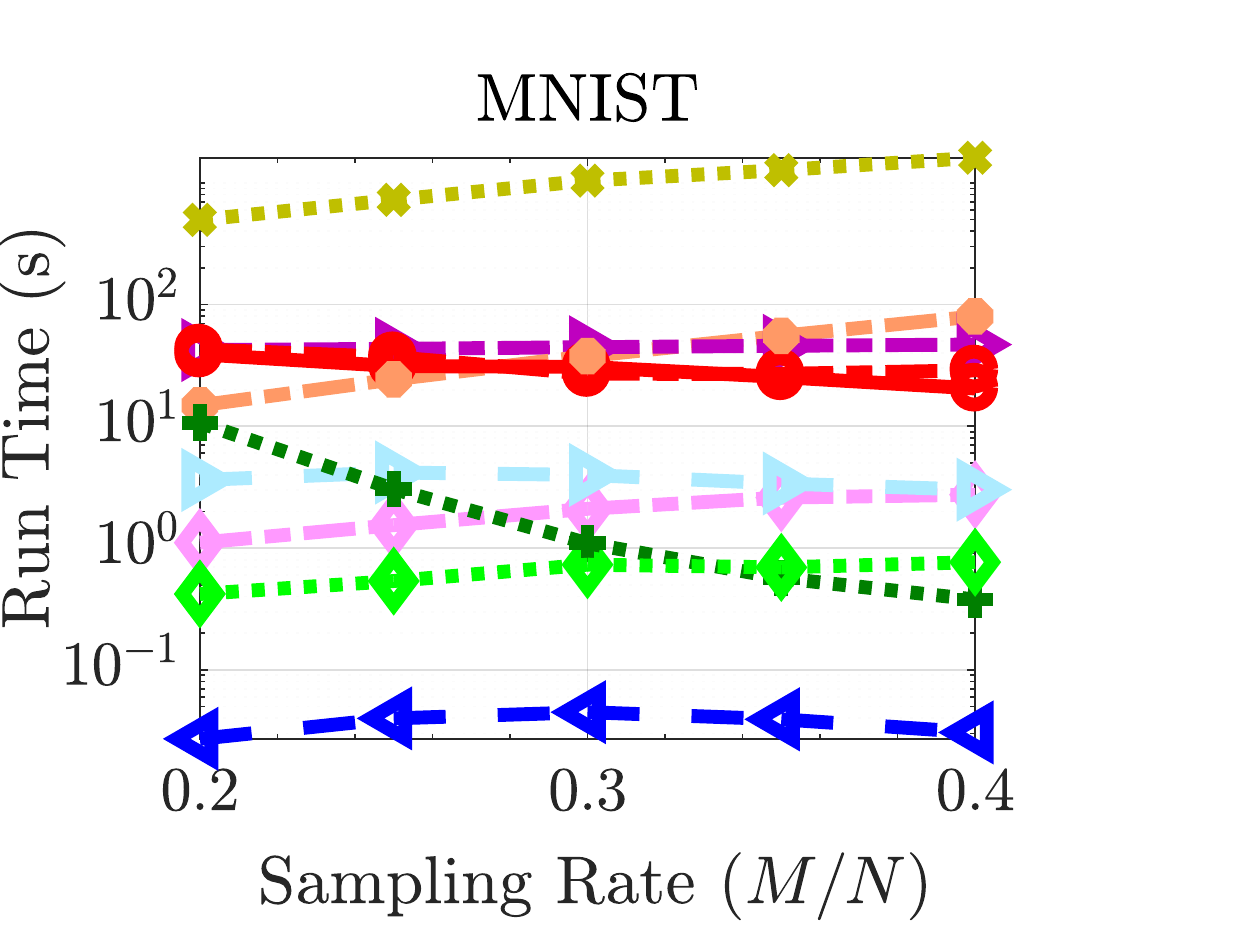}}  
		&
		\hspace{-1.25cm}  
		{\includegraphics[width =1.45in,trim=4.7cm 8.15cm 4cm 12.0cm,clip] {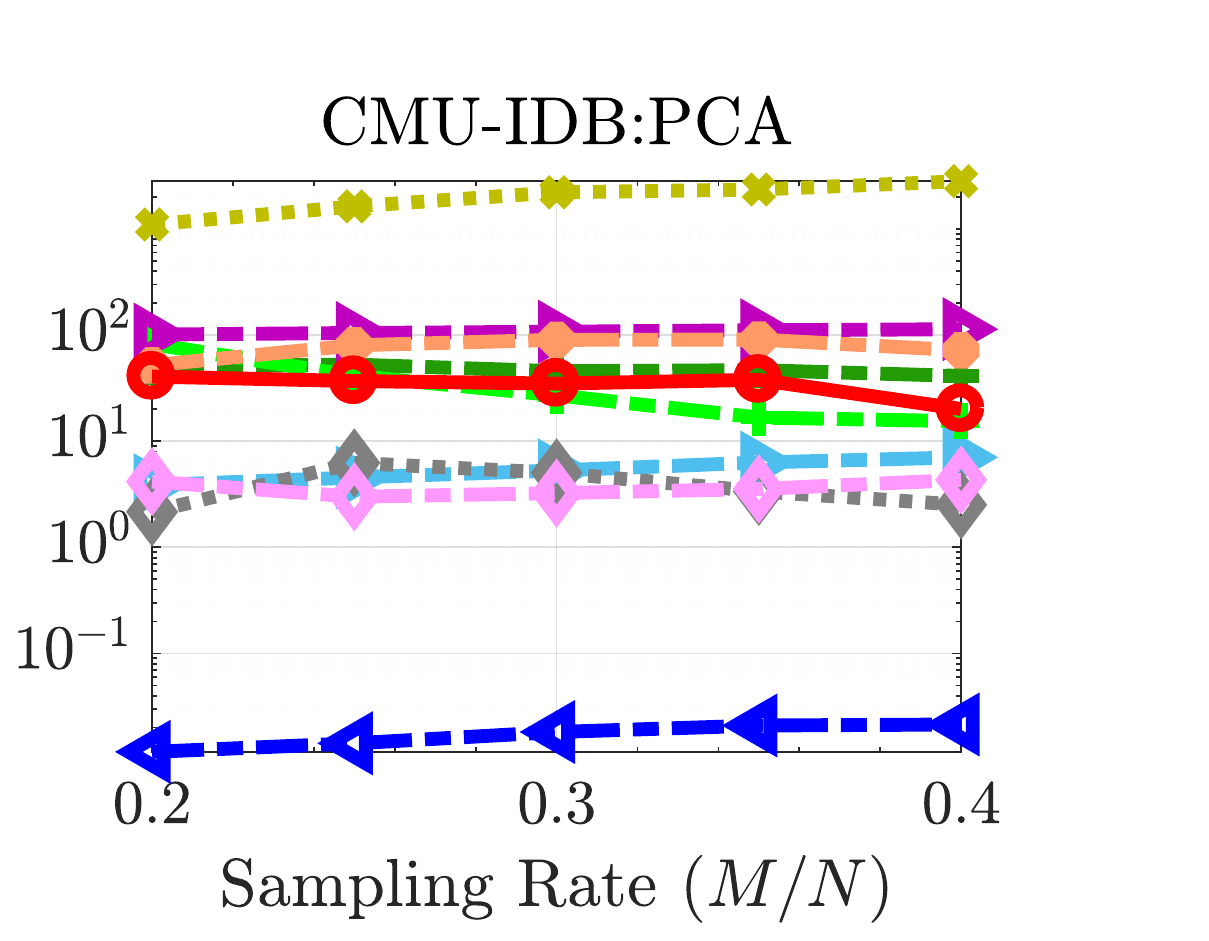}}     
		&
		\hspace{-1.22cm}  
		{\includegraphics[width =1.45in,trim=4.7cm 8.15cm 4cm 10.5cm,clip]{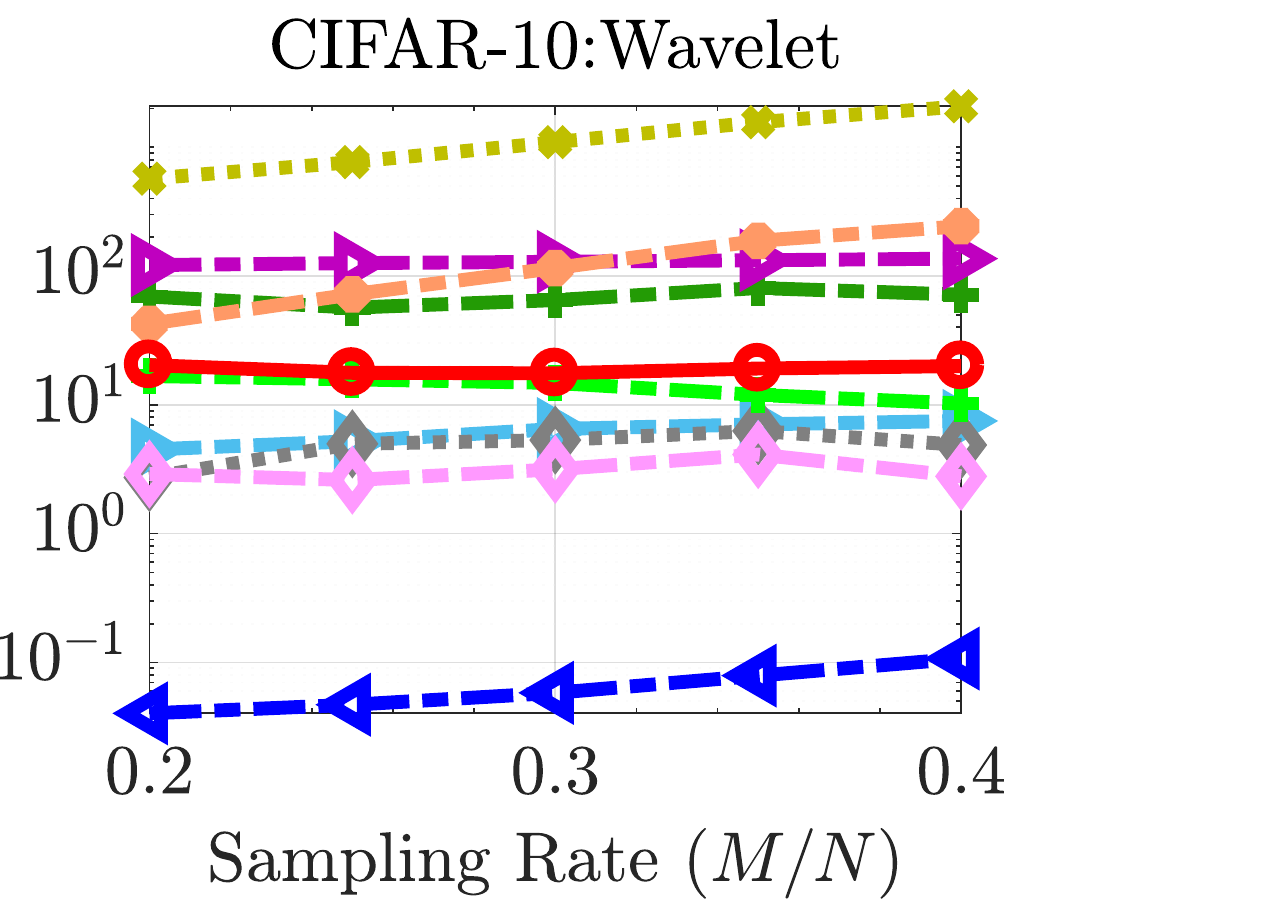}} 
	\end{tabular}
	\caption{Computational Cost. Runtime curves across different sampling rates on three datasets at noise level (SNR) of 30~dB. } \label{Run_Time_1}  
\end{figure}

 Figure \ref{Run_Time_1}  provides runtime performance at different sampling rates ($M/N$) on the three datasets. Noise level (in SNR) is 30 dB.Table~\ref{Tabel_long} provides the numerical result Adaptive-MRF shows to require a moderate runtime which is stable across different sampling rates. On MINST, the average runtime of our Adaptive-MRF is lower than StructOMP, comparable to MAP-OMP and Gibbs,  but higher  than Pairwise MRF, LAMP, MBCS-LBP, and GCoSAMP.  For CMU-IDB and CIFAR-10 datasets, our Adaptive-MRF  is faster than MAP-OMP, Gibbs, and StructOMP; is comparable to MBCS-LBP and Pairwise MRF; and is  slower than LAMP, GCOSAMP, OMP, and RLPHCS. Note that OMP and RLPHCS require less computation because they do not exploit signal structure.

\begin{table*}[t] 
	\centering
	\begin{tabular}{c|| c c c c c  c c c c c }  
		& \multicolumn{10}{c}{Comparison methods}\\ 
		Dataset & 
		\scriptsize{OMP} &  \scriptsize{RLPHCS} &  
		\scriptsize{StructOMP} &  \scriptsize{GCoSamp}   & 
		\scriptsize{MBCS-LBP} &    \scriptsize{Pairwise MRF} &   \scriptsize{LAMP} &  
		\scriptsize{Gibbs} &  \scriptsize{MAP-OMP}   & 
		\scriptsize{Adaptive-MRF (ours)}  \\  
		\hline \hline 
		MNIST & 0.045 & 3.984 & 1041.700 & 0.730 & 1.094 & 17.762 & 2.100 & 44.616 & 37.668 & \textbf{51.851}\\  
		CMU-IDB & 0.018 & 5.303 & 2216.700 & 5.033 & 45.664 & 27.100 & 3.245   & 107.650 & 90.306 & \textbf{34.941}  \\ 
		CIFAR-10 & 0.058 & 6.491 & 1095.000 & 5.326 & 65.010 & 14.781 & 3.157   & 128.930 & 115.360 & \textbf{17.568} \\ 
		\hline
	\end{tabular}
	\caption{Runtime comparison in seconds  at the sampling rate of 0.3$M$ corresponding to the performance reported in Figure~\ref{Run_Time_1}.\label{Tabel_long}}
\end{table*}

 \begin{figure}[t]   
	\vspace{-0.2cm} 
	\begin{tabular}{ccc}
			\multicolumn{3}{l}{ 
        	 {\includegraphics[width =3in,trim=3cm 14.5cm 0cm 13.cm,clip]{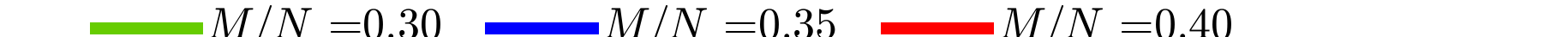}} }\\ 
			\hspace{-0.5cm}
			{\includegraphics[width =1.25in,trim=3cm 8.0cm 6.6cm 12cm,clip]{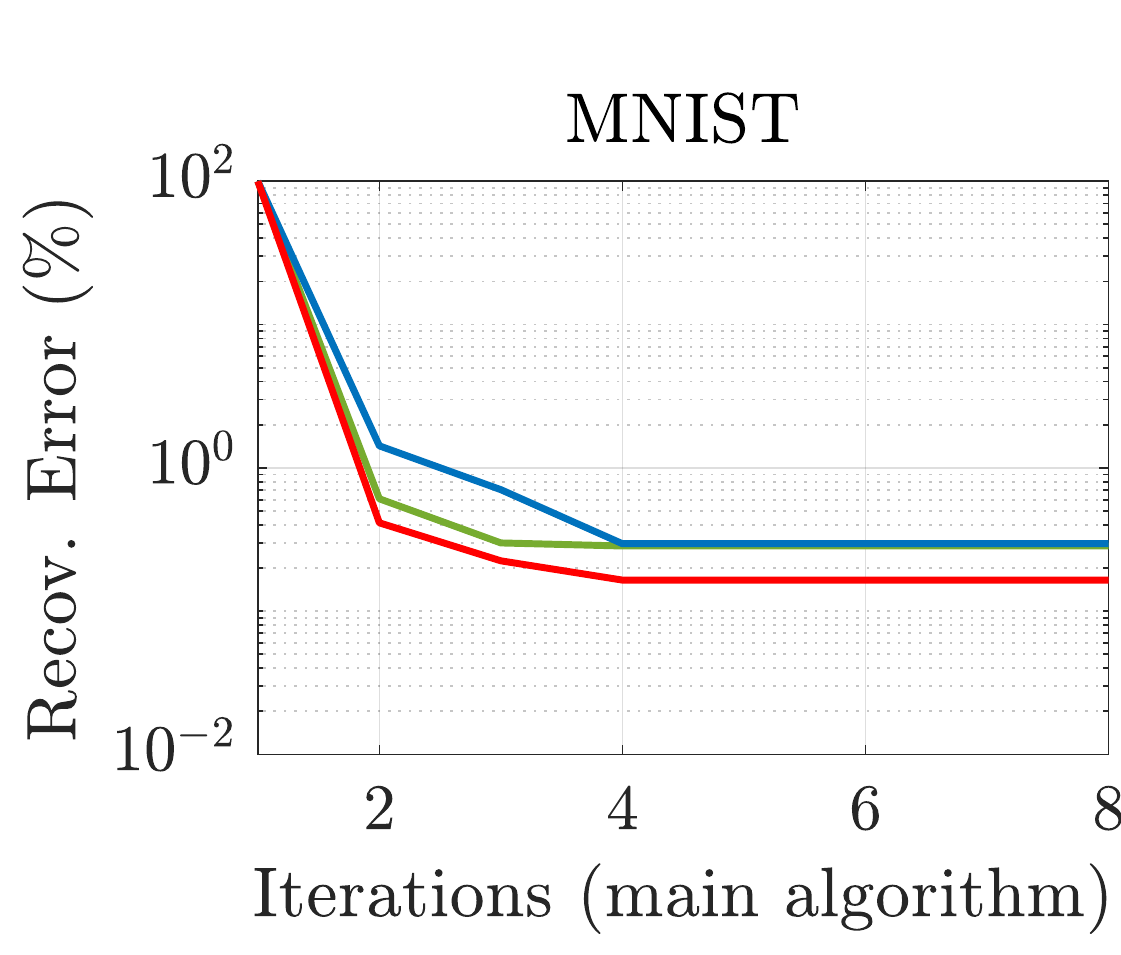}}  
			&
			\hspace{-0.5cm}   
			{\includegraphics[width=1.125in,trim=4cm 8.0cm 6.6cm 12cm,clip]{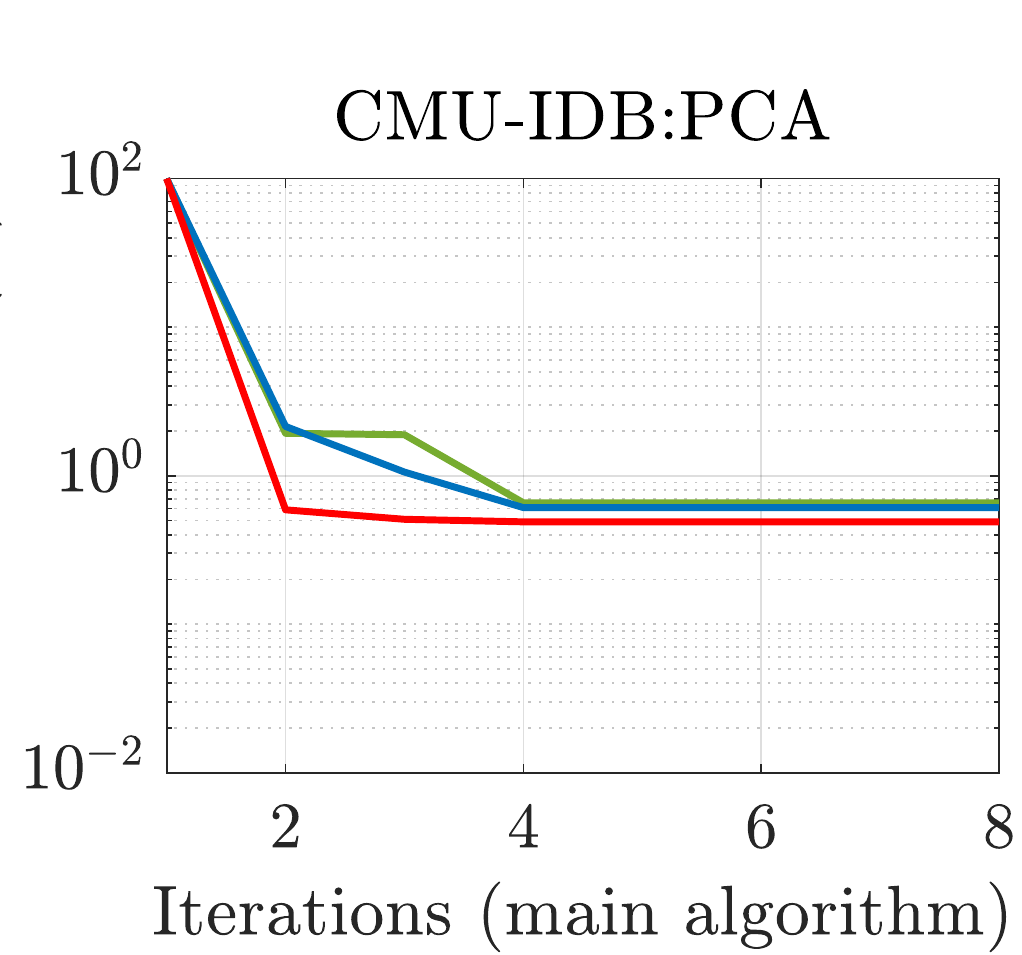}}     
			&
			\hspace{-0.5cm}   
			{\includegraphics[width =1.1in,trim=4.3cm 8.0cm 6.6cm 12cm,clip]{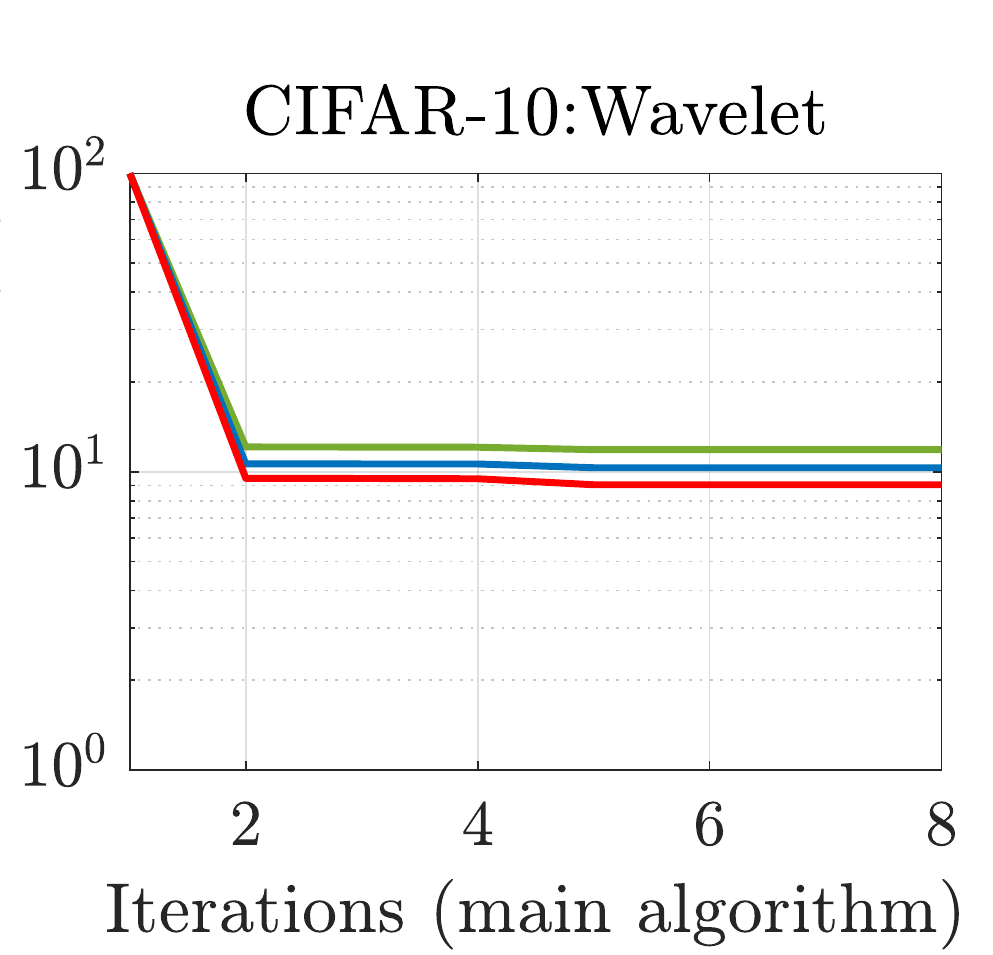}}
	\end{tabular}
\caption{ Empirical convergence of the Adaptive-MRF on MNIST dataset, CMU-IDB dataset, and CIFAR-10 dataset. Noise level (SNR) of 30 dB.} \label{Emp_convergence}
\end{figure}

\subsection{Convergence} 
\label{empirical_convergence}
We study the empirical convergence of the proposed method through the error of sparse signal estimation Eq.~\eqref{eq1.6.1}. From the point of view of Eq.~\eqref{eq1.6.1}, one may notice that the algorithm can converge within a few iteration, if the estimated support mostly resembles the high energy coefficients in sparse signal. The empirical convergence is provided in Figure~\ref{Emp_convergence}. We can see that the proposed method converges within a few iterations. Our Adaptive-MRF converges within 5 iterations on MNIST, CMU-IDB face images, and CIFAR-10 natural images. We provide empirical results of the empirical convergence on the rest of datasets in  \textit{ Section I-A,  supplementary material}. 

Notice we cannot provide the theoretical convergence of the proposed method due to several reasons. Firstly, from Eq.~\eqref{eq1.6.1}, the convergence of sparse signal estimation depends on the estimated supports from Eq.~\eqref{eq1.5}, but we cannot precisely define the convergence of the estimated supports that also depends on the estimation of other unknowns such as the adaptive MRF and noise and signal variances. Another possible direction is to analyze the convergence through the optimization problem Eq.~\eqref{eq.2.8} that has been reduced into two sub-problems, \ie optimizing Eq.~\eqref{eq1.4} and Eq.~\eqref{eq.2.11} in alternative minimization scheme. Still, theoretical convergence is difficult to analyze, since several coordinate spaces are involved with the two sub-problems.

\section{Conclusion}  
 We propose a novel graphical compressive sensing model to better capture the structure of sparse signals. Through imposing a graphical sparsity prior on the sparse signal, and adjusting all involved model parameters according to the observed measurements with a newly developed adaptive MRF inference framework, our model exhibits two important qualities in representing the structure of sparse signal, namely generality and adaptability. Experiments on three real-world datasets reveal that our model outperforms the state-of-the-art competitors in recovery accuracy and noise tolerance with stable runtime  across different sampling rates and signal sparsity. 
 
 The proposed method can tackle many types of sparse signals, \eg wavelet, DCT, and PCA signals, as evidenced by extensive experiments; thus, it can be exteded to many applications such as image restoration~\cite{zhang2018cluster,zhang2017beyond}, super-resolution~\cite{zhang2018exploiting,zhang2018adaptive}, and more. To carry out this work in future, the readers may follow our guideline of the methodology and algorithm settings as they are designed to be flexible for many datasets. To implement our work, a few parameters, \ie the maximum iteration and the neighboring coverage, are needed to be appropriately turned. Meanwhile, some theoretical questions regarding the algorithm convergence and performance bounds of signal recovery are difficult to analyze without precise knowledge about the class of sparse signals. With our best attempt, we answered   these questions with empirical results (\ie, the recovery performance in Section~\ref{Result_Comparison} and the empirical convergence in Section~\ref{empirical_convergence}). Additionally, we provided the theoretical result on the sample complexity in Section~\ref{Theoretical_Result}.    

 However, in most actual implementation, the class of sparse signal of the given task is known. Thus, the theoretical convergence as well as performance bounds of signal recovery can be analyzed by using  techniques such as those in~\cite{Baraniuk2010mbcs, 5967912}. In addition, the MRF parameter estimation and support estimation could be further customized to efficiently extract the underlying structure of sparse signals. Examples of how the MRF parameter estimation can be customized for a specific application  can be found in~\cite{zhang2018cluster,LeiECCV}.

\appendices
\section{Derivation of the sub-optimization problems}  
 
\subsection{Optimization over  $\vecb{s}$} \label{ApproxSupport}
In this section, we provide derivation to Eq.\eqref{eq1.7.1}. According to Eq.~\eqref{sup_eq_sub2.1_determinant}, the fourth term in Eq.~\eqref{sub_eq1.7} can be approximated with  summation of the logarithm of determinant of diagonal entries in  $(\sigma_n \vecb{I}  +   \vecb{A}_s \vecb{\Sigma}_{{x,s}}  \vecb{A}_s^{T})$ that is  positive definite. The optimization problem  Eq. \eqref{eq1.7}, given $\vecb{x}$, ${\sigma}_{n}$, and $\vecb{\Sigma}_{x,s}$, can be equivalently formulated as 
\begin{equation}
\begin{aligned}
& \min\limits_{\vecb{s}}   \quad
\frac{1}{2 \sigma_n}   \vecb{x}_{s} ^{T}   \vecb{A}_s   ^{T}  \vecb{A}_s  \vecb{x}_{s}
- \frac{1} {\sigma_n}  \vecb{y}^{T}  \vecb{A}_s \vecb{x}_s
+ \frac{1}{2} \vecb{x}_{s}^{T}  \vecb{\Sigma}_{{x,s}}^{-1}   \vecb{x}_{s}   \\
&+ \sum_{i \in \mathcal{V} }  { \log[ \vecb{\Sigma}_x]_{i,i}}  
+ \log [(\sigma_n  \vecb{\Sigma}_{{x}}^{-1}   +    \vecb{A}^{T} \vecb{A})]_{i,i}
-\log{p(\vecb{s};\hat{\vecb{\Theta}}_{\hat{{\mathcal{G}}}})}
\end{aligned}
\label{sub_eq1.7}
\end{equation}
Let $\vecb{v} \in \lbrace 0,1\rbrace^N$ be a binary variable vector that is the result from mapping each coefficient of  $\vecb{s}$ to binary value $0$ and $1$, \ie, if  ${s}_i > 0$, then $v_i = 1$; otherwise, $v_i = 0$. Then, we exploit Hadamard product properties to extract $\vecb{v}$ by transforming the following terms:  
\begin{equation}
\begin{aligned}
\vecb{x}_{s} ^{T}   \vecb{A}_s   ^{T}  \vecb{A}_s  \vecb{x}_{s}   &=  (\vecb{x}  \odot \vecb{v})^{T} \vecb{A}^T \vecb{A} (\vecb{x} \odot \vecb{v})  
&&= \vecb{v}^T \vecb{X}^T \vecb{A}^{T} \vecb{A} \vecb{X}  \vecb{v} \\       \vecb{x}_{s}^{T}  \vecb{\Sigma}_{{x,s}}^{-1}   \vecb{x}_{s}  &=  (\vecb{x}  \odot \vecb{v})^{T} \vecb{\Sigma}_{x}^{-1} (\vecb{x} \odot \vecb{v}) 
&&= \vecb{v}^T \vecb{X}^T \vecb{\Sigma}_{x}^{-1}\vecb{X}  \vecb{v}. \\
\frac{1} {\sigma_n}  \vecb{y}^{T}  \vecb{A}_s \vecb{x}_s  &=   \frac{1} {\sigma_n}  \vecb{y}^{T}   \vecb{A}  ( \vecb{x} \odot \vecb{v} ) 
&&= \frac{1} {\sigma_n}  \vecb{y}^{T}   \vecb{A} \vecb{X} \vecb{v}.   
  \end{aligned}
\label{sub_eq_hardarmard}
\end{equation}
Using the Hadarmard property, we decompose $\vecb{v}$ from $(\sigma_n \vecb{I}  +   \vecb{A}_s \vecb{\Sigma}_{{x,s}}  \vecb{A}_s^{T})$  as~\cite{Peleg2012}  
\begin{equation}
\begin{aligned}
\sum_{i \in \mathcal{V} }    \log [(\sigma_n \vecb{\Sigma}_{{x}}^{-1}   +     \vecb{A}^{T} \vecb{A} )]_{i,i}   
=  \sum_{i = 1 }^N  v_i \log [(\sigma_n   \vecb{\Sigma}_{{x}}^{-1}   +  \vecb{Q} )]_{i,i} ,      \end{aligned}
\label{sup_eq_log}
\end{equation}
where $\vecb{Q}$  is a diagonal matrix whose entries are  the diagonal entries of  $\vecb{A}^{T} \vecb{A}$. Combining all the transformations Eq.~\eqref{sub_eq_hardarmard} and Eq.~\eqref{sup_eq_log}, we obtain the MAP problem as in Eq.\eqref{eq1.7.1}.
 
\subsection{Optimization over  $\sigma_n$}   In this section, we provide derivation of the update formulation  Eq.~\eqref{eq1.9.5} for $\sigma_n$. From the sub-optimization over $\sigma_n$ Eq.\eqref{eq1.9}. Let $\vecb{\lambda} = \sigma_n \vecb{1}$ be a vector where each element is noise variance $\sigma_n$. Given  $\vecb{\Sigma}_{x,s}$, $\vecb{x}$, and $\vecb{s}$, the optimization Eq.\eqref{eq1.9}  is reformulated  as
\begin{equation}
\begin{aligned}
\min\limits_{\vecb{\lambda} } 
& \frac{1}{2 \sigma_n} ||\vecb{y} -   \vecb{A}_s \vecb{x}_s  ||^2   + \frac{1}{2} \log{| \textbf{diag}\lbrace \vecb{\lambda} \rbrace +   \vecb{A}_s \vecb{\Sigma}_{x,s} \vecb{A}_s^{T}| }.
\end{aligned}
\label{sub_eq1.9}
\end{equation}
The concave function $h(\vecb{\lambda}) = \log{|\textbf{diag}\lbrace \vecb{\lambda} \rbrace + \vecb{A}_s  \vecb{\Sigma}_{{x,s}}    \vecb{A}_s^{T}| } $  is transformed into a convex function which is its upper bound, using a conjugate function. Let  $h^{*}(\vecb{\lambda})$ be the concave conjugate function of $h(\vecb{\lambda})$  as follows: 
\begin{equation}
\begin{aligned}
h( \vecb{\lambda} ) = \log{| \textbf{diag}\lbrace \vecb{\lambda} \rbrace + \vecb{A}_s  \vecb{\Sigma}_{{x,s}}    \vecb{A}_s^{T}| }  
\leq  \vecb{\eta}^{T} \vecb{\lambda}    -   h^{*}(\vecb{\lambda} ),  \forall \mathbf{\eta} \geq 0. 
\end{aligned}
\label{sub_eq1.9.1}
\end{equation} 
Eq.~\eqref{sub_eq1.9.1} holds when 
\begin{equation} 
\begin{aligned}
\eta_k 
&= \nabla_{\lambda_k } 
\log{|\textbf{diag}\lbrace \vecb{\lambda} \rbrace  + \vecb{A}_s  \vecb{\Sigma}_{{x,s}}   \vecb{A}^{T}_s| }
\\
&= \Tr { \left[  
\mathbf{e}_k^{T}( \textbf{diag}\lbrace \vecb{\lambda} \rbrace + \vecb{A}_s  \vecb{\Sigma}_{{x,s}}   \vecb{A}^{T}_s  )^{-1} 
\mathbf{e}_k  \right] } .
\end{aligned}\label{sub_eq1.9.2} 
\end{equation}
Thus, we have $\vecb{\eta} =  \textbf{diag} \lbrace ( \textbf{diag}\lbrace \vecb{\lambda} \rbrace  + \vecb{A}_s  \vecb{\Sigma}_{{x,s}}   \vecb{A}^{T}_s ) ^{-1}  \rbrace.$
 
Substituting \eqref{sub_eq1.9.1} into \eqref{sub_eq1.9}, we obtain the following reformulated sub-problem over $\mathbf{\sigma}_n$: 
\begin{equation}
\begin{aligned}
\min\limits_{ \vecb{\lambda} } 
\frac{1}{\sigma_n}( \vecb{y} - \vecb{A}_s \vecb{x}_s)^{T} ( \vecb{y} - \vecb{A}_s \vecb{x}_s)  + \vecb{\eta}^{T} \vecb{\lambda}
= \sum_{i=1}^{M} \left( \frac{d_i^2}{ {\lambda}_i } +  {\lambda}_i  {\eta}_i \right),  
\end{aligned}
\label{eq1.9.4}
\end{equation}
where $d_i$ denotes the $i-$th entry of $\vecb{d} =   \vecb{y}-  \vecb{A}_s \vecb{x}_s.$ Because ${\vecb{\lambda} } >  0$, we obtain $  {\lambda}_i^{new} = \sqrt{\frac{  d_i^2}{ \eta_{i}}}.$ Thus, $\sigma_n^{new} = \frac{1}{M} \sum_{i=1}^{M} \sqrt{\frac{  d_i^2}{ \eta_{i}}}.$

\subsection{Optimization over $\vecb{\Sigma}_{x,s}$}   
  
In this section, we provide derivation of the update formulation Eq.~\eqref{eq1.8.1.7} for $\vecb{\Sigma}_{x,s}$.  From the sub-optimization over $\vecb{\Sigma}_{x}$ Eq. \eqref{eq1.8}, we let $\vecb{\nu}$ be a vector of the diagonal entry in $\vecb{\Sigma}_{x}$.  Given $\vecb{x},\vecb{s}$, and $\sigma_n$, we have the following optimization problem over $\vecb{\Sigma}_{x}$ 
\begin{equation}
\begin{aligned}
\min\limits_{\vecb{\nu}}
\frac{1}{2}  \vecb{x}^{T}   \vecb{\Sigma}_{x}^{-1} \vecb{x}
+ \frac{1}{2} \log{|\sigma_n \vecb{I}  +  \vecb{A'}  \vecb{\Sigma}_{x}  \vecb{A'}^{T}|},
\end{aligned}
\label{sub_eq1.8}
\end{equation} 
where $\vecb{A'}=\vecb{AV}$ is the product between $\vecb{A}$ and  $\vecb{V}$ to suppress the columns associated with zero elements in $\vecb{x}$. The first term in \eqref{sub_eq1.8} is convex over $\vecb{\nu}$, while the second term is concave over $\vecb{\nu}$. We will transform the second term into a convex function, by, first, decomposing the logarithm term as follows: 
\begin{equation} 
\begin{aligned}
\log{|\sigma_{n} \vecb{I}  +  \vecb{A'}\vecb{\Sigma}_{x} \vecb{A'}^{T}| } 
&= \log{| \mathbf{\Sigma}_{\mathbf{x}} ^{-1}   +  \frac{1}{\sigma_{n}} \mathbf{A'}^{T} \mathbf{A'} | } \\
&+ \log{|{\sigma}_n \vecb{I} |} + \log{|   \vecb{\Sigma}_{x} |}.  
\end{aligned}\label{sub_eq1.8.2} 
\end{equation}

Let $\vecb{\beta}$ be a point-wise inverse of the vector  $\vecb{\nu}$, i.e., $ \vecb{\beta} = \vecb{\nu}^{\odot -1}$.  We use a conjugate function to find a strict upper bound of the concave function  $g(\vecb{\beta}) = \log{|\vecb{\Sigma}_{x}^{-1} + \frac{1}{{\sigma}_n} \vecb{A'}^{T} \vecb{A'} | }$, as follows, $\forall \mathbf{\alpha} \geq 0$,  
\begin{equation} 
g(\vecb{\beta})  \leq   \vecb{\alpha}^{T}\vecb{\beta} - g^{*}(\vecb{\beta}),  
\label{sub_eq1.8.3}  
\end{equation} 
where $g^{*}(\vecb{\beta})$ is the concave conjugate function of $g(\vecb{\beta})$ and $\vecb{\alpha} = [\alpha_1,...,\alpha_K]^{T}$. The equation \eqref{sub_eq1.8.3} holds when  
\begin{equation} 
\begin{aligned}
&  \vecb{\alpha}_k 
= \nabla_{\vecb{\beta}_{k}} 
\log{|\vecb{\Sigma_{x}}^{-1} + \frac{1}{{\sigma}_n}  \vecb{A'}^{T} \vecb{A'} | } \\  
& = \Tr { \left[  
\vecb{e}_k^{T}(  \vecb{\Sigma}_{x}^{-1} +  \frac{1}{{\sigma}_n} \vecb{A'}^{T} \vecb{A'} )^{-1}
\vecb{e}_k  \right] }. 
\end{aligned}\label{sub_eq1.8.4} 
\end{equation} 
Thus,   $\vecb{\alpha} = \textbf{diag} \lbrace (  \vecb{\Sigma}_{x}^{-1} +  \frac{1}{{\sigma}_n} \vecb{A'}^{T} \vecb{A'} )^{-1}  \rbrace$.   Substituting \eqref{sub_eq1.8.3} into \eqref{sub_eq1.8} and using Eq\eqref{sub_eq1.8.2}, we have the sub-problem as follows: 

\begin{equation}
\begin{aligned}
\min\limits_{\vecb{\nu}}  
 \vecb{x}^{T} \vecb{\Sigma}_{x}^{-1}\vecb{x}  
+ \vecb{\alpha}^{T}\vecb{\beta}   + &  \log{|\vecb{\Sigma}_{x}|} =  \\
& \sum_{i = 1}^{N} \left( \left(x_i^2 + {\alpha}_i  \right)  {\nu}_i^{-1}  
+    \log{  {\nu}_i} \right)  .
\end{aligned}
\label{eq1.8.1.6}
\end{equation} 
Because $ {\nu}_i> 0$, the update of $ {\nu}_i$ is  ${\nu}_i^{new} =  x_i^2 + {\alpha}_i.$

{
\bibliographystyle{IEEEtran}
\bibliography{egbib}
}

\begin{IEEEbiography}
[{\includegraphics[width=1in,height=1.25in,keepaspectratio]{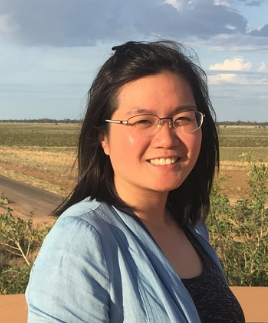}}]{Suwichaya Suwanwimolkul}
 is a Ph.D. candidate in the School of Computer Science, The University of Adelaide. She was a research engineer at Geo-Informatics and Space Technology agency, Thailand. Her research interests include image processing and machine learning.
\end{IEEEbiography}  
\vspace{-1.5cm}

\begin{IEEEbiography}
[{\includegraphics[width=1in,height=1.25in,keepaspectratio]{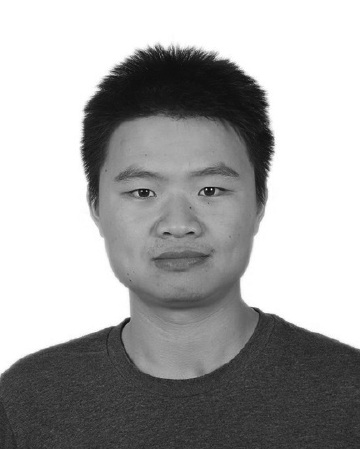}}]{Lei Zhang}
 received the Ph.D. degree in computer science and technology from Northwestern Polytechnical University, Xi'an, in 2018. He is currently a research staff in the school of computer science, the University of Adelaide, Australia. His research interests include image processing and machine learning.
\end{IEEEbiography}  

\vspace{-1.5cm}
\begin{IEEEbiography}[{\includegraphics[width=1in,height=1.25in,clip,keepaspectratio]{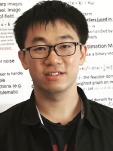}}]{Dong Gong}
received the Bachelor degree in computer science from the Northwestern Polytechnical University, Xi'an, China. He is currently pursuing the PhD degree with the School of Computer Science, Northwestern Polytechnical University. He was a joint-training Ph.D student in The University of Adelaide in 2015 and 2016. His current research interests include machine learning and optimization techniques and their applications in image processing and computer vision.
\end{IEEEbiography}  

\vspace{-1.5cm}
\begin{IEEEbiography}[{\includegraphics[width=1in,height=1.25in,clip,keepaspectratio]{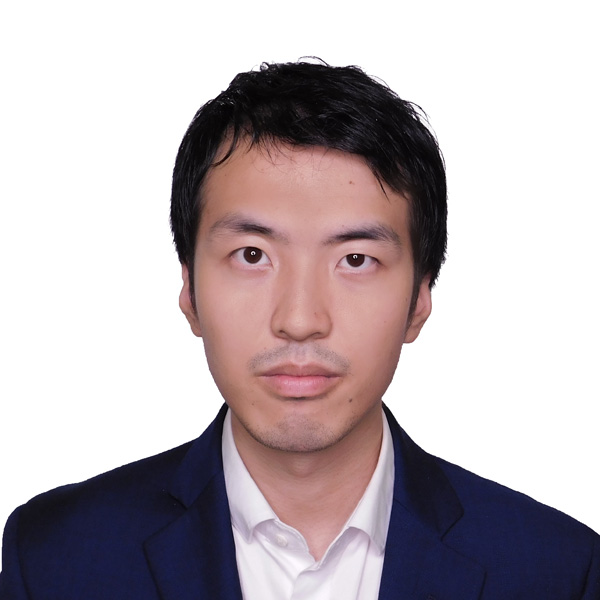}}]{Zhen Zhang} is a postdoctoral research fellow in the Department of Computer Science at the National University of Singapore under the supervision of Prof. Lee Wee Sun. His research interests are mainly computer vision and machine learning, particularly in probabilistic graphical model and its applications in computer vision. 
Previously, he received a PhD in Computer Science from Northwestern Polytechnical University (Xi'an, China) under the supervision of Prof. Yanning Zhang. During Nov. 2012 to Dec. 2014, he was a visiting student at Australian Centre for Visual Technologies, University of Adelaide, under the supervision of Prof. Anton van den Hengel and Dr. Qinfeng (Javen) Shi.
\end{IEEEbiography}

\vspace{7.5cm}
 
\begin{IEEEbiography}[{\includegraphics[width=1in,height=1.25in,clip,keepaspectratio]{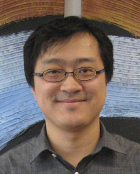}}]{Chao Chen}  is an Assistant Professor of Biomedical Informatics at the
Stony Brook University, United States. He received his B.S from Peking
University, and completed his Ph.D program at Rensselaer Polytechnic
Institute. His research interests include machine learning, biomedical
image analysis, and topological data analysis.
\end{IEEEbiography}

\vspace{-8.0cm} 
\begin{IEEEbiography}[{\includegraphics[width=1in,height=1.25in, clip,keepaspectratio]{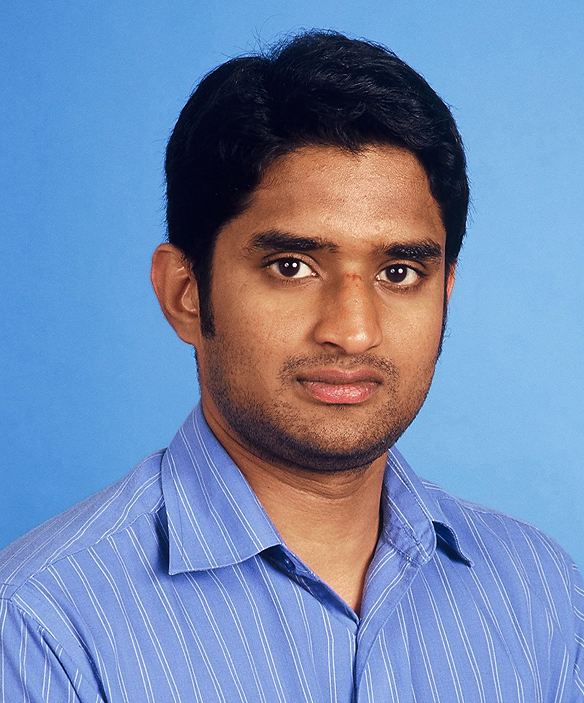}}]{Damith C. Ranasinghe} received the Ph.D. degree
in electrical and electronic engineering from The
University of Adelaide, Australia, in 2007. From
2005 to 2006, he was a Visiting Scholar with the
Massachusetts Institute of Technology and a Post-
Doctoral Research Fellow with the University of
Cambridge from 2007 to 2009. He joined The
University of Adelaide in 2010, and is currently a
Tenured Senior Lecturer with the School of Computer
Science, where he leads the Research Group with the
Adelaide Auto-ID Laboratory. His research interests
include sensing and ubiquitous computing, wearable computing, human activity recognition, and physical cryptography.
\end{IEEEbiography}

\vspace{-8.0cm}
\begin{IEEEbiography}[{\includegraphics[width=1in,height=1.25in,clip,keepaspectratio]{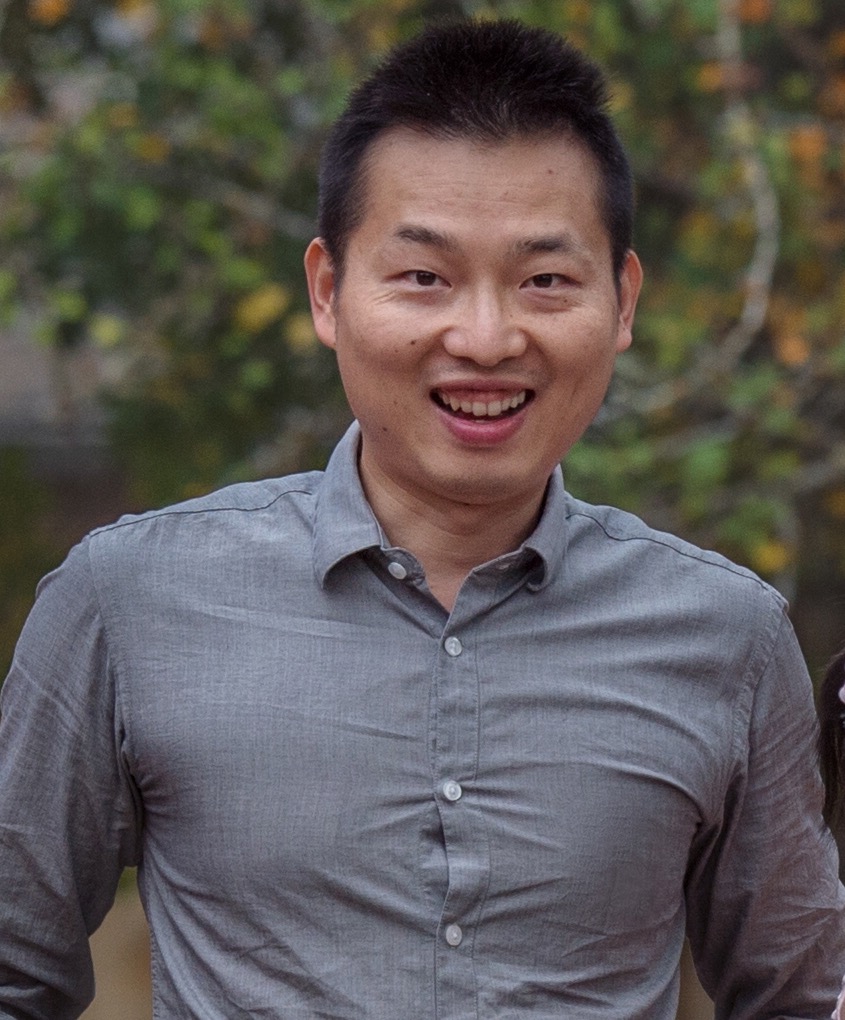}}]{Javen Qinfeng Shi}  is the director and founder of Probabilistic Graphical Model Group, and Associate Professor (Reader) at The University of Adelaide. His research interests include Machine Learning and Computer Vision, particularly Probabilistic Graphical Models, Deep Learning, and PAC-Bayes Bounds Analysis. He has published over 40 top tier papers, and has been a chief investigator (CI) for multiple Australian Research Council (ARC) grants (3 as lead CI, 2 as co-CI). He is the first ARC Discovery Early Career Researcher Awardee (DECRA) in Machine Learning (awarded in 2011, funded for 2012-2014), and has been invited twice to Prime Minister’s Science Prizes Dinner. His PGM group develops efficient algorithms and systems that can evolve and learn from data in almost any domain ranging from computer vision, water utility, health, smart agriculture, smart manufacturing (Industry 4.0), and automated trades etc. 
\end{IEEEbiography}

\end{document}